\pdfoutput=1
\documentclass[12pt,a4paper]{article}

\usepackage{ifthen} 
\newboolean{pdflatex}
\setboolean{pdflatex}{true} 

\newboolean{articletitles}
\setboolean{articletitles}{true} 

\newboolean{uprightparticles}
\setboolean{uprightparticles}{false} 

\newcommand{\X}{\theX}
\newcommand{\PII}{\psitwos}

\newcommand{\zt}{\ensuremath{z}\xspace}
\newcommand{\tz}{\ensuremath{t_z}\xspace}

\newcommand{\pz}{\ensuremath{p_z}\xspace}
\newcommand{\qtag}{\ensuremath{\text{tag}}\xspace}
\newcommand{\jet}{\ensuremath{\text{jet}}\xspace}
\newcommand{\dimu}{\ensuremath{\mu^+\mu^-}\xspace}
\newcommand{\dipi}{\ensuremath{\pi^+\pi^-}\xspace}

\newcommand{\dec}{\ensuremath{\mu^+\mu^-\pi^{+}\pi^{-}}\xspace}
\newcommand{\pp}{\ensuremath{\Pp\Pp}\xspace}
\newcommand{\akt}{\ensuremath{\text{anti-}k_\text{T}}\xspace}
\newcommand{\fastjet}{\textsc{FastJet}\xspace}
\newcommand{\vzero}{\ensuremath{V^0}\xspace}
\newcommand{\vzeros}{\ensuremath{V^0}s\xspace}
\usepackage{comment} 
\usepackage[section]{placeins} 

\def\paperauthors{LHCb collaboration} 
\def\paperasciititle{Measurements of Psi(2S) and X(3872) production within fully reconstructed jets} 
\def\papertitle{Measurements of \PII and \theX production within fully reconstructed jets} 
\def\paperkeywords{{High Energy Physics}, {LHCb}} 
\def\papercopyright{\the\year\ CERN for the benefit of the LHCb collaboration} 
\def\paperlicence{CC BY 4.0 licence}
\def\paperlicenceurl{https://creativecommons.org/licenses/by/4.0/}

\usepackage[top=1in, bottom=1.25in, left=1in, right=1in]{geometry}

\usepackage{microtype}
\usepackage{lineno}  
\usepackage{xspace} 
\usepackage{caption} 

\usepackage{graphicx}  
\usepackage{color}
\usepackage{colortbl}
\graphicspath{{./figs/}} 

\usepackage{amsmath} 
\usepackage{amssymb}
\usepackage{amsfonts}
\usepackage{upgreek} 

\newcommand*\patchAmsMathEnvironmentForLineno[1]{%
\expandafter\let\csname old#1\expandafter\endcsname\csname #1\endcsname
\expandafter\let\csname oldend#1\expandafter\endcsname\csname
end#1\endcsname
 \renewenvironment{#1}%
   {\linenomath\csname old#1\endcsname}%
   {\csname oldend#1\endcsname\endlinenomath}%
}
\newcommand*\patchBothAmsMathEnvironmentsForLineno[1]{%
  \patchAmsMathEnvironmentForLineno{#1}%
  \patchAmsMathEnvironmentForLineno{#1*}%
}
\AtBeginDocument{%
\patchBothAmsMathEnvironmentsForLineno{equation}%
\patchBothAmsMathEnvironmentsForLineno{align}%
\patchBothAmsMathEnvironmentsForLineno{flalign}%
\patchBothAmsMathEnvironmentsForLineno{alignat}%
\patchBothAmsMathEnvironmentsForLineno{gather}%
\patchBothAmsMathEnvironmentsForLineno{multline}%
\patchBothAmsMathEnvironmentsForLineno{eqnarray}%
}

\usepackage{hyperxmp}

\usepackage[pdftex,
            pdfauthor={\paperauthors},
            pdftitle={\paperasciititle},
            pdfkeywords={\paperkeywords},
            pdfcopyright={Copyright (C) \papercopyright},
            pdflicenseurl={\paperlicenceurl}]{hyperref}
\usepackage[colorinlistoftodos,textsize=scriptsize]{todonotes}

\usepackage[bottom,flushmargin,hang,multiple]{footmisc}

\usepackage[all]{hypcap} 

\usepackage{xspace} 
\usepackage{upgreek}

\def\lhcb   {\mbox{LHCb}\xspace}

\def\MagUp {\mbox{\em Mag\kern -0.05em Up}\xspace}

\ifthenelse{\boolean{uprightparticles}}%
{

 \def\Pchi        {\ensuremath{\upchi}\xspace}                 
 \def\Ppsi        {\ensuremath{\uppsi}\xspace}

 \def\PDelta      {\ensuremath{\Delta}\xspace}                 
 \def\PXi         {\ensuremath{\Xi}\xspace}                 
 \def\PLambda     {\ensuremath{\Lambda}\xspace}                 
 \def\PSigma      {\ensuremath{\Sigma}\xspace}                 
 \def\POmega      {\ensuremath{\Omega}\xspace}                 
 \def\PUpsilon    {\ensuremath{\Upsilon}\xspace}
 \let\oldPi\Pi
 \def\PPi         {\ensuremath{\oldPi}\xspace}

 \def\PB      {\ensuremath{\mathrm{B}}\xspace}                 
                  
 \def\PD      {\ensuremath{\mathrm{D}}\xspace}

 \def\PJ      {\ensuremath{\mathrm{J}}\xspace}                 
 \def\PK      {\ensuremath{\mathrm{K}}\xspace}

 \def\Pb      {\ensuremath{\mathrm{b}}\xspace}                 
 \def\Pc      {\ensuremath{\mathrm{c}}\xspace}

 \def\Pi      {\ensuremath{\mathrm{i}}\xspace}

 \def\Pp      {\ensuremath{\mathrm{p}}\xspace}

 \def\Ps      {\ensuremath{\mathrm{s}}\xspace}

 \def\thebaroffset{0.0em}
}
{

 \def\Pchi        {\ensuremath{\chi}\xspace}                 
 \def\Ppsi        {\ensuremath{\psi}\xspace}                 
                  
 \mathchardef\PDelta="7101
 \mathchardef\PXi="7104
 \mathchardef\PLambda="7103
 \mathchardef\PSigma="7106
 \mathchardef\POmega="710A
 \mathchardef\PUpsilon="7107
 \mathchardef\PPi="7105
                  
 \def\PB      {\ensuremath{B}\xspace}                 
                  
 \def\PD      {\ensuremath{D}\xspace}

 \def\PJ      {\ensuremath{J}\xspace}                 
 \def\PK      {\ensuremath{K}\xspace}

 \def\Pb      {\ensuremath{b}\xspace}                 
 \def\Pc      {\ensuremath{c}\xspace}

 \def\Pi      {\ensuremath{i}\xspace}

 \def\Pp      {\ensuremath{p}\xspace}

 \def\Ps      {\ensuremath{s}\xspace}

 \def\thebaroffset{0.18em}
}
\newcommand{\offsetoverline}[2][\thebaroffset]{\kern #1\overline{\kern -#1 #2}}%

\makeatletter
\ifcase \@ptsize \relax
  \newcommand{\miniscule}{\@setfontsize\miniscule{4}{5}}
\or
  \newcommand{\miniscule}{\@setfontsize\miniscule{5}{6}}
\or
  \newcommand{\miniscule}{\@setfontsize\miniscule{5}{6}}
\fi
\makeatother

\DeclareRobustCommand{\optbar}[1]{\shortstack{{\miniscule (\rule[.5ex]{1.25em}{.18mm})}
  \\ [-.7ex] $#1$}}

\def\squark    {{\ensuremath{\Ps}}\xspace}

\def\cquark    {{\ensuremath{\Pc}}\xspace}

\def\bquark    {{\ensuremath{\Pb}}\xspace}

\def\kaon    {{\ensuremath{\PK}}\xspace}

\def\KorKbar {\kern \thebaroffset\optbar{\kern -\thebaroffset \PK}{}\xspace}

\def\KS      {{\ensuremath{\kaon^0_{\mathrm{S}}}}\xspace}

\def\D       {{\ensuremath{\PD}}\xspace}

\def\DorDbar {\kern \thebaroffset\optbar{\kern -\thebaroffset \PD}\xspace}
\def\Dz      {{\ensuremath{\D^0}}\xspace}

\def\Dp      {{\ensuremath{\D^+}}\xspace}
\def\Dm      {{\ensuremath{\D^-}}\xspace}

\def\DpDm    {\ensuremath{\Dp {\kern -0.16em \Dm}}\xspace}

\def\Dstarpm {{\ensuremath{\D^{*\pm}}}\xspace}

\def\B       {{\ensuremath{\PB}}\xspace}

\def\BorBbar {\kern \thebaroffset\optbar{\kern -\thebaroffset \PB}\xspace}

\def\Bd      {{\ensuremath{\B^0}}\xspace}

\def\BdorBdbar {\kern \thebaroffset\optbar{\kern -\thebaroffset \Bd}\xspace}

\def\Bpm     {{\ensuremath{\B^\pm}}\xspace}

\def\Bs      {{\ensuremath{\B^0_\squark}}\xspace}

\def\BsorBsbar {\kern \thebaroffset\optbar{\kern -\thebaroffset \Bs}\xspace}

\def\jpsi     {{\ensuremath{{\PJ\mskip -3mu/\mskip -2mu\Ppsi}}}\xspace}
\def\psitwos  {{\ensuremath{\Ppsi{(2S)}}}\xspace}
\def\psiprpr  {{\ensuremath{\Ppsi(3770)}}\xspace}

\def\chic     {{\ensuremath{\Pchi_\cquark}}\xspace}

\def\Upsilonres  {{\ensuremath{\PUpsilon}}\xspace}
\def\Y#1S{\ensuremath{\PUpsilon{(#1S)}}\xspace}

\def\chib     {{\ensuremath{\Pchi_{b}}}\xspace}

\def\theX     {{\ensuremath{\Pchi_{c1}(3872)}}\xspace}

\def\Lz          {{\ensuremath{\PLambda}}\xspace}

\def\LorLbar     {\kern \thebaroffset\optbar{\kern -\thebaroffset \PLambda}\xspace}

\def\Lc          {{\ensuremath{\Lz^+_\cquark}}\xspace}

\def\to                 {\ensuremath{\rightarrow}\xspace}

\def\AT#1     {\ensuremath{A_{\mathrm{T}}^{#1}}\xspace}           

\def\C#1      {\ensuremath{\mathcal{C}_{#1}}\xspace}                       
\def\Cp#1     {\ensuremath{\mathcal{C}_{#1}^{'}}\xspace}                    
\def\Ceff#1   {\ensuremath{\mathcal{C}_{#1}^{\mathrm{(eff)}}}\xspace}        
\def\Cpeff#1  {\ensuremath{\mathcal{C}_{#1}^{'\mathrm{(eff)}}}\xspace}       
\def\Ope#1    {\ensuremath{\mathcal{O}_{#1}}\xspace}                       
\def\Opep#1   {\ensuremath{\mathcal{O}_{#1}^{'}}\xspace}                    

\newcommand{\nospaceunit}[1]{\ensuremath{\text{#1}}}       
\newcommand{\aunit}[1]{\ensuremath{\text{\,#1}}}       

\newcommand{\tev}{\aunit{Te\kern -0.1em V}\xspace}
\newcommand{\gev}{\aunit{Ge\kern -0.1em V}\xspace}
\newcommand{\mev}{\aunit{Me\kern -0.1em V}\xspace}
\newcommand{\kev}{\aunit{ke\kern -0.1em V}\xspace}
\newcommand{\ev}{\aunit{e\kern -0.1em V}\xspace}
 
\newcommand{\mevc}{\ensuremath{\aunit{Me\kern -0.1em V\!/}c}\xspace}
\newcommand{\gevc}{\ensuremath{\aunit{Ge\kern -0.1em V\!/}c}\xspace}
\newcommand{\mevcc}{\ensuremath{\aunit{Me\kern -0.1em V\!/}c^2}\xspace}
\newcommand{\gevcc}{\ensuremath{\aunit{Ge\kern -0.1em V\!/}c^2}\xspace}

\def\mm   {\aunit{mm}\xspace}

\def\mum  {\ensuremath{\,\upmu\nospaceunit{m}}\xspace}

\def\fb   {\ensuremath{\aunit{fb}}\xspace}
\def\invfb   {\ensuremath{\fb^{-1}}\xspace}

\def\ps   {\ensuremath{\aunit{ps}}\xspace}

\def\gsim{{~\raise.15em\hbox{$>$}\kern-.85em
          \lower.35em\hbox{$\sim$}~}\xspace}
\def\lsim{{~\raise.15em\hbox{$<$}\kern-.85em
          \lower.35em\hbox{$\sim$}~}\xspace}

\def\sqs   {\ensuremath{\protect\sqrt{s}}\xspace}

\def\pt         {\ensuremath{p_{\mathrm{T}}}\xspace}

\def\ptot       {\ensuremath{p}\xspace}

\def\evtgen     {\mbox{\textsc{EvtGen}}\xspace}

\def\geant      {\mbox{\textsc{Geant4}}\xspace}

\def\photos     {\mbox{\textsc{Photos}}\xspace}

\def\pythia     {\mbox{\textsc{Pythia}}\xspace}

\def\tell1  {TELL1\xspace}
\def\ukl1   {UKL1\xspace}

\newcommand{\lhcborcid}[1]{\href{https://orcid.org/#1}{\hspace*{0.1em}\raisebox{-0.45ex}{\includegraphics[width=1em]{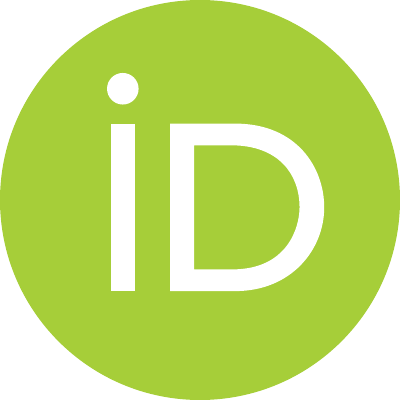}}}}

\usepackage{cite} 
\usepackage{mciteplus}

\usepackage{multirow}

\usepackage{longtable} 
\usepackage[capitalise]{cleveref}

\begin{document}

\renewcommand{\thefootnote}{\fnsymbol{footnote}}
\setcounter{footnote}{1}


\begin{titlepage}
\pagenumbering{roman}

\vspace*{-1.5cm}
\centerline{\large EUROPEAN ORGANIZATION FOR NUCLEAR RESEARCH (CERN)}
\vspace*{1.5cm}
\noindent
\begin{tabular*}{\linewidth}{lc@{\extracolsep{\fill}}r@{\extracolsep{0pt}}}
\ifthenelse{\boolean{pdflatex}}
{\vspace*{-1.5cm}\mbox{\!\!\!\includegraphics[width=.14\textwidth]{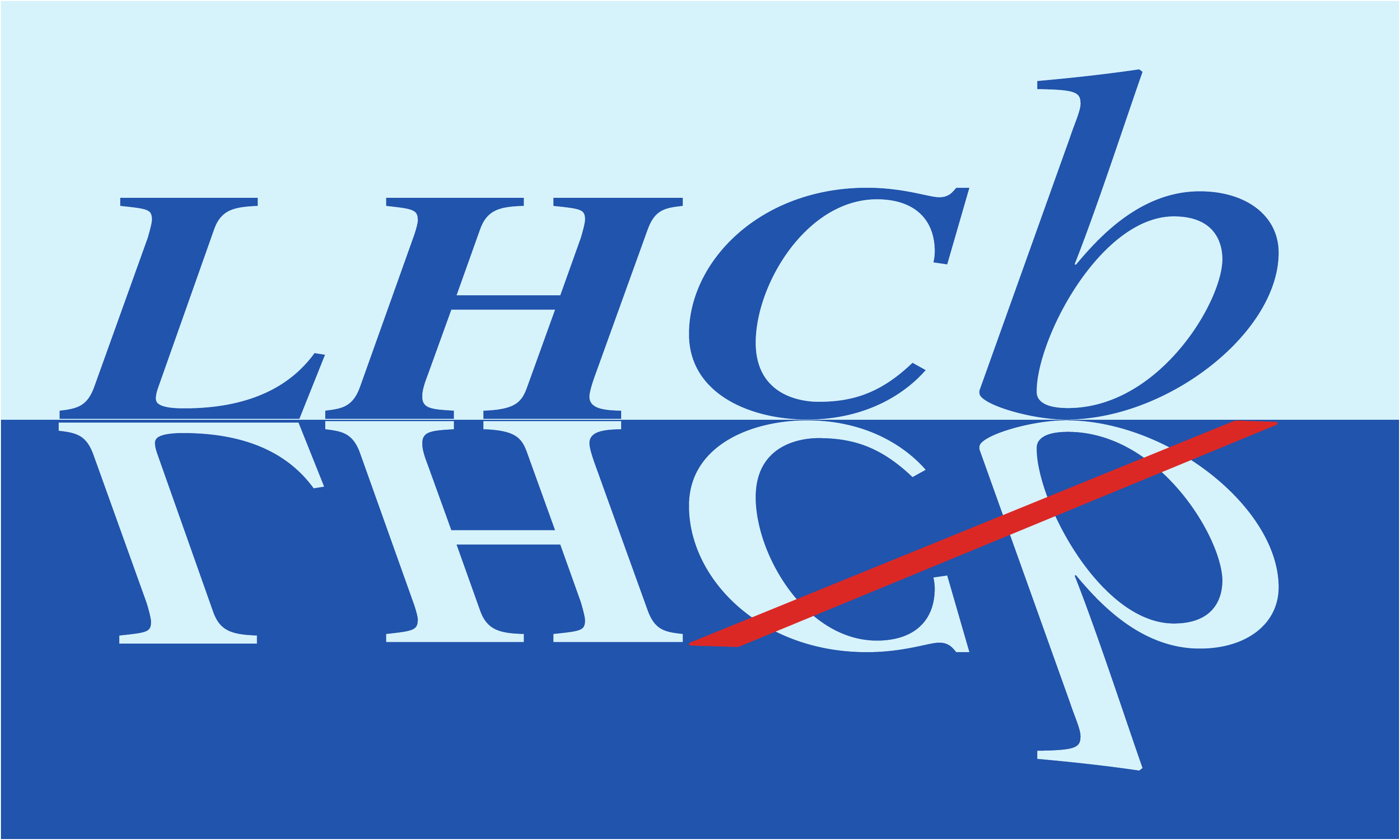}} & &}%
{\vspace*{-1.2cm}\mbox{\!\!\!\includegraphics[width=.12\textwidth]{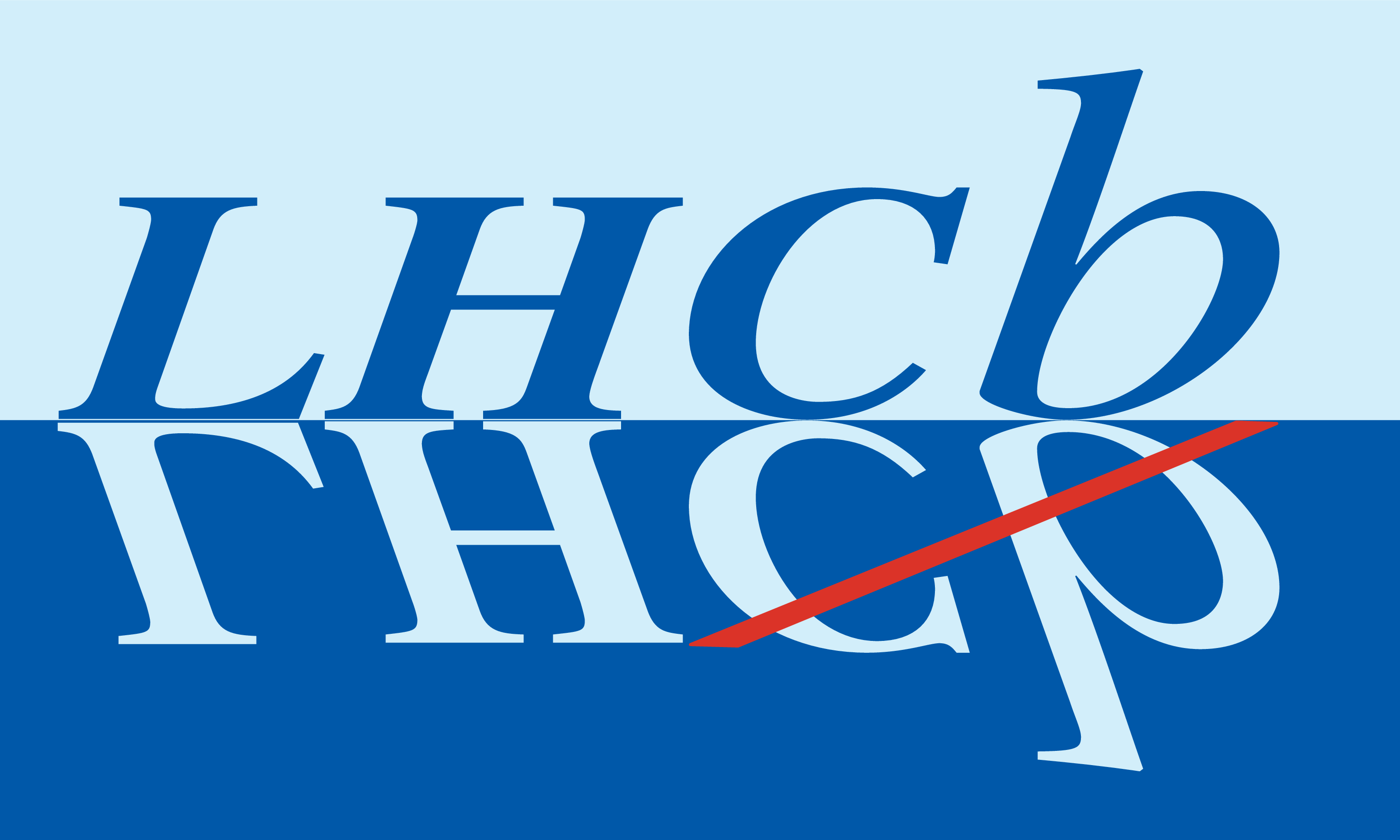}} & &}%
\\
 & & CERN-EP-2024-241 \\  
 & & LHCb-PAPER-2024-021 \\  
 & & \today \\ 
 & & \\
\end{tabular*}

\vspace*{4.0cm}

{\normalfont\bfseries\boldmath\huge
\begin{center}
  \papertitle 
\end{center}
}

\vspace*{1.5cm}

\begin{center}
\paperauthors\footnote{Authors are listed at the end of this paper.}
\end{center}

\vspace{\fill}

\begin{abstract}
  \noindent This paper presents the first measurement of \psitwos and \theX meson production within fully reconstructed jets. Each quarkonium state (tag) is reconstructed via its decay to the $\jpsi$($\rightarrow$\dimu)$\pi^+\pi^-$ final state in the forward region using proton-proton collision data collected by the LHCb experiment at the center-of-mass-energy of $13\tev$ in 2016, corresponding to an integrated luminosity of $1.64\invfb$. The fragmentation function, presented as the ratio of the quarkonium-tag transverse momentum to the full jet transverse momentum ($\pt(\qtag)/\pt(\jet)$), is measured differentially in $\pt(\jet)$ and $\pt(\qtag)$ bins. The distributions are separated into promptly produced quarkonia from proton-proton collisions and quarkonia produced from displaced $b$-hadron decays. While the displaced quarkonia fragmentation functions are in general well described by parton-shower predictions, the prompt quarkonium distributions differ significantly from fixed-order non-relativistic QCD (NRQCD) predictions followed by a QCD parton shower.
\end{abstract}

\vspace*{2.0cm}

\begin{center}
  Published in The European Physical Journal C 85 (2025) 562
\end{center}

\vspace{\fill}

{\footnotesize 
\centerline{\copyright~\papercopyright. \href{\paperlicenceurl}{\paperlicence}.}}
\vspace*{2mm}

\end{titlepage}


\newpage
\setcounter{page}{2}
\mbox{~}
%
%
%
%


\renewcommand{\thefootnote}{\arabic{footnote}}
\setcounter{footnote}{0}

\cleardoublepage


\pagestyle{plain} 
\setcounter{page}{1}
\pagenumbering{arabic}


\section{Introduction}\label{sec:intro}

The production of quarkonium states, flavourless mesons composed of at least a $c$ or $b$ quark and its corresponding antiquark, provides a unique probe for studying quantum 
chromodynamics (QCD)~\cite{Martin:166575}. The masses of the observed quarkonium states, ranging from $3$ to $11\gevcc$, cover the transition between the perturbative and nonperturbative regimes of QCD\@. 
In proton-proton (\pp) collisions at the LHC, these quarkonium states can be produced at center-of-mass energies which are orders of magnitude larger than their relatively low masses. Consequently, any model attempting to successfully predict the production of quarkonia must span a wide range of energy scales.
\par
Extensive differential cross-section measurements have been performed for the production of the \jpsi meson, 
the lightest vector quarkonium state, 
in hadron-hadron collisions at a variety of experiments~\cite{Clark:1978mg, Ueno:1978vr, UA1:1987azc, UA1:1988pyx, UA1:1990eni, CDF:1997uzj, CDF:2004jtw, D0:1996awi, PHENIX:2011gyb} including those at the LHC~\cite{LHCb-PAPER-2015-037, LHCb-PAPER-2012-039, LHCb-PAPER-2011-003, ALICE:2012vup,ATLAS:2011aqv,CMS:2010nis}. Production cross-sections for the heavier excited states such as the \psitwos, \psiprpr, \chic, \Upsilonres, and \chib resonances have also been measured at the LHC~\cite{LHCb-PAPER-2018-049,ATLAS:2015zdw,ATLAS:2023qnh,CMS:2015lbl,ATLAS:2014zpz,ATLAS:2016kwu,LHCb-PAPER-2011-019,ATLAS:2014ala,CMS:2012qwg,LHCb-PAPER-2018-002,LHCb-PAPER-2013-016,20119,PhysRevD.87.052004,PhysRevD.83.112004,2013101,CMS:2017dju}. Many of these measurements can be described by nonrelativistic QCD (NRQCD)~\cite{Bodwin:1994jh, Cho:1995vh, Cho:1995ce}, an effective field theory where the production of the quarkonium state is factorized into a perturbative (short-distance) and nonperturbative (long-distance) component.
\par
However, most NRQCD-based calculations for quarkonia production at the Tevatron and LHC~\cite{Campbell:2007ws, Lansberg:2008gk, Gong:2008sn, Leibovich:1996pa, Braaten:2000gw, Beneke:1996yw, Braaten:1999qk, Gong:2012ug, Chao:2012iv} predict large transverse polarization of the \jpsi and \psitwos mesons, which is not observed in data~\cite{LHCb-PAPER-2013-008, LHCB-PAPER-2013-067, ALICE:2011gej, CDF:2007msx, CMS:2013gbz, ALICE:2018crw, ALICE:2020iev}, with similar results observed at RHIC~\cite{STAR:2020igu,PHENIX:2020dqu}. Indeed, 
NRQCD fails to simultaneously describe all measured quarkonium observables~\cite{Lansberg:2019adr, Chung:2018lyq}. Additionally, fixed-order NRQCD predicts \jpsi mesons to be produced mostly isolated, while jet fragmentation measurements at the LHC demonstrate that \jpsi mesons are predominantly produced surrounded by nearby particles, within jet radii of $R=0.3$ and $R=0.5$~\cite{LHCb-PAPER-2016-064,CMS:2019ebt,CMS:2021puf}. Developments in calculating quarkonium fragmentation functions~\cite{Bain:2016rrv, Bain:2017wvk} and quarkonium production in parton showers~\cite{Cooke:2023ldt} may help resolve these 
isolation discrepancies observed in jets while also consistently describing other observables. 
\par
This incomplete knowledge about quarkonia production mechanisms has wide implications on the study of the medium formed in ultrarelativistic heavy-ion collisions~\cite{Matsui:1986dk}. Indeed, quarkonia not produced from particle decays are commonly expected to form in the very early stages of particle collisions and so they 
are regularly used as probes of the hot and dense medium created in heavy-ion collisions.
\par
The reconstruction of identified particles within fully reconstructed jets is a unique way to study the underlying mechanisms involved in the production of these particles. Despite these prospects, only a few measurements of jet fragmentation functions of 
identified particles have been performed. Among them are studies of \Dz~\cite{ALICE:2022mur,ALICE:2019cbr}, \Dstarpm~\cite{UA1:1990mkp, PhysRevD.79.112006,ATLAS:2011chi,H1:2011myz,ZEUS:2008axw} and 
\Bpm mesons~\cite{ATLAS:2021agf}, and \Lc baryons~\cite{ALICE:2023jgm}. The production of quarkonia as components of jets remains largely unexplored and provides an important handle on understanding the mechanisms behind quarkonia production in hadron-hadron collisions. To date, measurements have been limited to the \jpsi state only~\cite{LHCb-PAPER-2016-064, CMS:2021puf}. 
In this paper, for the first time, measurements of fragmentation functions of \psitwos and \theX mesons are presented.
Both mesons (referred to as tags from here on) are reconstructed using the $\jpsi$($\rightarrow$\dimu)$\pi^+\pi^-$ decay channel. 
The fragmentation functions are measured through the transverse momentum ratio $\zt \equiv \pt(\qtag)/\pt(\jet)$, with \pt the component of the 
momentum transverse to the beam. The fraction of prompt \psitwos states that are produced from feed-down of higher-mass resonances is significantly smaller compared to the \jpsi meson~\cite{Faccioli:2008ir,ATLAS:2014zpz,LHCb-PAPER-2011-045}, making it an ideal candidate for the theoretical interpretation of 
quarkonium formation. 
These measurements are performed differentially in both the transverse momentum of the quarkonium, $\pt(\qtag)$, and of the jet, $\pt(\jet)$.

\section{Detector and dataset}\label{sec:data}

The \lhcb detector~\cite{LHCb-DP-2008-001,LHCb-DP-2014-002} is a single-arm forward spectrometer 
covering the \mbox{pseudorapidity} range $2.0<\eta <5.0$, designed for the study of particles containing \bquark 
and \cquark quarks. The detector includes a high-precision tracking system consisting of a silicon-strip 
vertex detector surrounding the $pp$ interaction region, a large-area silicon-strip detector located upstream of 
a dipole magnet with a bending power of about $4\,{\mathrm{T\,m}}$, and three stations of 
silicon-strip detectors and straw drift tubes placed downstream of the magnet. The tracking system measures the 
momentum, \ptot, of charged particles with a relative uncertainty that varies from $0.5\%$ at low momentum to 
$1.0\%$ at 200\gevc. The minimum distance of a track to a primary \pp collision vertex (PV), called the impact 
parameter (IP),  is measured with a resolution of $(15+29/\pt)\mum$, where \pt is in units of\,\gevc. Different types of charged hadrons are distinguished using 
information from two ring-imaging Cherenkov detectors (RICH). Photons, electrons and hadrons are identified by a 
calorimeter system consisting of scintillating-pad and preshower detectors, an electromagnetic and a hadronic calorimeter. Muons are identified by a system composed of alternating layers of iron and multiwire proportional chambers~\cite{LHCB-DP-2012-002}.

The data sample used in this analysis is collected from \pp collisions at a center-of-mass energy of $\sqs = 13\tev$, and corresponds to an integrated luminosity of $1.64\invfb$ recorded during 
2016. The online event selection is performed by a trigger~\cite{LHCb-DP-2012-004} that consists of a hardware stage using information from the calorimeter and muon systems, followed 
by a software stage that performs the $\mu^+\mu^-$ candidate reconstruction. The hardware stage selects events with at least one \dimu candidate with \mbox{$\sqrt{\pt(\mu^+)\pt(\mu^-)} 
> 1.3-1.5\gevc$}, where the threshold varied during the 2016 data taking period. In the software stage, two oppositely charged muon candidates with $\pt(\mu) > 
0.5\gevc$ are required to form a \dimu candidate with a mass within $\pm\,120\mevcc$ of the known \jpsi mass~\cite{PDG2024}.

This analysis uses the same data-taking scheme that was first introduced in Ref.~\cite{LHCb-PAPER-2016-064}. Here, offline-quality track reconstruction~\cite{Aaij:2016rxn} is achieved in the online system through real-time alignment and 
calibration~\cite{LHCb-PROC-2015-011}. All fully online reconstructed high-level objects such as particle candidates are retained, while the lower-level detector information is discarded. At the high-level trigger (HLT) stage, this strategy allows the online selection of this analysis to record all events with a \dimu candidate consistent with a \jpsi meson, but with no $\pt$ or displacement requirements on the \dimu candidate with respect to the PV.

Simulation is required to model the effects of the detector response and the imposed selection 
requirements on the \dec final state. In the simulation, \pp collisions are generated using \pythia~8.186~\cite{Sjostrand:2007gs} with a specific \lhcb configuration~\cite{LHCb-PROC-2010-056}. Decays of unstable particles are described by \evtgen~\cite{Lange:2001uf}, in which 
final-state radiation is generated using \photos~\cite{davidson2015photos}. The interaction of the generated particles with the detector, and its response, are implemented using the \geant 
toolkit~\cite{Allison:2006ve, *Agostinelli:2002hh} as described in Ref.~\cite{LHCb-PROC-2011-006}.

\section{Event selection}\label{sec:sel}

The event selection requires two distinct objects, the quarkonium tag and the jet associated with that tag. 
Because the tag is a constituent of its associated jet, all possible tags in the event must be selected prior 
to constructing the jets in the event. Both the \psitwos and \theX meson constituents are reconstructed using 
the \dec final state, which is expected to be produced primarily from the intermediate resonance decay $\jpsi 
\to \mu^+\mu^-$ in association with two charged pions. The muon and pion candidates are selected from well-reconstructed charged tracks. Tracks matched with hits in the muon stations 
downstream of the calorimeters that satisfy additional muon-identification criteria from the calorimeters and 
RICH are identified as muon candidates. Tracks without muon station hits that fulfil pion-identification criteria are taken as charged pion candidates. Both the muon and pion candidates must be 
reconstructed within the fiducial region of this 
analysis: their pseudorapidity range is limited to $2.0 < \eta < 4.5$, their transverse momenta to $\pt > 
0.5\gevc$, the momenta of the muons are limited to $p(\mu) > 6\gevc$, and the momenta of the pions to 
$p(\pi) > 3\gevc$.
\par
The \dimu candidates that pass the online criteria should be consistent with a \jpsi meson. Additionally, the two muons originating from the \jpsi candidate must form a good-quality vertex with a distance-of-closest approach (DOCA) between the muon candidates less than 
$0.2\mm$. If the \jpsi candidate is located within \mbox{$2.0 < \eta(\jpsi) < 4.5$}, it is combined 
with two opposite-sign pion candidates to form a \psitwos or \theX tag candidate. 
The DOCA between the \jpsi candidate and 
both pion candidates must be less than $0.5\mm$. In the first step, tag candidates with a mass $>4100\mevcc$ are discarded from further analysis. 
For the remaining candidates, the momentum vectors of the four decay particles are fitted, constraining all four tracks to a common vertex and the 
mass of the \dimu candidate to the known mass of the \jpsi meson~\cite{PDG2024}. This significantly improves the mass resolution of the tag candidates.
If the refitted mass of a tag candidate, $m(\text{tag})$, is \mbox{$3635 < m(\text{tag}) <3730\mevcc$}, the tag is considered to be a 
\psitwos candidate. Alternatively, if the mass is \mbox{$3830 < m(\text{tag}) < 3920\mevcc$} and the decay energy release ($Q$-value), \mbox{$Q 
\equiv m(\dec) - m(\dimu) - m(\dipi) < 150\mevcc$}, the tag is considered a \theX candidate. Requiring this $Q$-value significantly improves the signal-to-background ratio of the \theX selection~\cite{LHCb-PAPER-2021-045}. The fiducial volume of the tags is limited to $2.0 < \eta(\qtag) < 4.5$ with $\pt(\qtag) > 2\gevc$.
\par
After all possible tag candidates are selected in the event, jets are built using charged and neutral particle 
candidates, all reconstructed online with the full detector 
information, using a particle-flow algorithm~\cite{LHCb-PAPER-2013-058}. The constituent \dec tracks of the tag 
candidates are excluded from this particle-flow input, and instead the tag candidate is considered a single object. 
If more than one \dec combination within the jet radius can form a viable tag candidate, one combination is pseudo-randomly selected and substituted for a single tag object, while the other combinations are left as individual particles in the particle flow for the full jet reconstruction.
The \akt sequential clustering algorithm~\cite{Cacciari:2008gp} with jet radius parameter $R = 0.5$, as implemented in the \fastjet package~\cite{Cacciari:2011ma}, is applied to this particle-flow input to produce jets. These jets are required to fall within the fiducial region of $2.5 < \eta(\jet) < 4.0$ with $\pt(\jet) > 5\gevc$.
\par
For events that contain a jet with one tag constituent, where the tag and the jet fulfil all the selection requirements, a final selection is applied that requires exactly one reconstructed primary vertex per event. 
Unlike charged particle-flow inputs, neutral 
inputs cannot be associated with a primary vertex. By requiring a single vertex, the jet candidate is not contaminated with 
neutral particle-flow input from other \pp interactions. This significantly improves the jet energy resolution and reduces systematic 
uncertainties arising from modelling the detector response.

\section{Signal determination}\label{sec:fits}

This analysis relies on the reconstruction of the transverse-momentum fraction of the tag with respect to the jet containing the tag, defined as
\begin{equation}
    \zt \equiv \frac{\pt(\qtag)}{\pt(\jet)}.
\end{equation}
The distribution of this ratio is measured differentially in intervals of $\pt(\qtag)$ and $\pt(\jet)$, and the distributions are presented separately for tags originating from the primary vertex of the event (prompt) and tags produced from a \bquark-hadron decay (displaced). 
\par
\begin{figure}
    \includegraphics[width=0.5\textwidth]{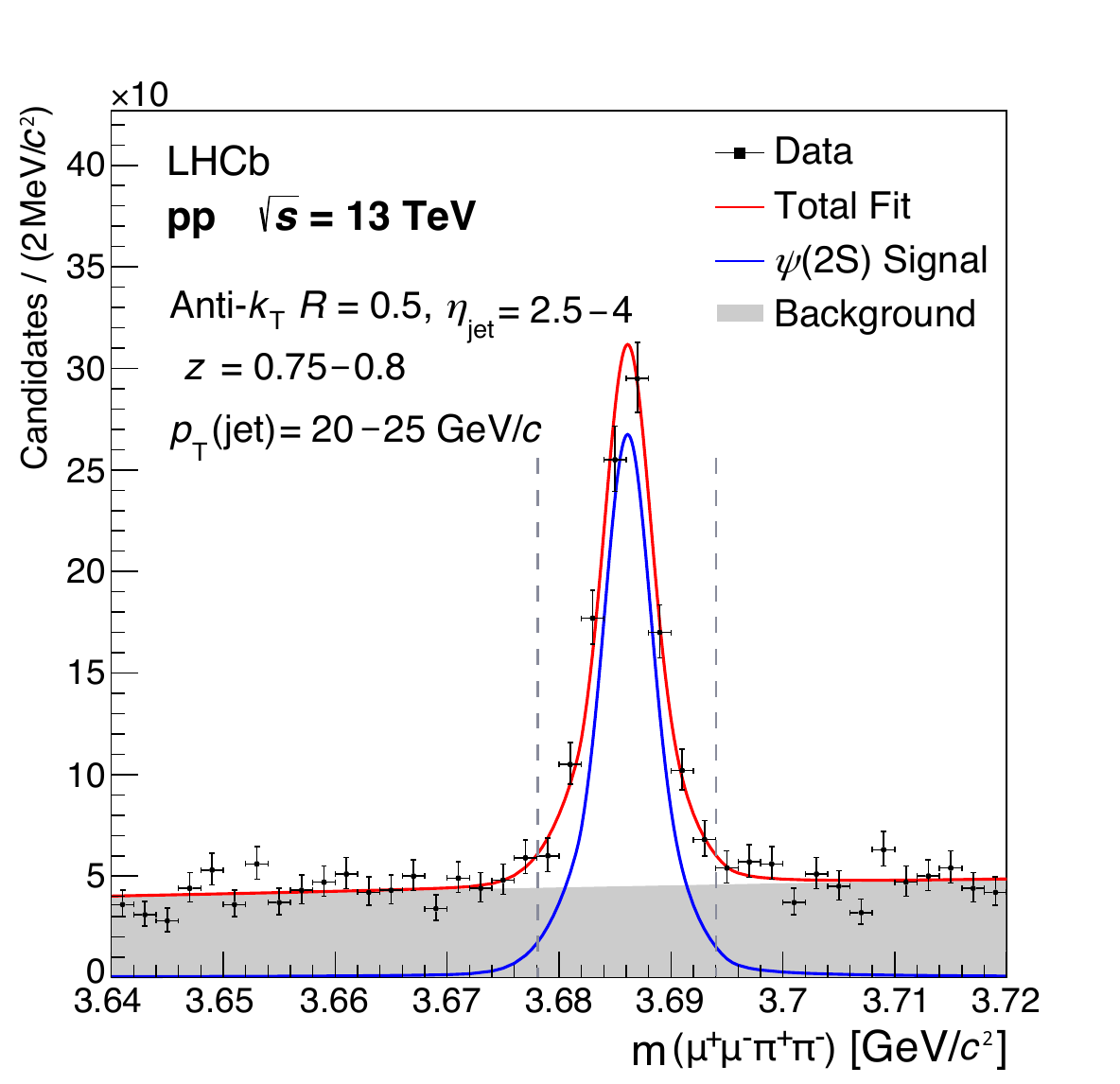}
    \includegraphics[width=0.5\textwidth]{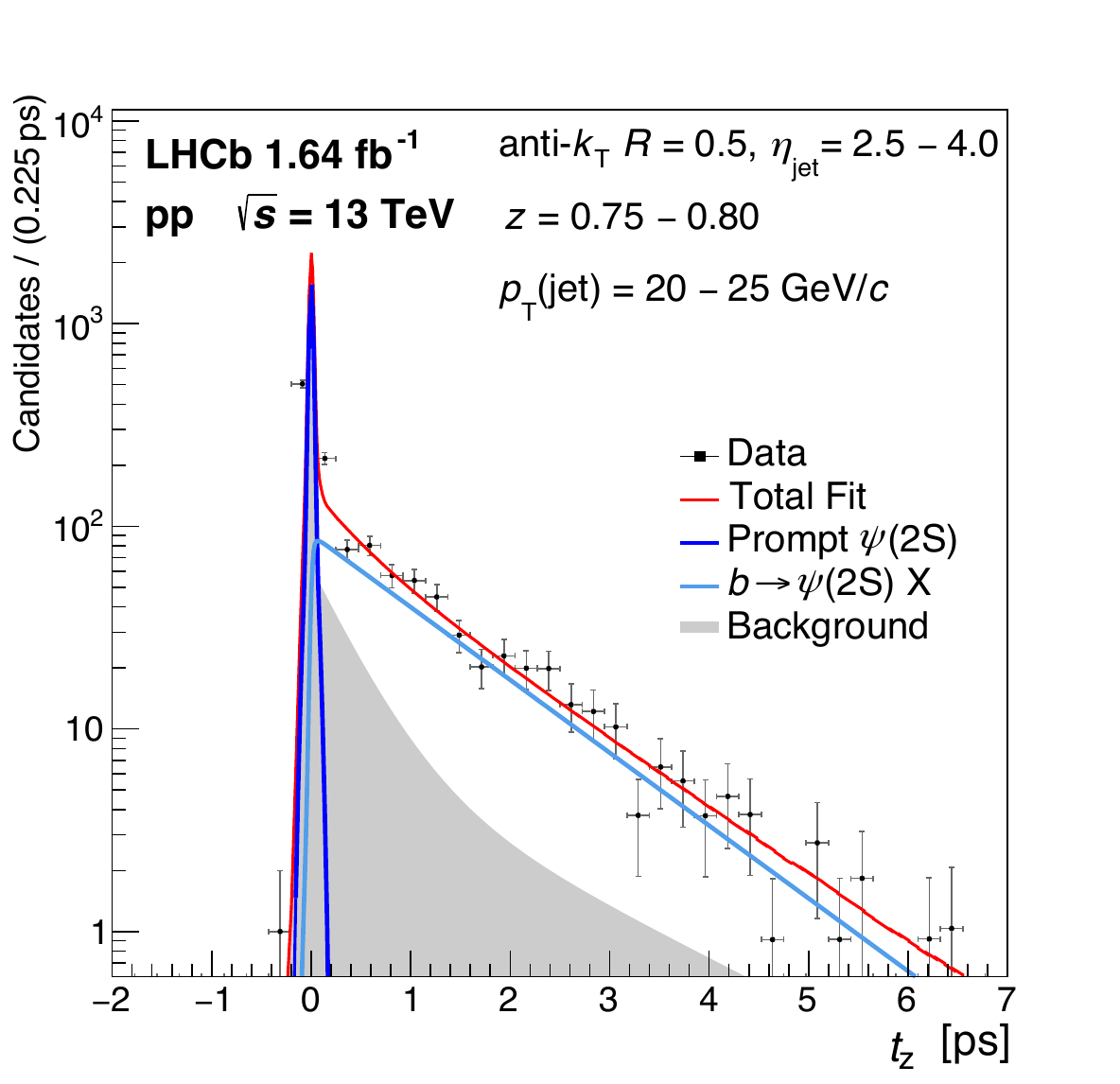}
    \caption{(Left) Invariant-mass distribution and (right) decay-time distribution for \ensuremath{\Ppsi{(2S)}} in a single \ensuremath{z}--\ensuremath{p_{\mathrm{T}}}(\jet) interval at mid \ensuremath{z}, superimposed with the results from unbinned maximum-likelihood fits. The displaced \ensuremath{\Ppsi{(2S)}} fraction in this interval is $(56.4\pm2.2)\%$.\label{fig:massPsi}}
\end{figure}
\begin{figure}
    \includegraphics[width=0.5\textwidth]{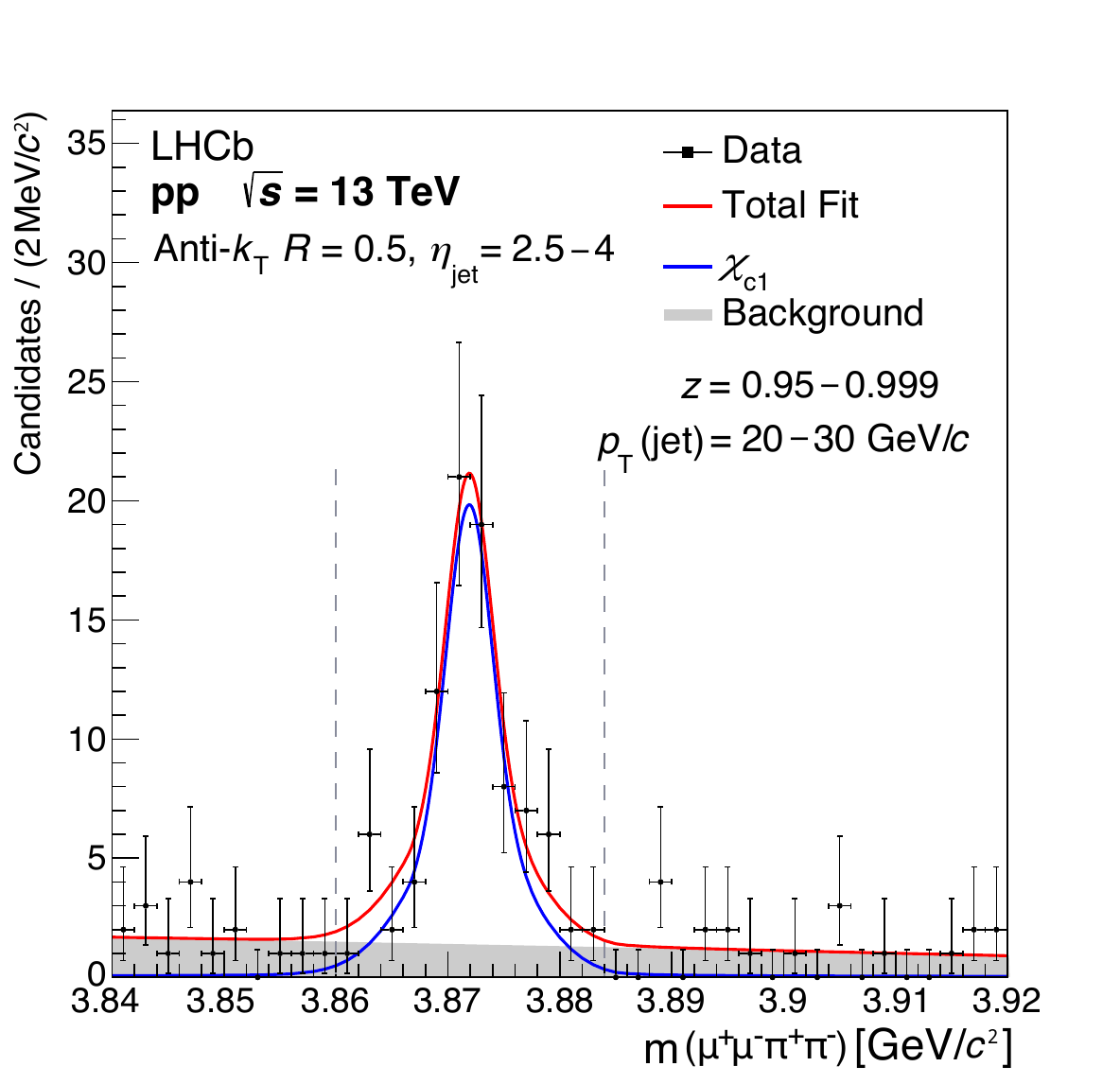}
    \includegraphics[width=0.5\textwidth]{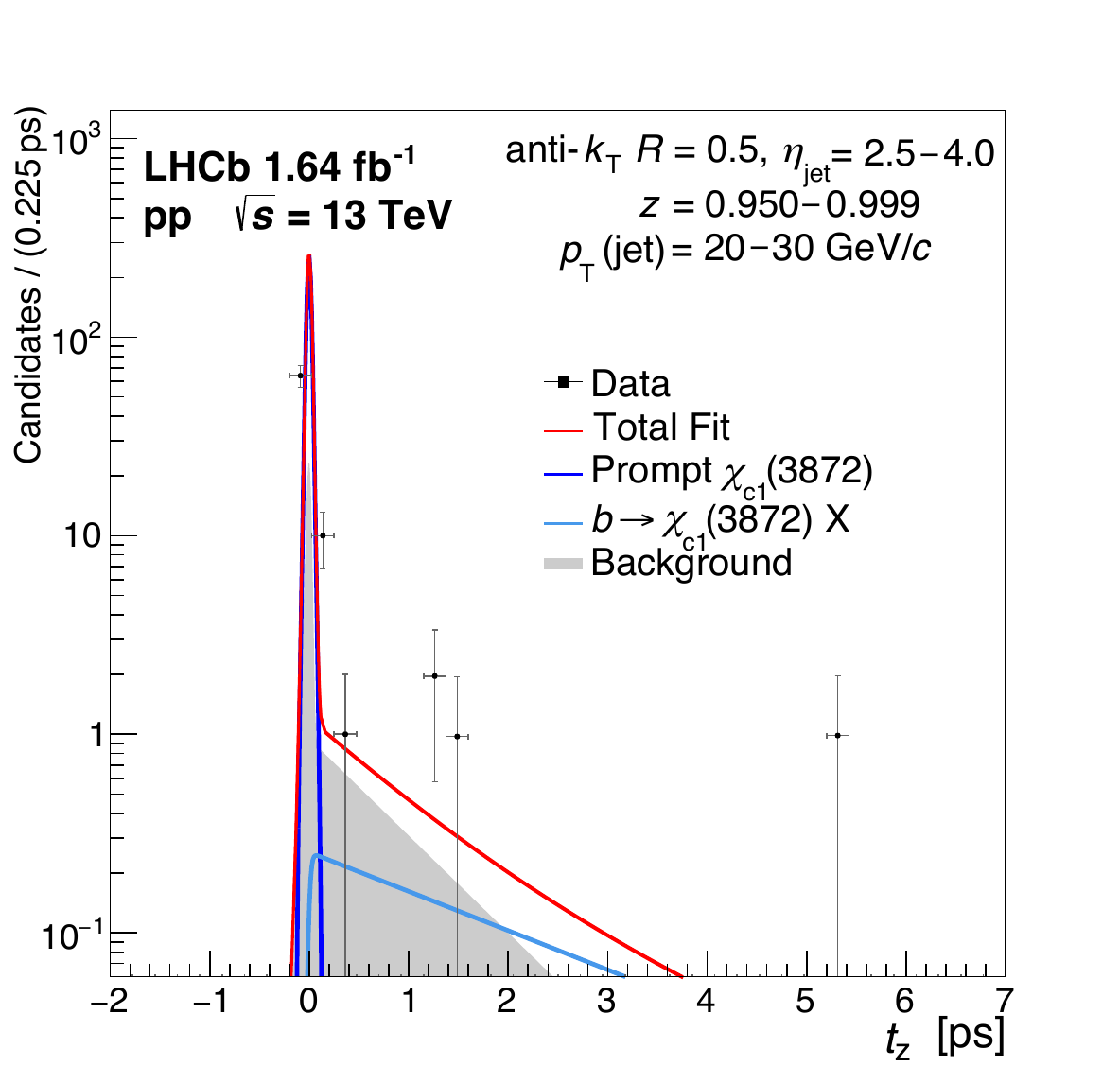}
    \caption{(Left) Invariant-mass distribution and (right) decay-time distribution for \ensuremath{\Pchi_{c1}(3872)} in a single \ensuremath{z}--\ensuremath{p_{\mathrm{T}}}(\jet) interval at high \ensuremath{z}, superimposed with the results from unbinned maximum-likelihood fits. The displaced \ensuremath{\Pchi_{c1}(3872)} fraction in this interval is $(3.5\pm2.9)\%$.\label{fig:massX}}
\end{figure}

The signal yields are determined from unbinned fits to the tag invariant-mass distribution for each \zt--\pt interval, where \pt here means either $\pt(\qtag)$ or $\pt(\jet)$. The signal distribution is modelled by the sum of two Crystal Ball (CB) functions~\cite{Skwarnicki:1986xj}, which share a common mean 
and power-law tail exponent, $n$. The parameter $\alpha$, which determines the onset of the power-law tail, is fixed to an equal value but with opposite sign for the two CBs. The start values for the $\alpha$ and $n$ parameters are taken from 
fits to simulation and the uncertainty from these fits are used to limit their parameter ranges. The background distribution is modelled with a first-degree polynomial.
The left panels of \cref{fig:massPsi,fig:massX} show the full mass range including signal and sideband regions overlayed with sample fits for \psitwos and \theX mass distributions in two particular kinematic bins. The signal-to-background ratio increases 
significantly as \zt increases, with an almost background-free signal for isolated \psitwos and \theX tags. 
\par
After performing the mass fits for a given \zt--\pt interval, the prompt and nonprompt signal fractions are determined with an unbinned fit to the pseudo decay-time distribution of the tag,
\begin{equation}\label{Eq:lifetime}
    \tz \equiv \lambda \: \frac{m}{\pz},
\end{equation} 
where $\lambda$ is the flight distance projected along the beam axis between the tag decay and primary vertices, $m$ is the nominal mass of the \psitwos or \theX meson~\cite{PDG2024}, and \pz is the component of the tag momentum along the beam axis~\cite{LHCb-PAPER-2011-003}.
\par
For the fit to the \tz distribution the sample is restricted to the \mbox{$m$(\dec)} signal region (\mbox{3678 $<$ $m$(\dec) $<$ 3694 \mevcc} for \psitwos and \mbox{3860 $<$ $m$(\dec) $<$ 3884 \mevcc} for \X), to enhance the signal contribution. These signal regions are shown in \cref{fig:massPsi,fig:massX} as the regions between the vertical dashed lines in the mass distributions. The background \tz distribution is modelled from the \tz distribution in the mass sideband regions around the known \psitwos or \theX meson mass~\cite{PDG2024}, and is fit empirically with a Gaussian core and exponential tails. This distribution is normalized using the background fraction contribution from the mass fit. The prompt signal distribution is modelled with a Dirac delta function and the displaced signal distribution with an exponential decay function. Both signal distributions are then convolved with the same double-Gaussian resolution function to account for detector effects. The five parameters extracted from the fit are the two Gaussian widths, the ratio between the two Gaussian normalizations, an exponential decay constant of the displaced signal yield, and the ratio of prompt and displaced signal contributions. 

\begin{figure}
    \includegraphics[width=0.5\textwidth]{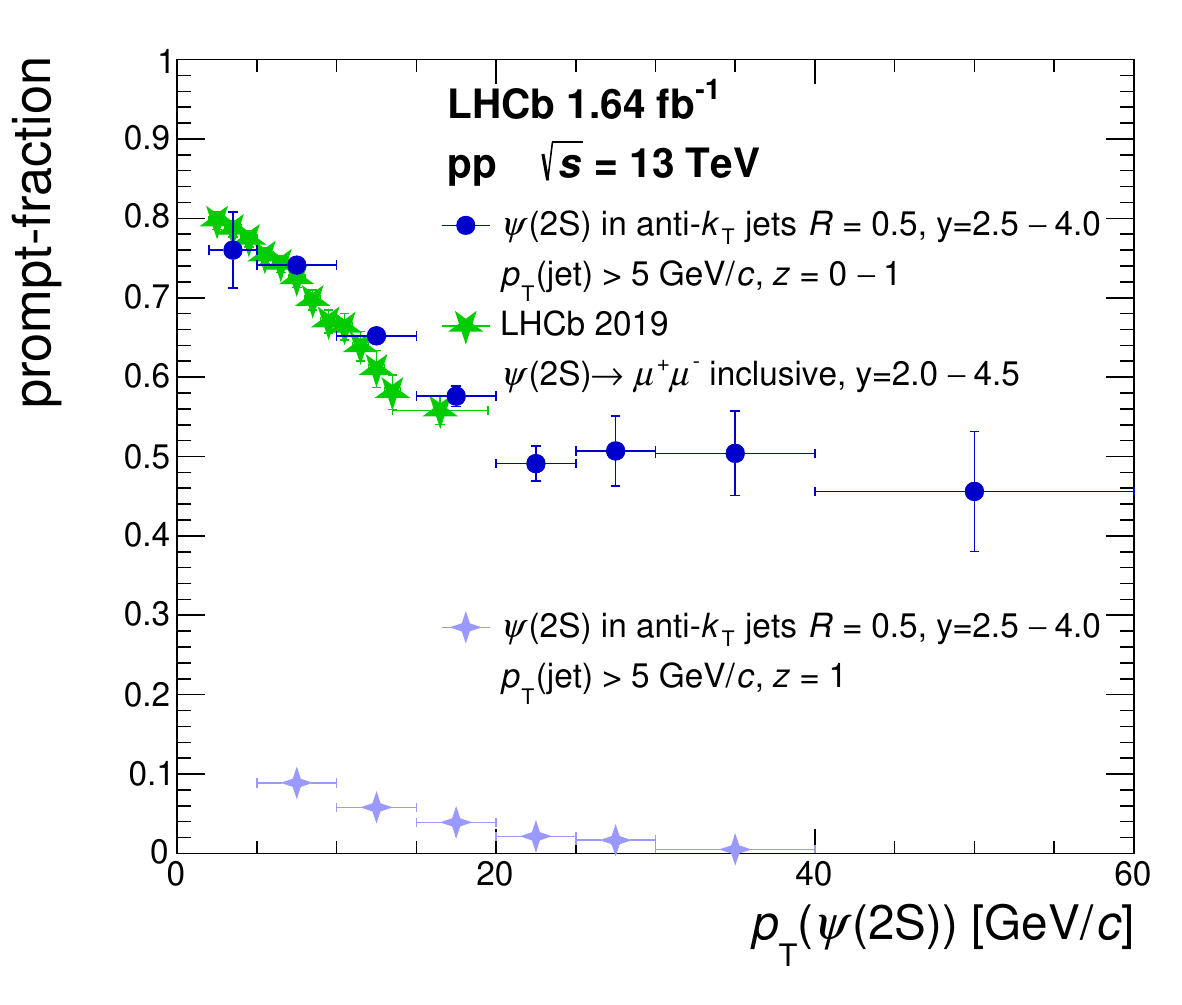}
    \includegraphics[width=0.5\textwidth]{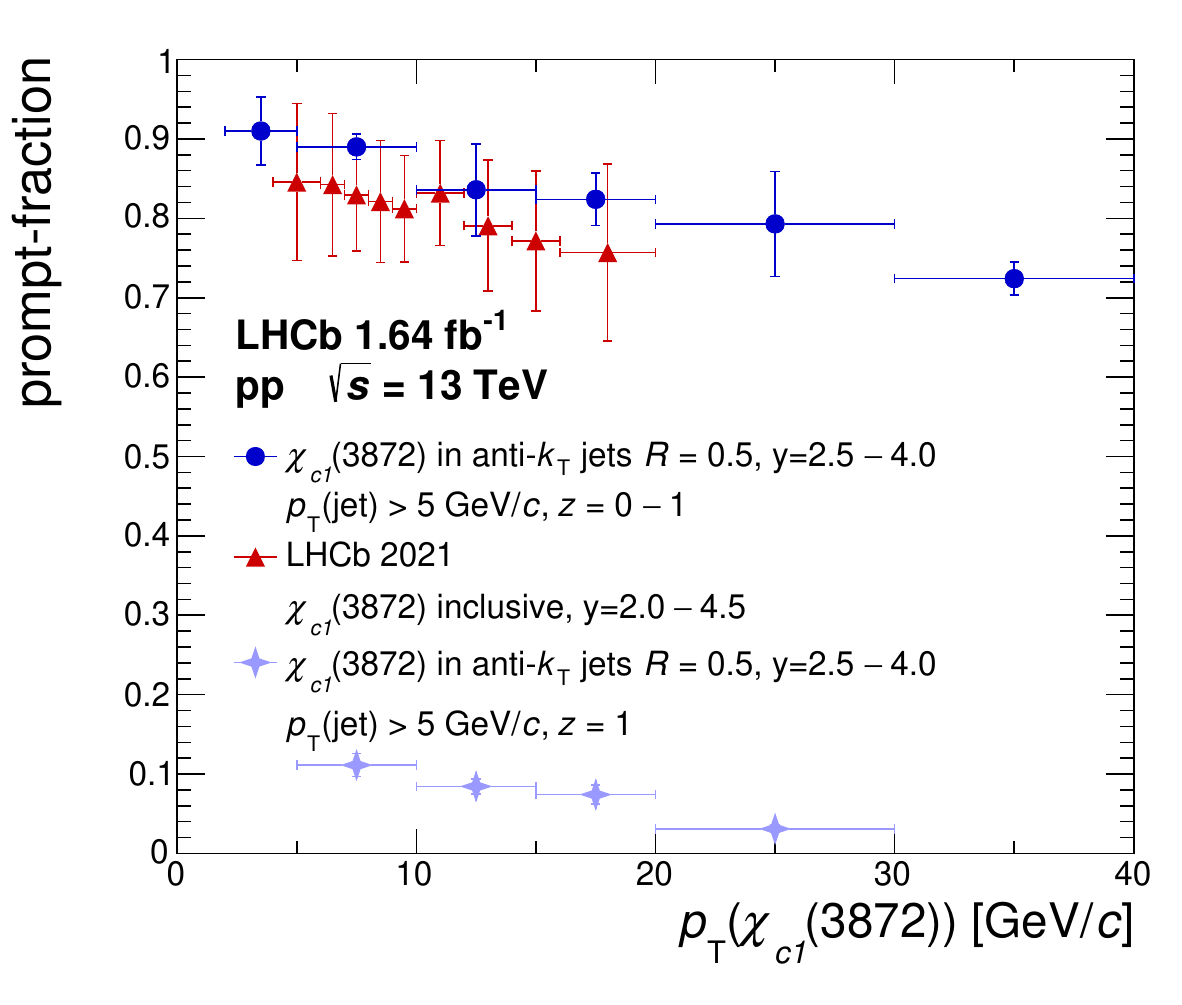}
    
    \caption{Fraction of prompt (blue circles) signal for (left) \ensuremath{\Ppsi{(2S)}} and (right) \ensuremath{\Pchi_{c1}(3872)} mesons as a function of their \ensuremath{p_{\mathrm{T}}} integrated over all \ensuremath{z} values. The prompt single-particle jets (\mbox{\zt = 1}) which are a subcomponent of the total prompt fraction are displayed separately (violet stars). The results of this analysis are compared to previous LHCb results from 2019 using \ensuremath{\mu^+\mu^-} final states~\cite{LHCb-PAPER-2018-049} (green stars) and results from 2021 using \ensuremath{\mu^+\mu^-\pi^{+}\pi^{-}} final states~\cite{LHCb-PAPER-2021-026} (red triangles), both at \ensuremath{\protect\sqrt{s}} = 13\,TeV.\label{fig:promptRatio}}
\end{figure}

The statistical uncertainties on the signal yields are propagated from the uncertainties of the mass and the \tz fits. The fraction of prompt signal integrated over \zt 
is shown in \cref{fig:promptRatio} as a function of the \psitwos and \theX \pt, and is consistent with previous LHCb results~\cite{LHCb-PAPER-2018-049, LHCb-PAPER-2021-026}.
The fraction of prompt fragmentation with \zt=1, which is not part of the reported fragmentation functions, is shown as well. These single-particle jets comprise only a tiny fraction of the total tag production.

\section{Signal efficiencies}\label{sec:eff}

The efficiency correction adjusts the signal yield in each \zt--\pt interval for inefficiencies in the reconstruction and selection procedure of the tag candidates and the jets. The total signal efficiency for the tags is factorized into the product of three efficiencies: online, pion, and tag.
\par
The online efficiency is the efficiency for a \dimu candidate to pass the online reconstruction and selection. The muon reconstruction efficiencies are determined from simulation as a function of $\eta(\mu)$ and $\pt(\mu)$. A tag-and-probe method using data is used to calibrate the simulation to data~\cite{LHCb:2014nio}. The average muon reconstruction efficiency is roughly $95\%$ and is largely independent of the \zt--\pt interval. The online selection efficiency is determined directly from the data by using an independent set of triggers to those used in this analysis~\cite{LHCb-PUB-2014-039}. This efficiency is evaluated as a function of $\eta(\dimu)$ and $\sqrt{\pt(\mu^+)\pt(\mu^-)}$. The online selection efficiency rises with $\pt(\jet)$ from roughly $50\%$ for $ 5 < \pt(\jet) < 8\gevc$, to $75\%$ for $40 < \pt(\jet) < 60\gevc$, and is largely independent of~\zt. 
\par
The pion efficiency is the efficiency for a pion to both be reconstructed and pass the analysis selection. The pion reconstruction efficiency is determined using the same method as the muon reconstruction efficiency, 
but with an additional hadronic-interaction correction derived from control samples. This efficiency increases with $\pt(\jet)$ and \zt. It increases with \zt from $50\%$ to $60\%$ for $ 5 < \pt(\jet) < 8\gevc$, and from $60\%$ to $75\%$ with \zt for $40 < \pt(\jet) < 60\gevc$. Simulation is used to determine the pion selection efficiency which corrects for the track-quality 
and particle-identification requirements. The efficiency of this selection increases with both \zt and $\pt(\jet)$. It increases with \zt from $35\%$ to $50\%$ for $ 5 < \pt(\jet) < 8\gevc$, and from $65\%$ to $80\%$ with \zt for $40 < \pt(\jet) < 60\gevc$.
\par
Finally, the tag efficiency is the efficiency for a reconstructed tag to pass either the \psitwos or \theX selection requirements. The efficiency of the $Q$ requirement for \theX tags is determined from simulation to be $88\%$ and found to be consistent over all \zt--\pt intervals. The efficiency of DOCA selection between the \dimu and pion candidates is found from simulation to be consistent with $100\%$.
\par
Since the efficiencies described above are calculated for prompt tag candidates, an additional correction is applied for displaced tag candidates. The decay-time-dependent tracking correction is roughly $105\%$ for tags at intermediate \tz values of $1 < \tz < 5\ps$ relative to \mbox{\tz = 0\ps}. These corrections are applied to the \tz distributions before the \tz fit is performed.
\par
The total efficiency ranges between $8\%$ and $25\%$, depending on the \zt--\pt interval. In 
general, the efficiency increases with both $\pt(\jet)$ and \zt. Since the efficiencies are 
primarily driven by the track reconstruction, when \zt--$\pt(\qtag)$ intervals are used, the 
efficiency is mostly constant over \zt and increases with $\pt(\qtag)$ from $5\%$ for \mbox{$2 < \pt(\qtag) < 5\gevc$}, to $28\%$ for $30 < \pt(\qtag) < 40\gevc$.
\par
One final efficiency correction is applied after the unfolding, described in Sec.~\ref{sec:unfold}. This is the jet reconstruction efficiency, which corrects for inefficiencies to reconstruct the jet object given that the tag was reconstructed in the detector. The jet reconstruction efficiency is largely constant 
as a function of \zt and \pt. It falls off for low jet \pt (outside the reported ranges) and for large \zt values close to 1. At large \zt values the reconstruction efficiencies drop by about 10\% to 25\% as compared to values at mid-\zt ranges. The unfolded spectra are corrected for these large-\zt inefficiencies.

\section{Unfolding}\label{sec:unfold}

\begin{figure}
    \centering
    \includegraphics[width=.56\textwidth]{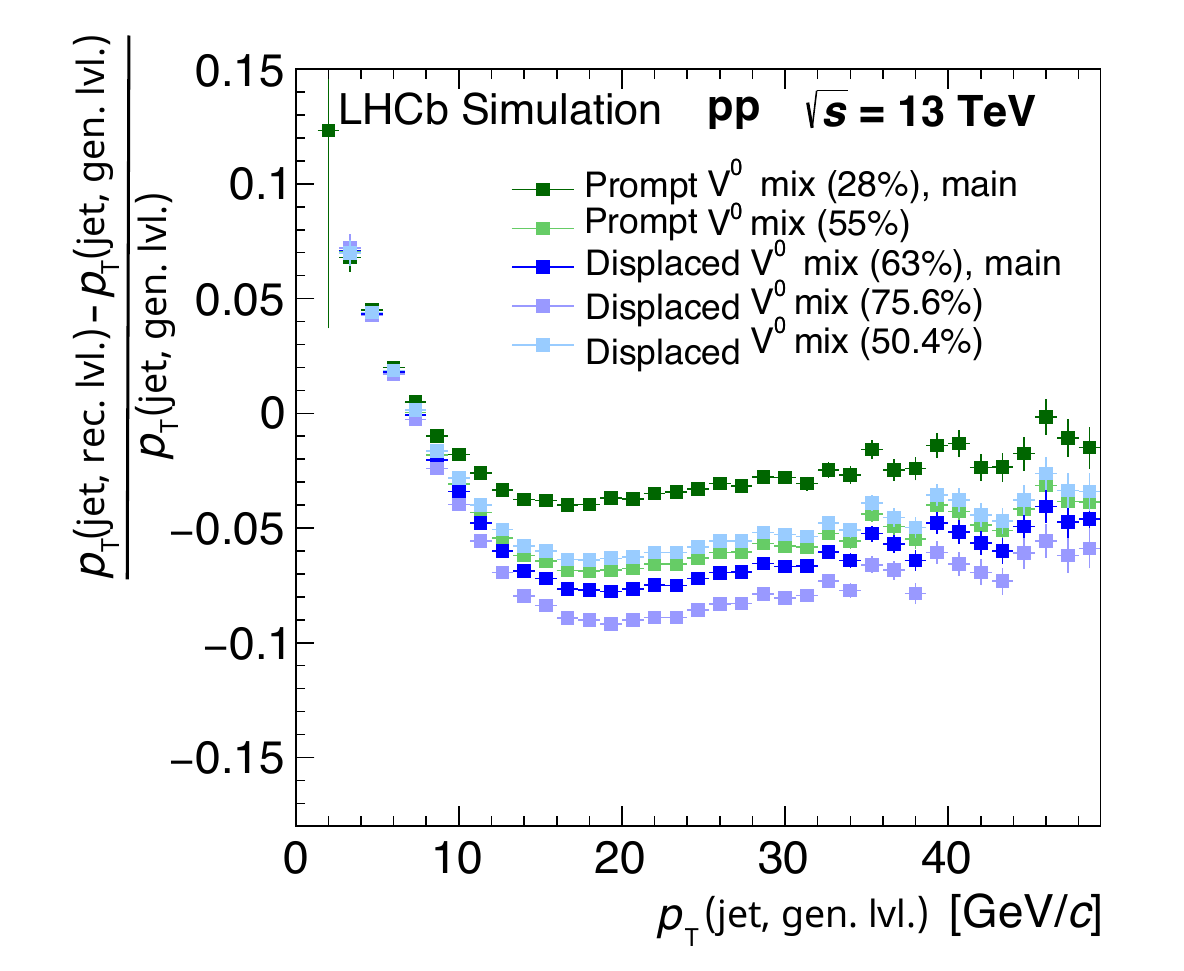}
    \caption{Jet energy scale    
    for different \ensuremath{V^0}s fraction variations, including prompt (green) and displaced (blue) variations, as determined from simulation. The percentage of events with \vzero content in these simulations is denoted in brackets. The jet energy scale (JES) is defined as the normalized difference in \ensuremath{p_{\mathrm{T}}}(\jet) retrieved at reconstruction level (rec. lvl.) and at MC generator level (gen. lvl.).\label{fig:jes}}
\end{figure}

Detector resolution effects result in the migration of candidates between \zt--\pt intervals. These migrations are primarily driven by the energy-scale shift and 
resolution of the $\pt(\jet)$, and consequently impact both \zt and $\pt(\jet)$ but not $\pt(\qtag)$. This analysis applies an iterative unfolding technique~\cite{DAgostini:1994fjx} based on the \textsc{RooUnfold} package~\cite{Adye:2011gm}. The detector response matrix maps 
the generator-level observables to the reconstructed-level observables and is defined using simulation. Current prompt-tag simulations do not provide an adequate description of basic jet properties in the data, such as the slope of the $\pt(\jet)$ 
distribution and the distribution of the number of constituents versus $\pt(\jet)$. As the number of constituents in jets in the prompt-tag simulation is significantly lower than in data, the simulation yields higher \zt values and thus offers insufficient coverage for lower \zt values. Hence, displaced-tag simulations of \psitwos and \theX meson production, which provide better coverage of the data, are used to define the prompt response matrix.
The distinction between the displaced and prompt response matrices is primarily driven by the expected difference in the relative fraction of long-lived neutral hadrons which are not reconstructed such as \KS mesons and \Lz baryons (\vzeros), in these two simulation samples. For example, some heavy-flavour decays include long-lived neutral hadrons that cannot be reconstructed. Samples with larger \vzero fractions result in a downward shift of the jet energy scale. From simulation, the fraction of events with \vzero content is estimated to be $63\%$ in displaced tag production and $28\%$ in prompt tag production. The shift in the jet energy scale for a variety of \vzero fractions, including the expected prompt and displaced fraction, is shown in \cref{fig:jes}. The jet energy resolution is independent of the \vzero fraction and is roughly $15\%$ for $\pt(\jet) = 10\gevc$ and $13\%$ for $\pt(\jet) = 40\gevc$.

\begin{figure}
    \includegraphics[width=0.99\textwidth]{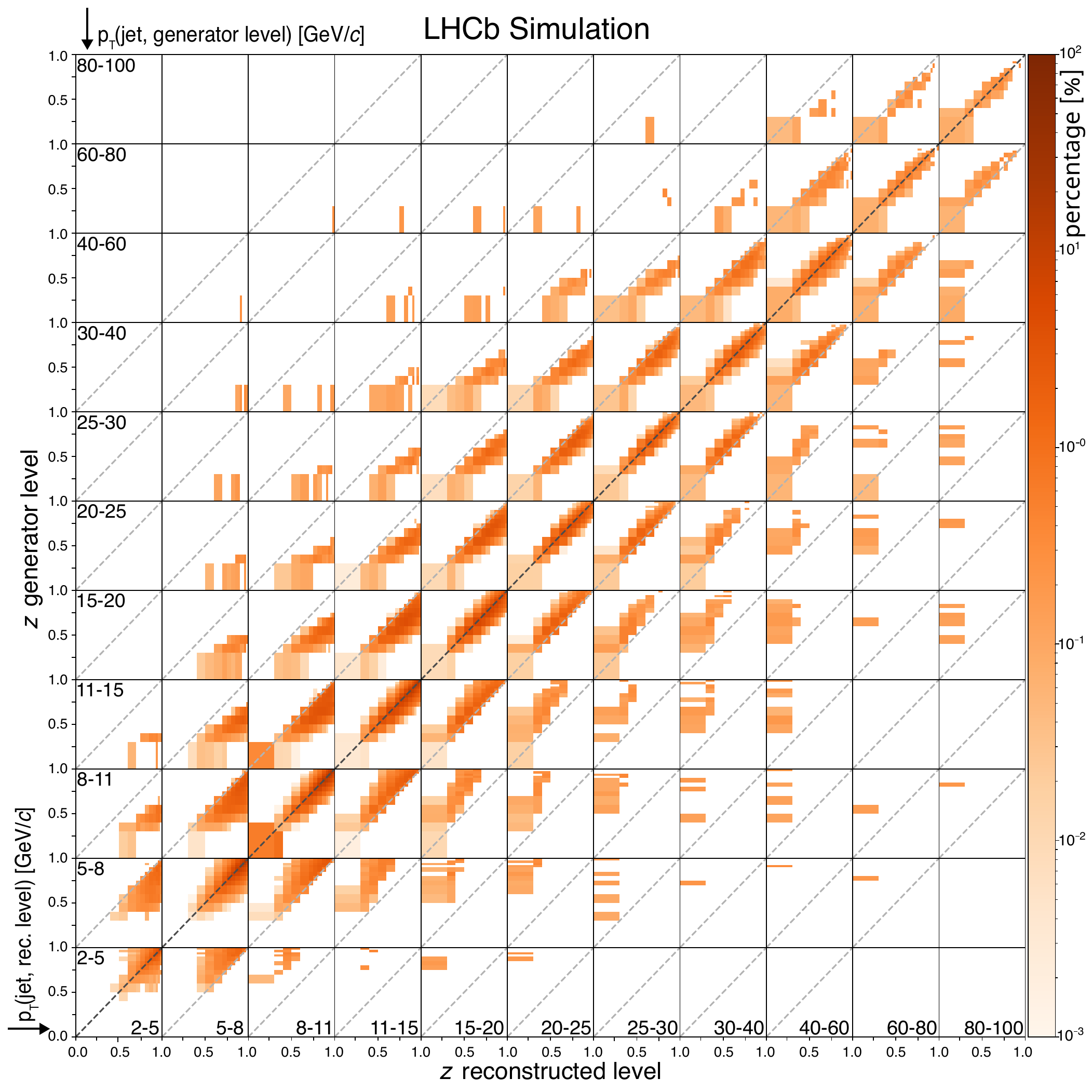}
    \caption{Four-dimensional prompt response matrix for unfolding the \ensuremath{\Ppsi{(2S)}} result differential in \ensuremath{z} and \ensuremath{p_{\mathrm{T}}}(\jet), assuming a percentage of events with \ensuremath{V^0} content of $28\%$. The colour axis represents the bin-to-bin migration probabilities between the generated observable and the reconstructed observable.\label{fig:rm}}
\end{figure}

Both the prompt and displaced response matrices are further corrected to match the data using a momentum-balance technique applied to $Z+\jet$ data~\cite{LHCb-PAPER-2013-058, LHCb-PAPER-2016-064}. Based on these studies the $\pt(\jet)$ value is shifted downwards by $4\%$ and the $\pt(\jet)$ 
resolution is increased by $7\%$. \Cref{fig:rm} shows the prompt response matrix for the \psitwos \zt--$\pt(\jet)$ measurement. Since the measurements of this analysis are doubly differential in \zt and $\pt(\jet)$ or $\pt(\qtag)$, we apply an iterative two-dimensional Bayesian unfolding procedure until the folding result becomes stable. The lowest measured $\pt(\jet)$ interval, $5 < \pt(\jet) < 8\gevc$, and the lowest measured $\pt(\qtag)$ interval, $2 < \ensuremath{p_{\mathrm{T}}}(\qtag) < 5\gevc$, are only used as input for the unfolding to stabilize the output of the second lowest \pt interval.

\section{Systematic uncertainties}\label{sec:sys}
There are two main sources of systematic uncertainties: those from the efficiency determination and those from the unfolding procedure (systematic uncertainties from the signal fits are negligible).
\par
The efficiency uncertainties are fully correlated between \zt--\pt intervals and are determined by varying the efficiency corrections. 
For the corrections based on simulation, the binning of the muon and pion efficiencies as a function of $\eta$ and \pt is changed. Additionally, the default muon and pion efficiencies are varied within their statistical uncertainties. 
For the data-driven corrections, the statistical uncertainties are used to vary the efficiencies. 
For the online (trigger) efficiency, an additional correction is obtained to take into account differences from the true online efficiency to the efficiencies obtained using the independent set of triggers method in this analysis. The sample used is an inclusive \jpsi sample. The total efficiency uncertainty is the quadratic sum of these contributions and is roughly $6\%$. This uncertainty is dominated by the variations from the data-driven corrections. For low $\pt(\jet)$ and $\pt(\qtag)$ the efficiency uncertainty increases to $\sim$10\% due to the online and pion reconstruction efficiency uncertainties. All other uncertainties are below the $1\%$ level.
\par
The unfolding uncertainties are uncorrelated between \zt--\pt intervals, and are evaluated by varying the unfolding procedure. They are determined by varying the 
unfolding regularization parameter and by changing the fraction of \vzeros in the simulation that is used to define the response matrix and thus effectively changing the jet energy scale. Another source of uncertainty 
arises from the prior of the generator-level distribution. To evaluate this contribution, three variations are performed. First, a 
flat prior in \zt is used. Second, the prompt prior for the displaced response matrix is used and the displaced prior 
for the prompt response matrix is used. Third, the prior is weighted to match the unfolded result and unfold again with this adapted prior. The unfolding procedure is 
repeated with all variations and the differences between the results are assigned as systematic uncertainties. Since all variations of the unfolding procedure quantify 
the same systematic source of uncertainty, the combined uncertainty of the unfolding is calculated by the variance of the differences and is around $10\%$. Changing the prior to a flat distribution and varying the \vzero content leads to the largest
deviations to the main result. Systematic uncertainties due to vertex confusion from pile-up and from reconstruction efficiency variations in dense jets are negligible due to requiring only one PV per event.

\section{Results}\label{sec:res}

This section presents the double-differential \zt--\pt distributions after the efficiency corrections of Sec.~\ref{sec:eff} and the unfolding of Sec.~\ref{sec:unfold} are applied. All the provided distributions contain the data together with the uncorrelated statistical uncertainties propagated from the mass and \tz fits (shown as black lines in the figures that follow), the correlated uncertainties due to the efficiency corrections (empty blue boxes) and the uncorrelated uncertainties from the unfolding procedure (shaded blue boxes). Where available, the results are compared to predictions made with the \pythia~8.309 event generator\footnote{The production of prompt $\psitwos$ is simulated using the \emph{Charmonium:all=on} flag with the default long-distance matrix element (LDME) values. Further details of the generation are provided in the supplementary material of Ref.~\cite{LHCb-PAPER-2016-064}. The \theX is produced using the same flag, except as the \theX is not modelled directly in \pythia, the generated $\chic_{1}$ state has its mass replaced with the mass of the \theX, and these are selected as they both have the same quantum numbers.}~\cite{Bierlich:2022pfr}, with the majority of the uncertainties being too small to be visible. \pythia predictions for the \zt distributions at low jet \pt are omitted, as the $2 \rightarrow 2$ NRQCD cross-sections diverge as $\pt \rightarrow 0$, so it is difficult to generate sufficient statistics with \pythia at this \pt regime when clustered in jets, leading to unreliable predictions. This is even more exaggerated in the tag \pt distributions, so \pythia predictions are omitted here also.

\begin{figure}
    \centering
    \includegraphics[width=0.397\textwidth]{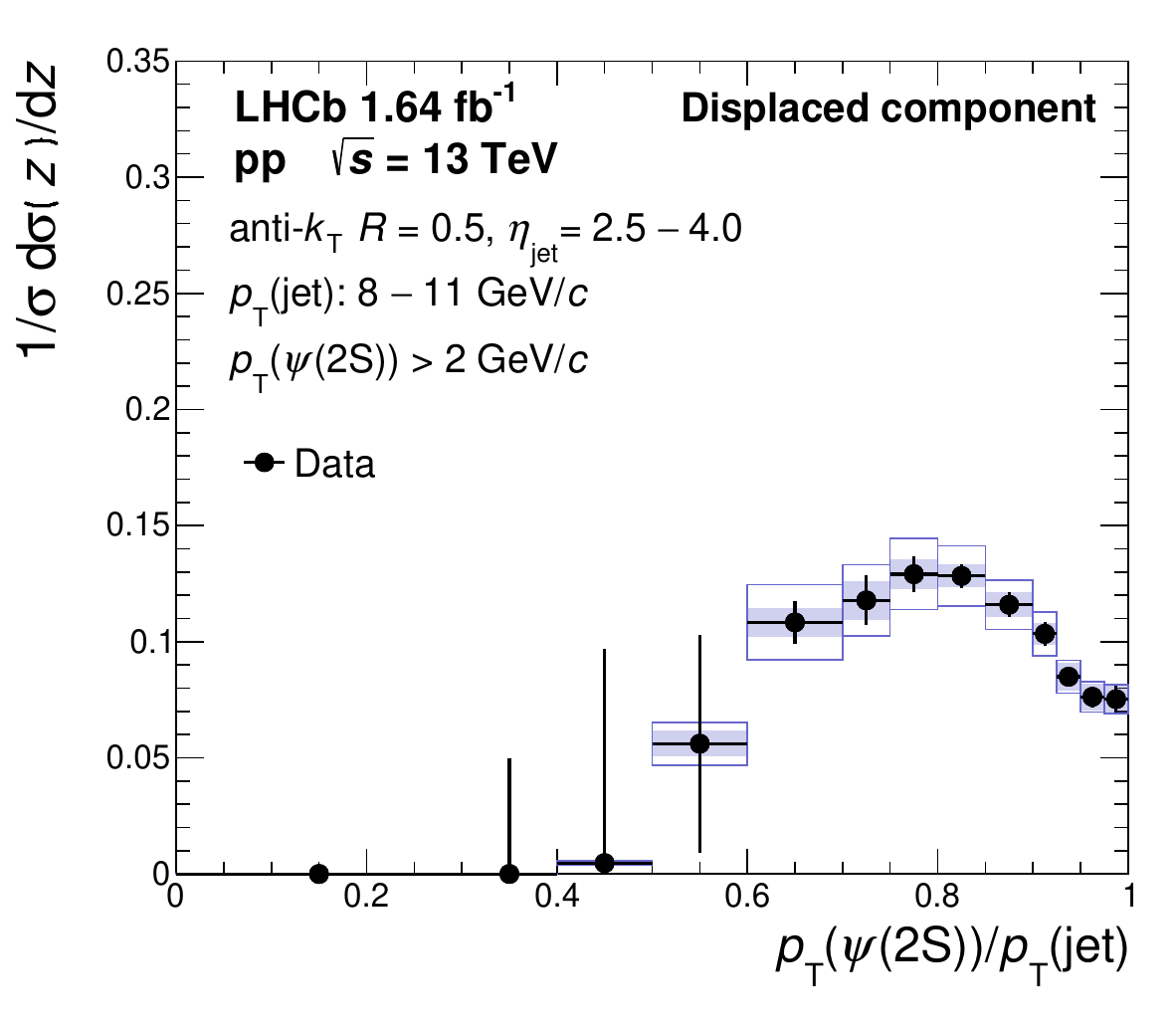}
    \includegraphics[width=0.397\textwidth]{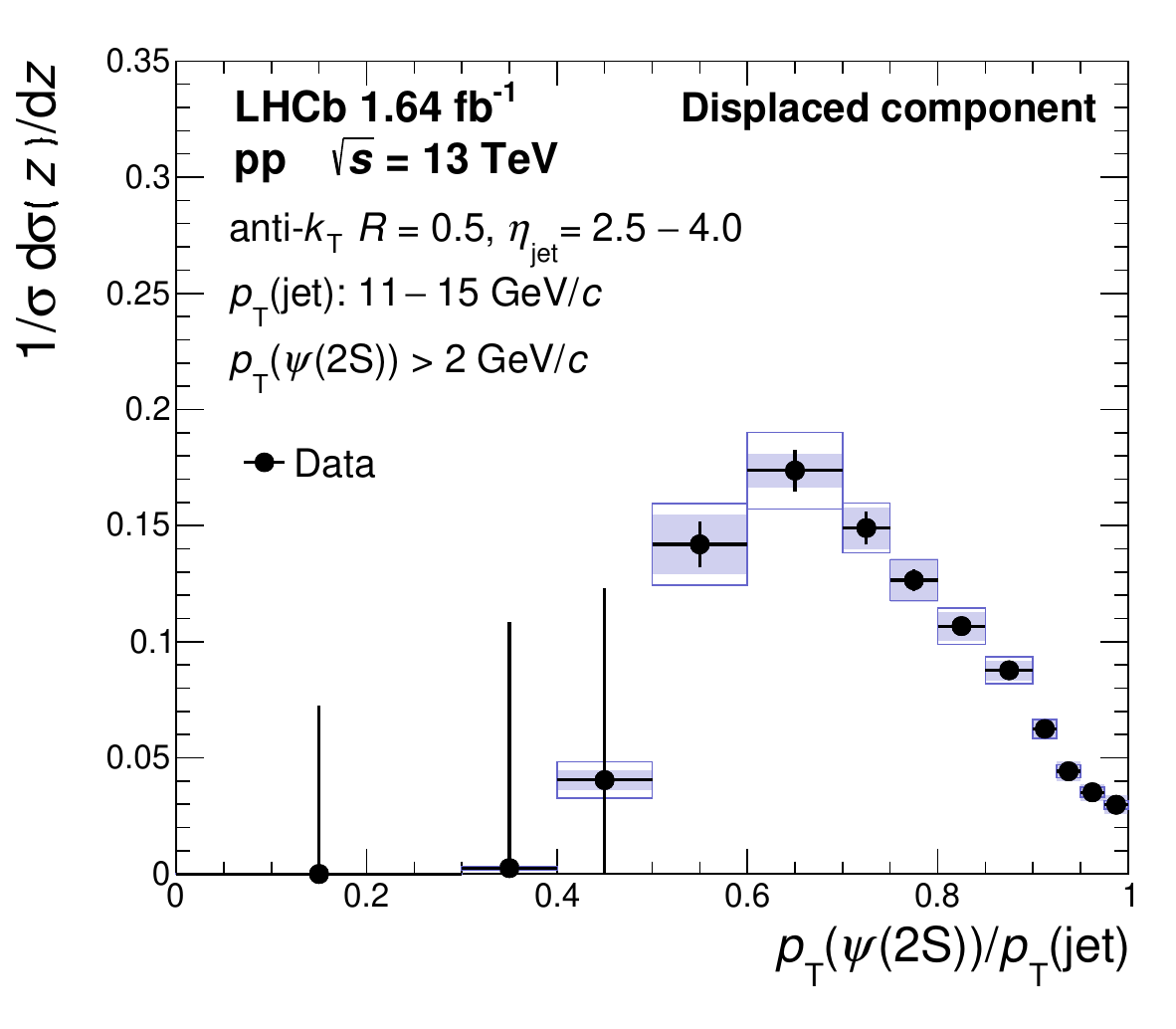}
    \\
    \includegraphics[width=0.397\textwidth]{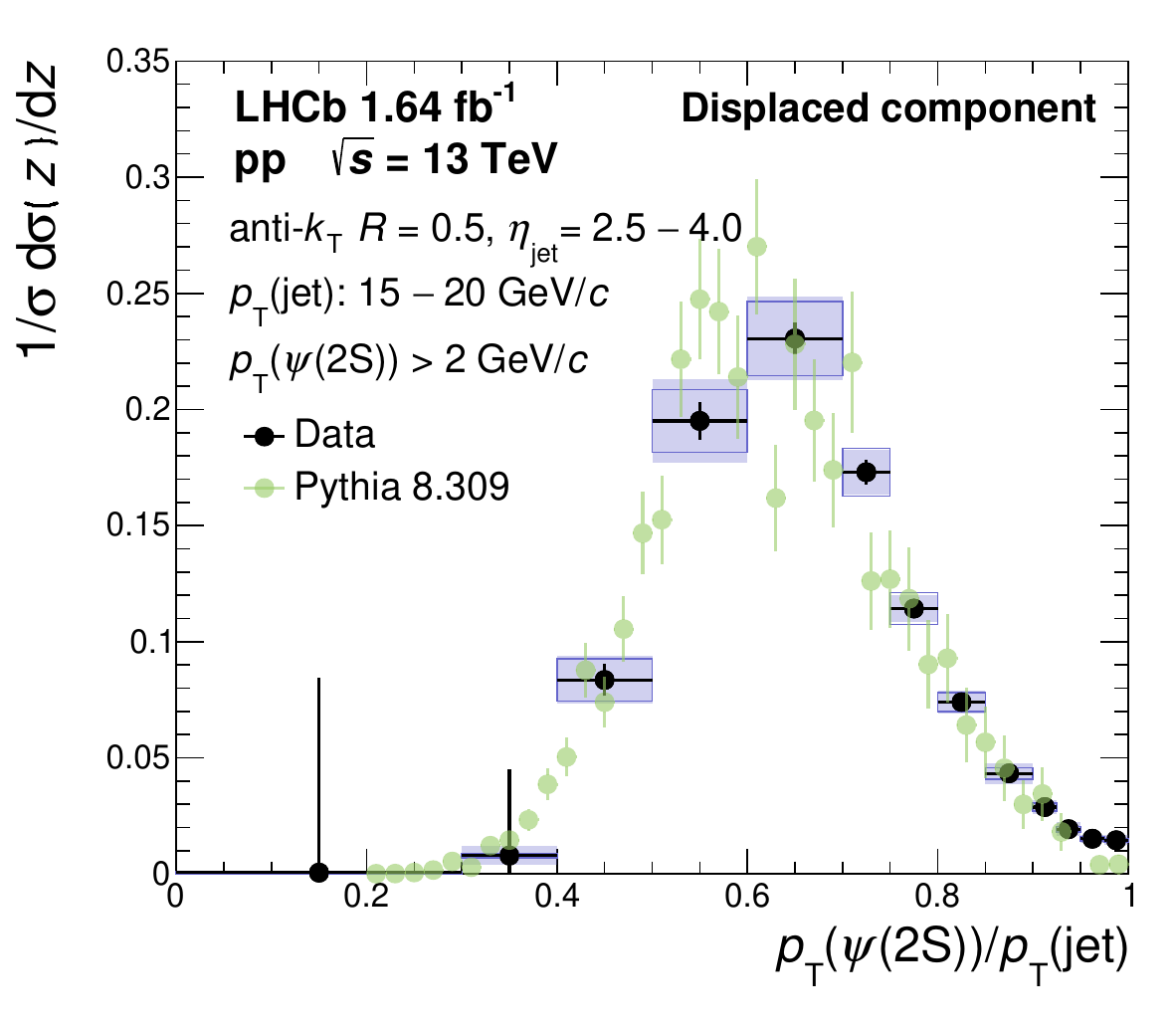}
    \includegraphics[width=0.397\textwidth]{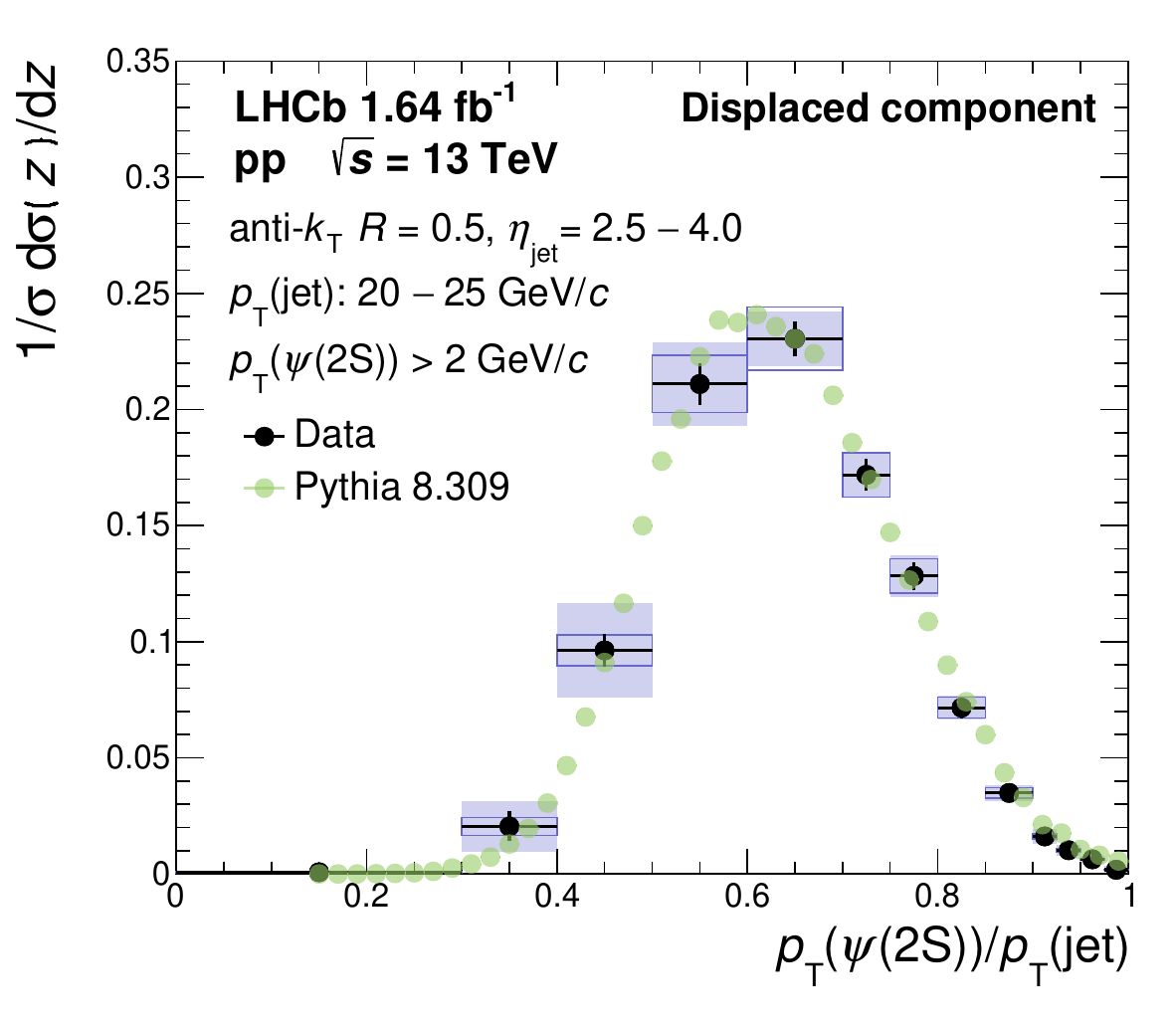}
    \\
    \includegraphics[width=0.397\textwidth]{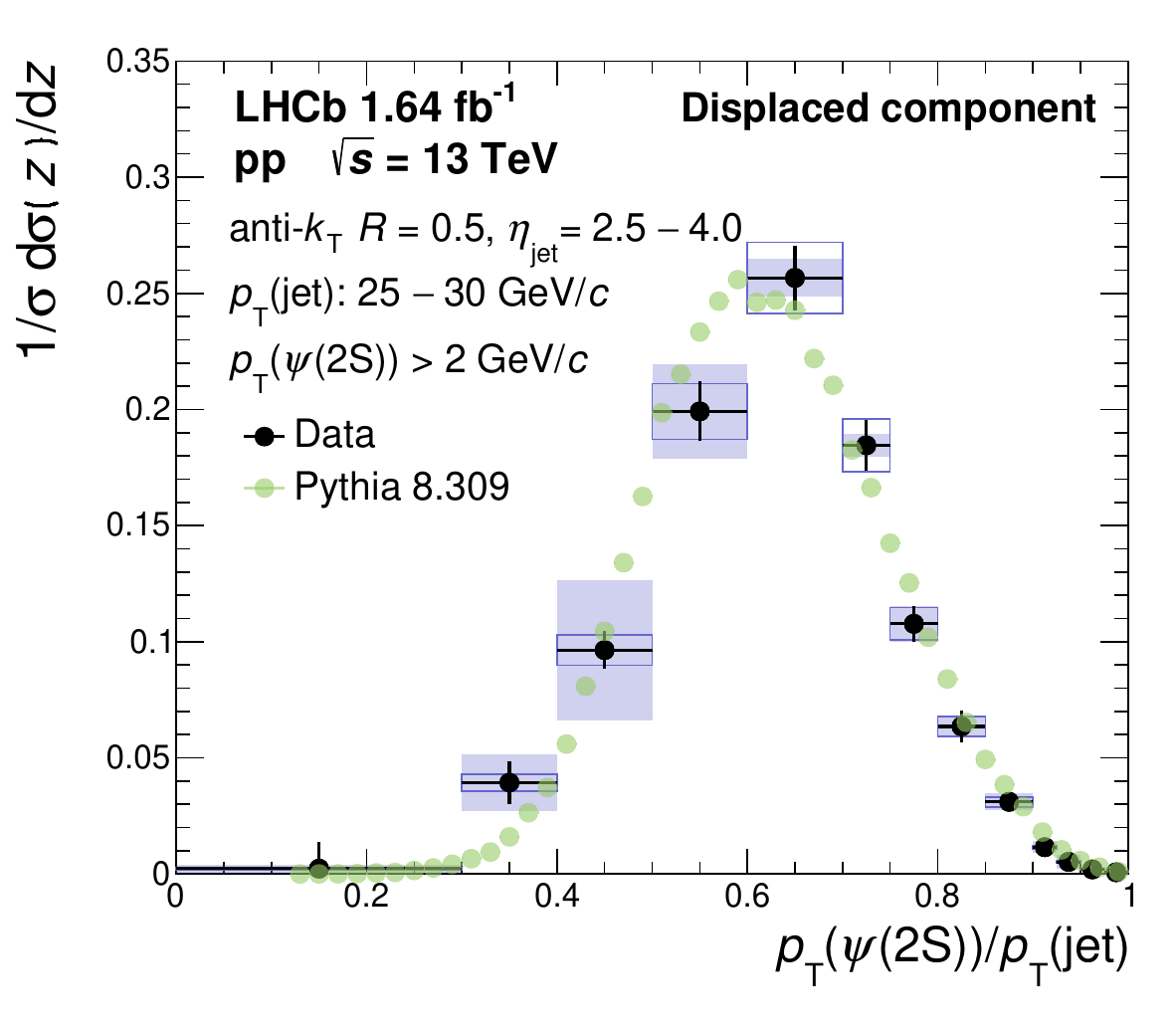}
    \includegraphics[width=0.397\textwidth]{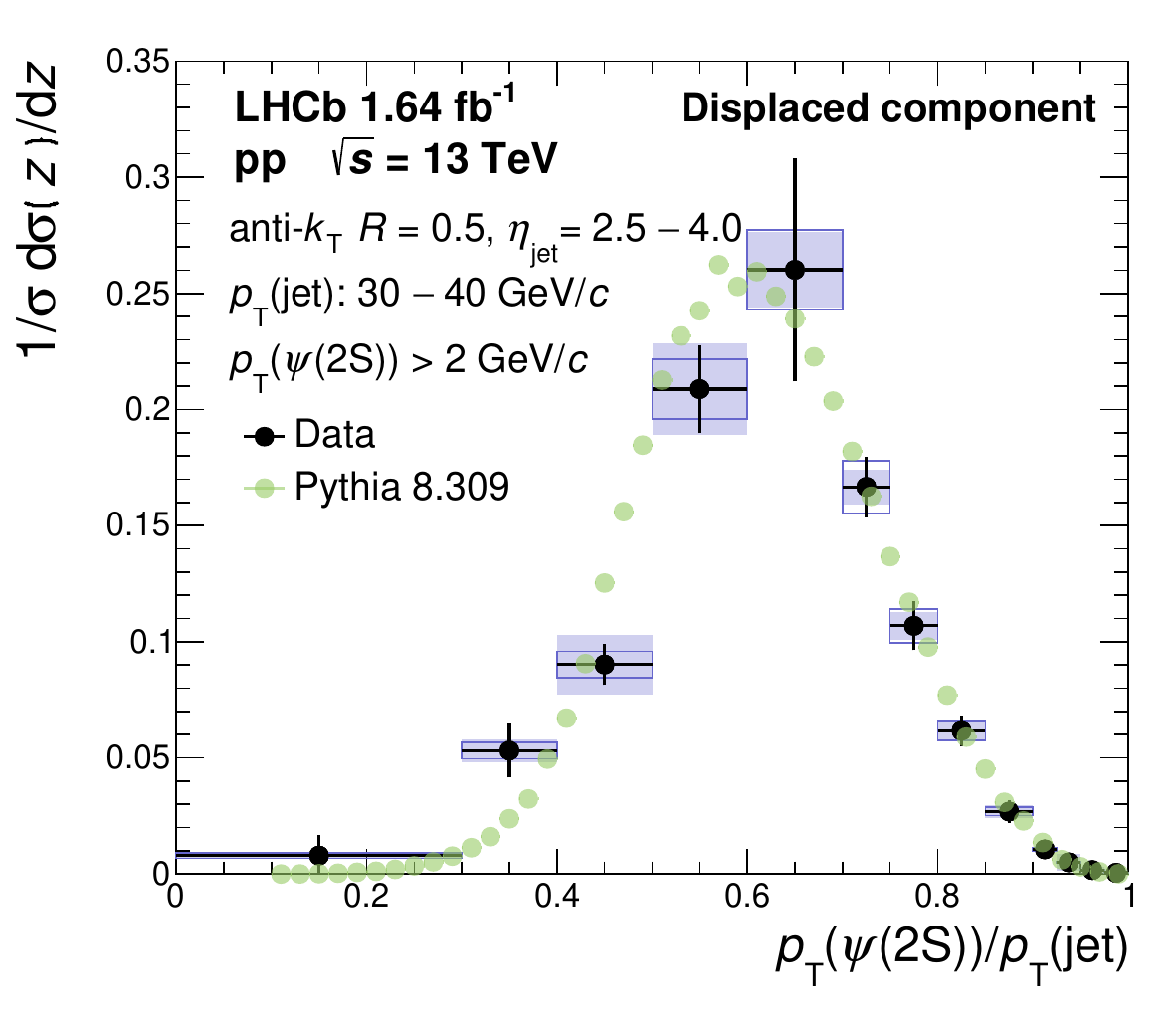}
    \\
    \includegraphics[width=0.397\textwidth]{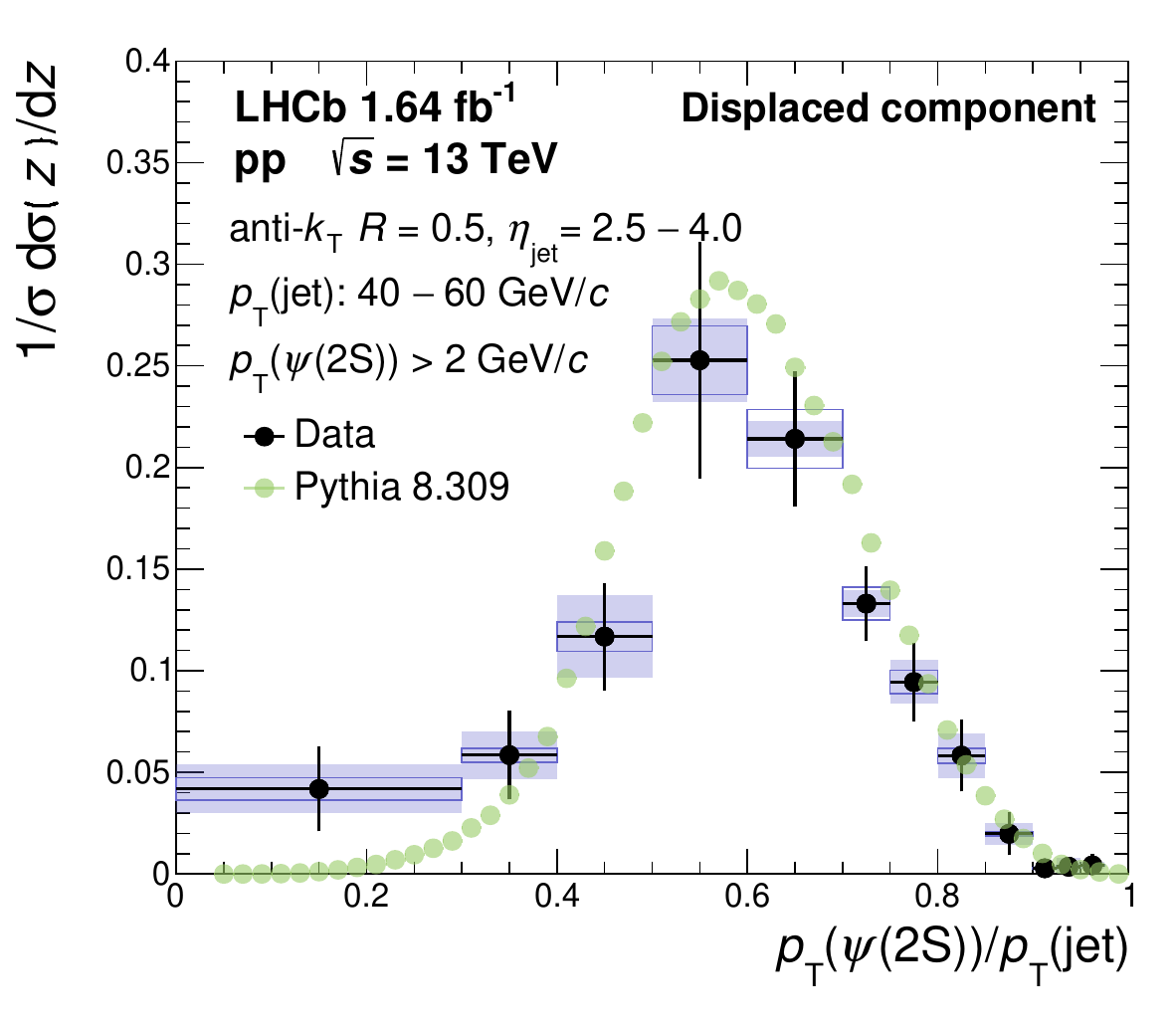}
    \caption{Distributions of the transverse momentum fraction \ensuremath{z} in intervals of jet \ensuremath{p_{\mathrm{T}}} for displaced \ensuremath{\Ppsi{(2S)}} production with statistical (black error bars), correlated efficiency (empty blue boxes), and uncorrelated unfolding (shaded blue boxes) uncertainties, compared to predictions obtained from \pythia~8.309.\label{fig:Results_P2_NP}}
\end{figure}

\begin{figure}
    \centering
    \includegraphics[width=0.397\textwidth]{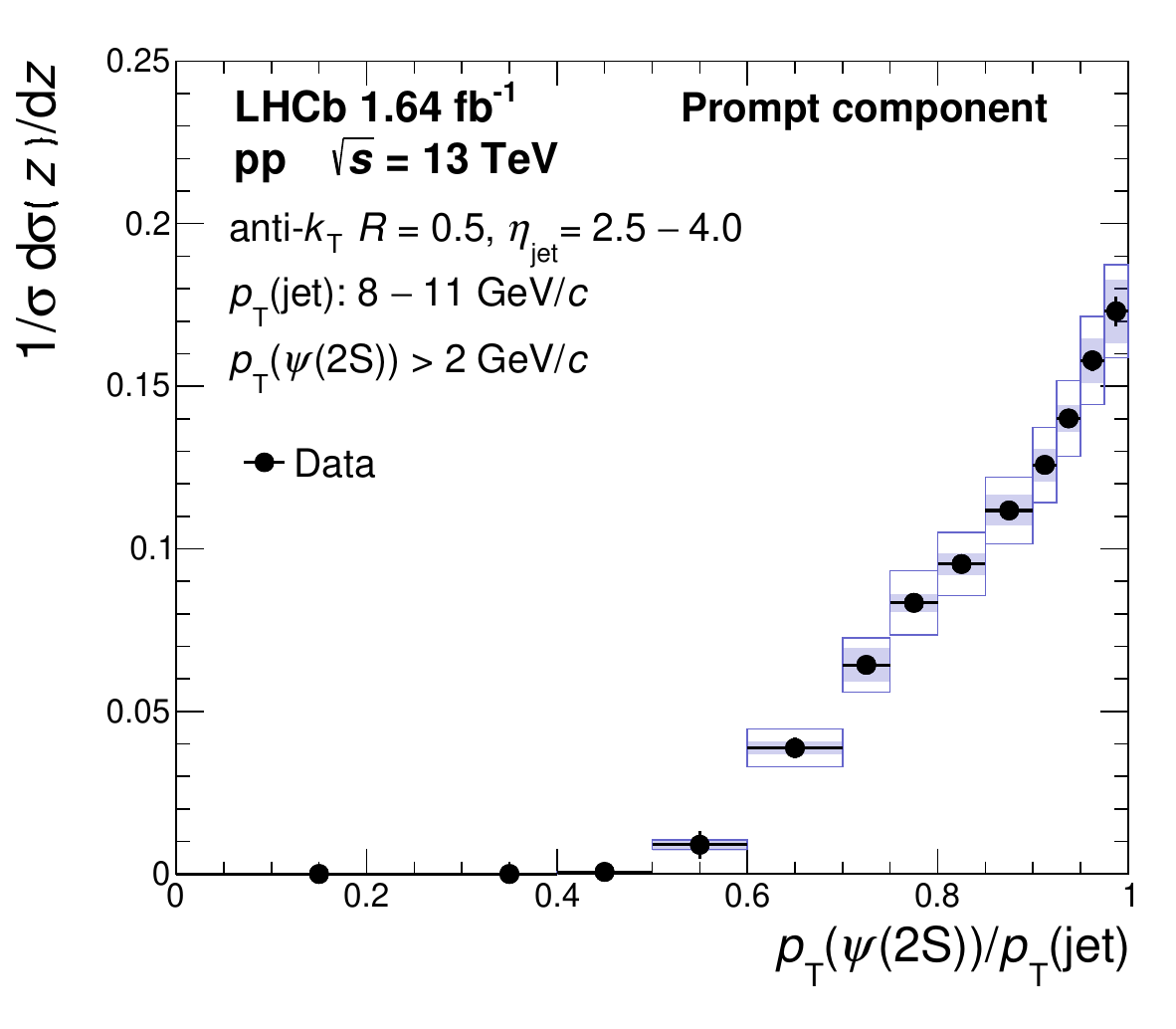}
    \includegraphics[width=0.397\textwidth]{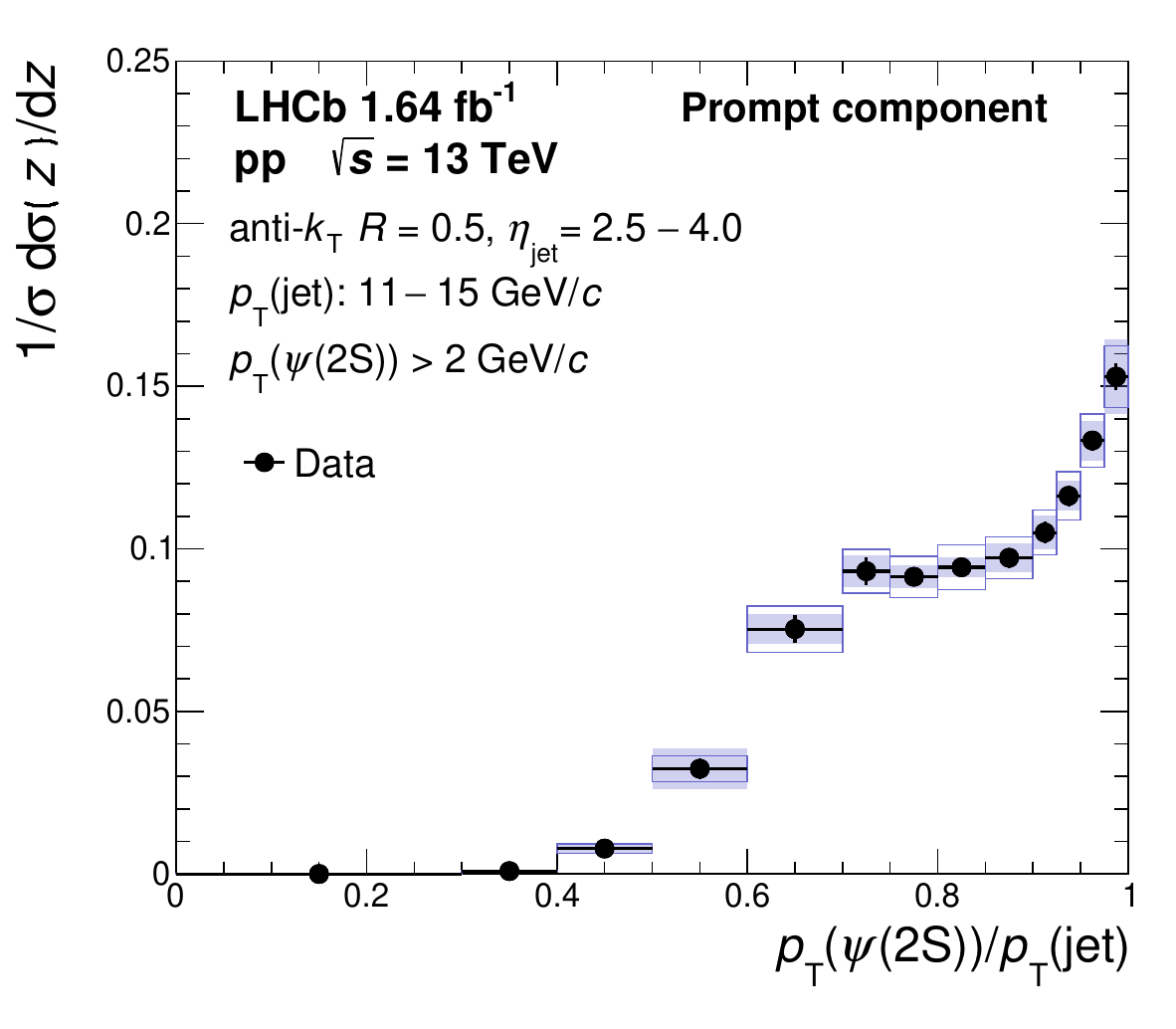}
    \\
    \includegraphics[width=0.397\textwidth]{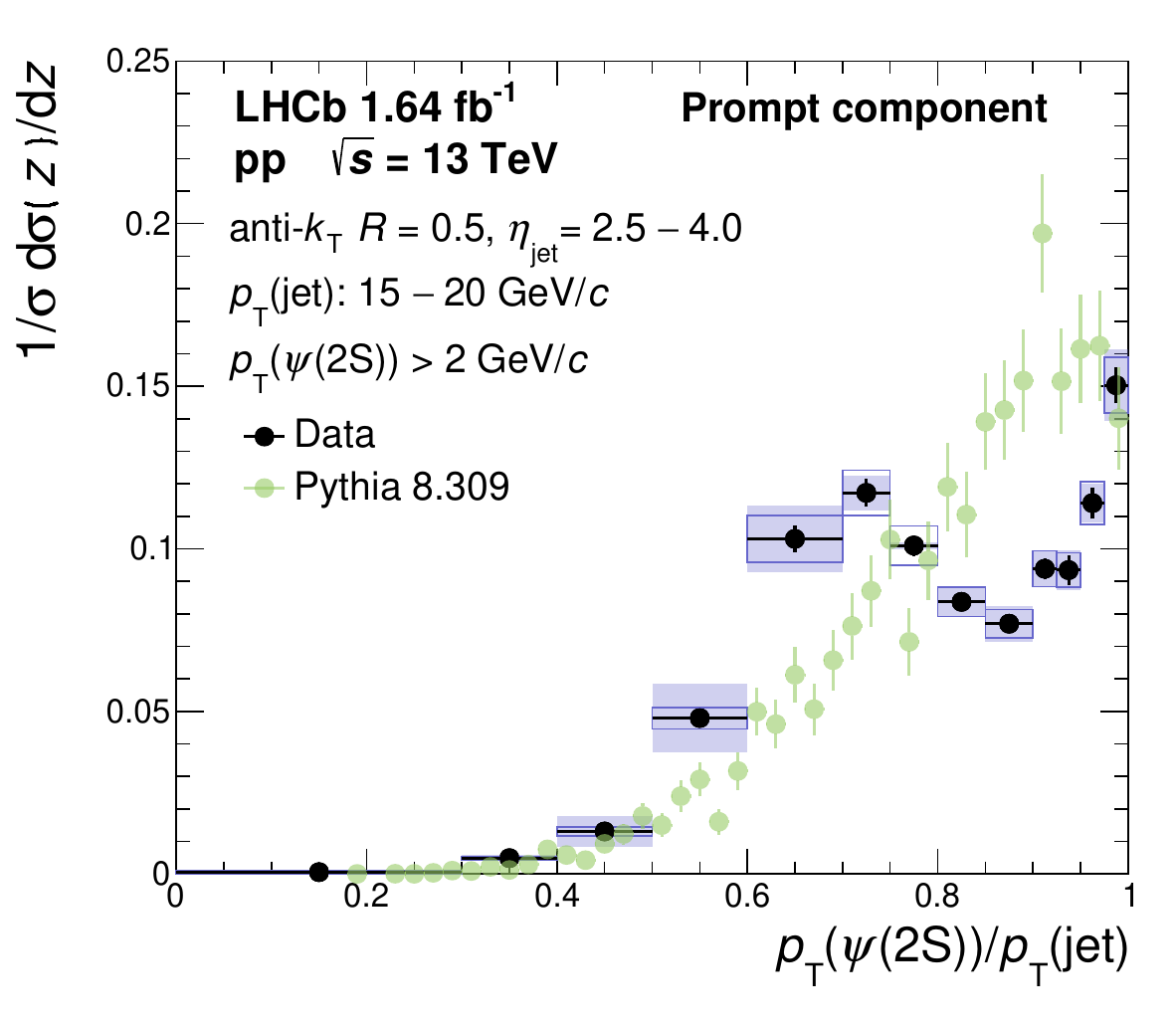}
    \includegraphics[width=0.397\textwidth]{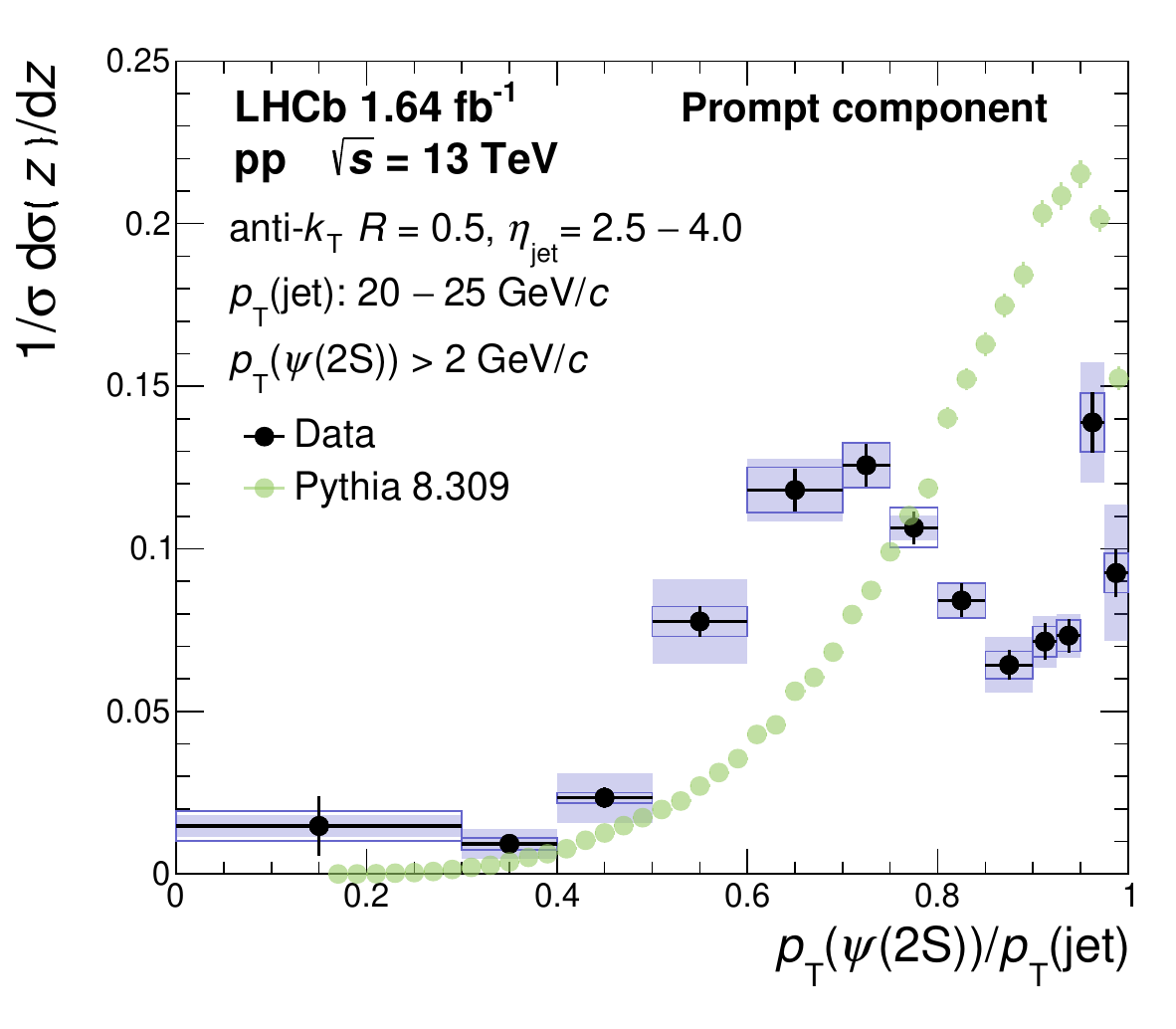}
    \\
    \includegraphics[width=0.397\textwidth]{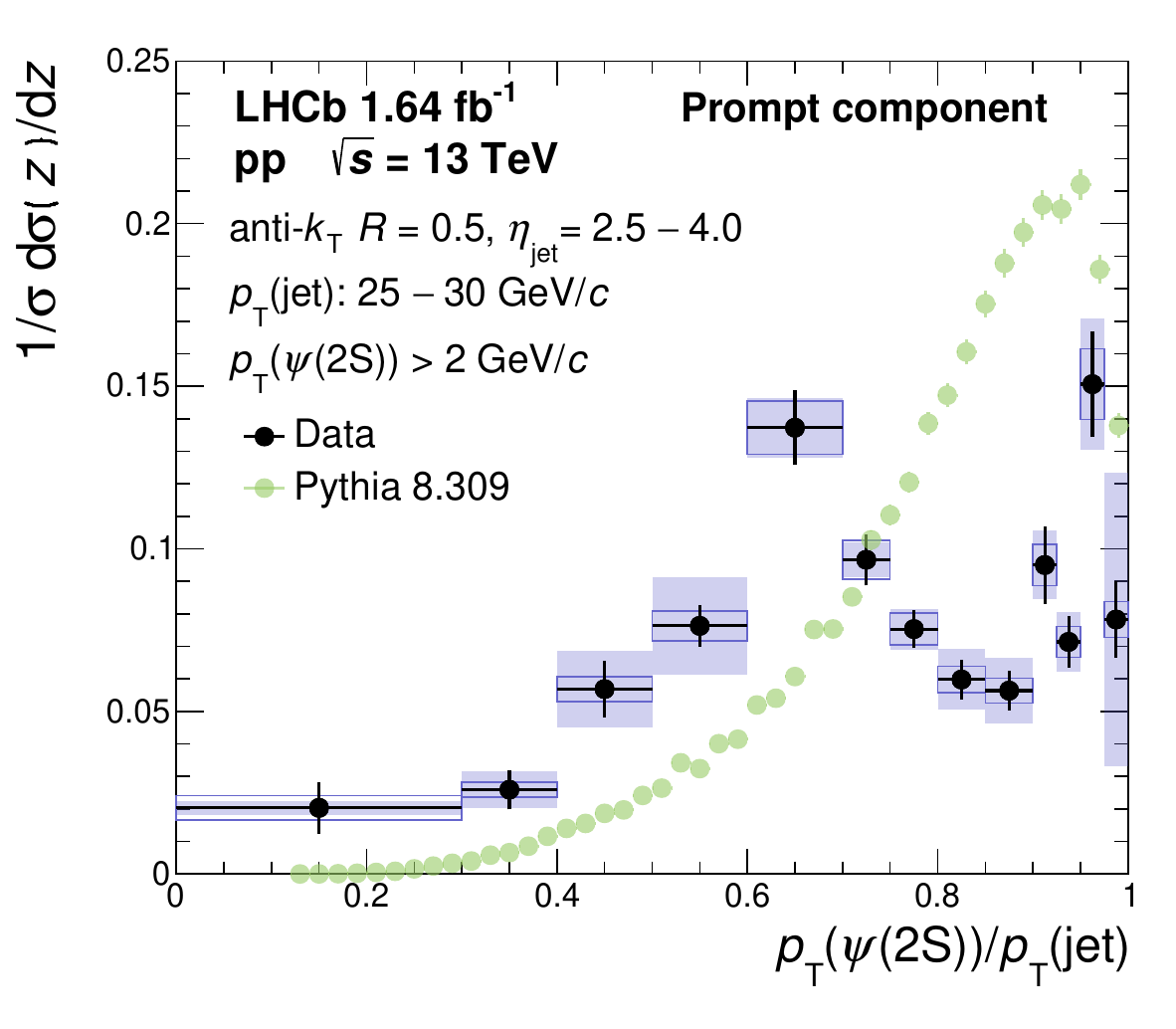}
    \includegraphics[width=0.397\textwidth]{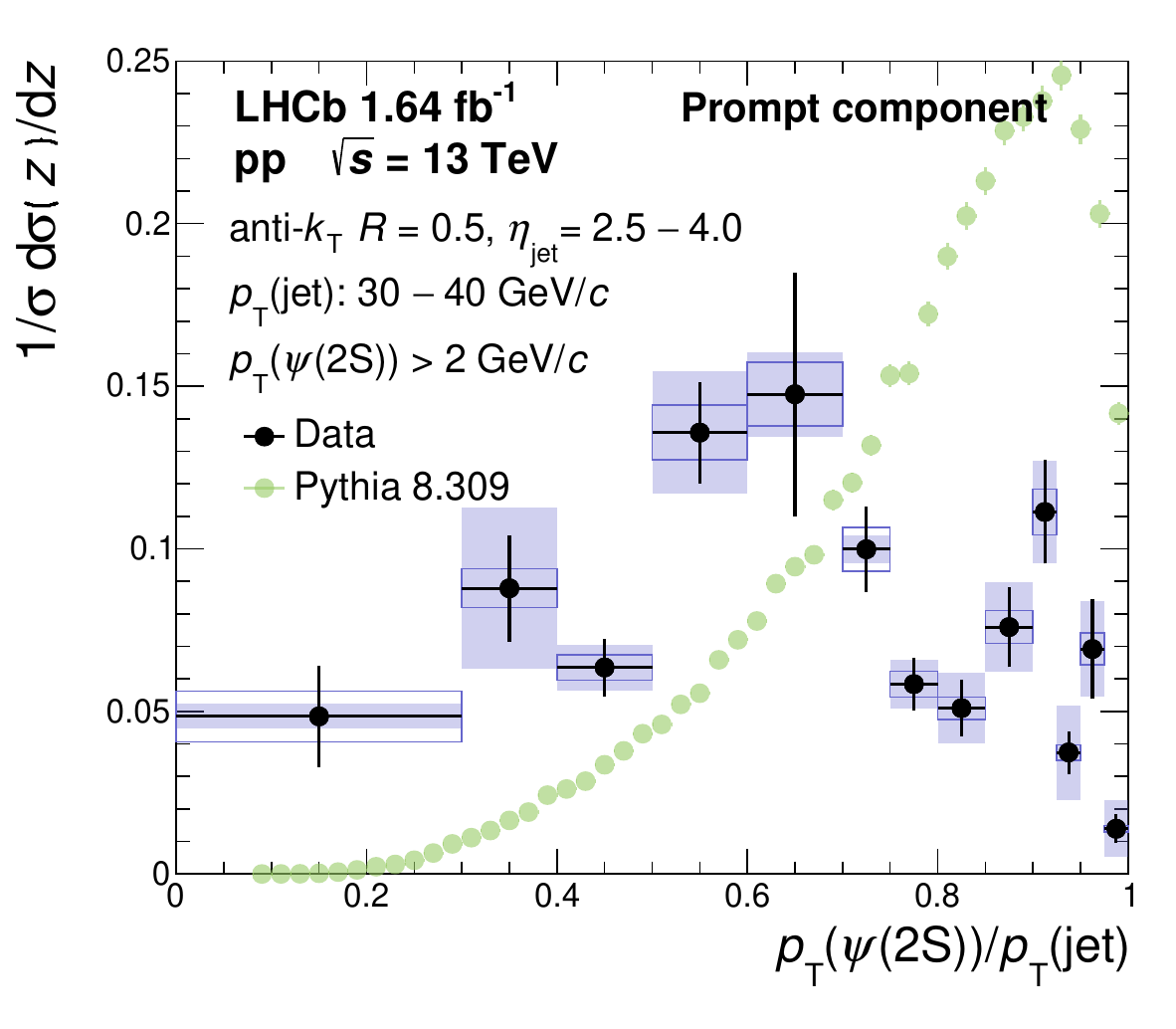}
    \\
    \includegraphics[width=0.397\textwidth]{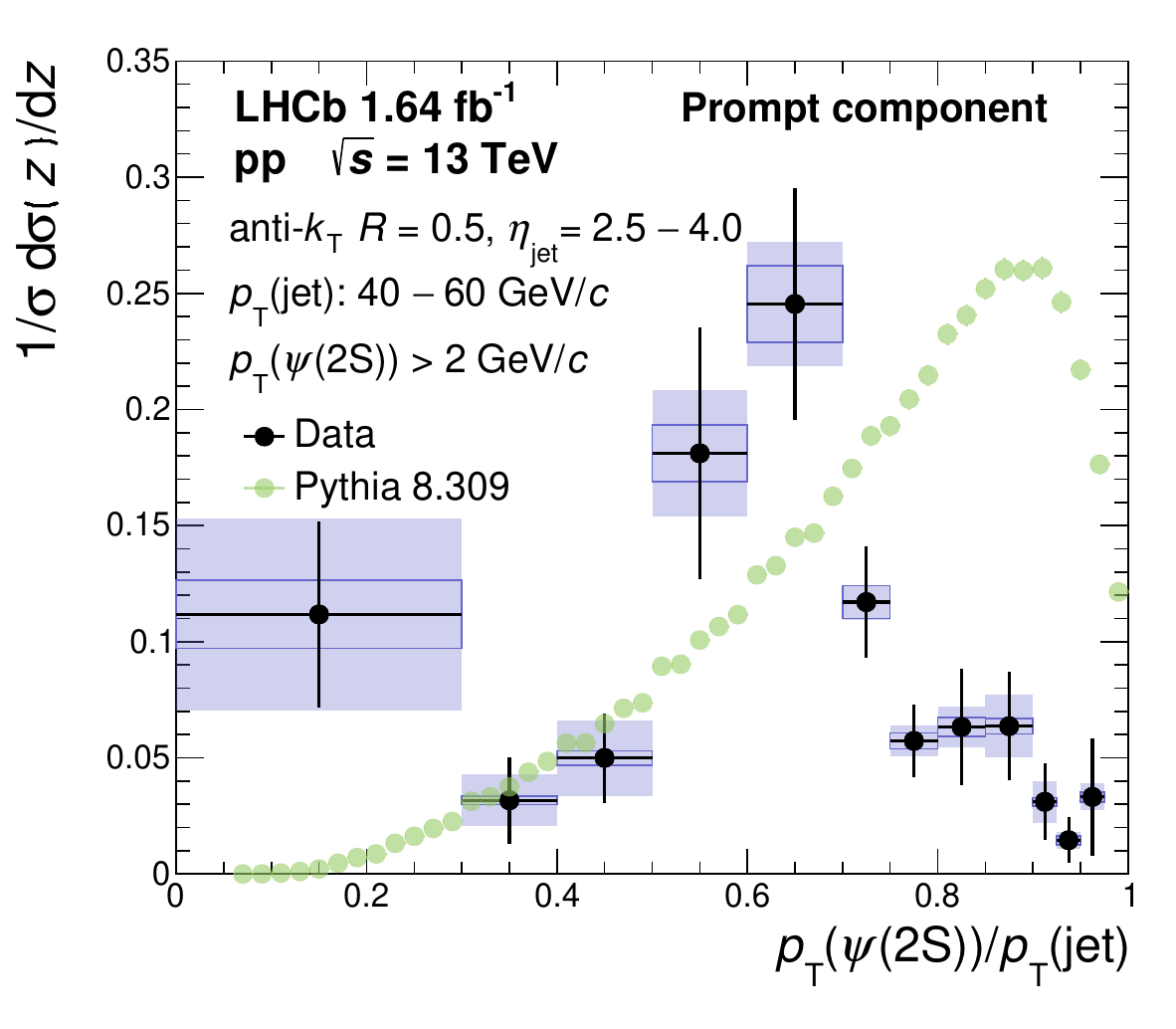}
    \caption{Distributions of the transverse momentum fraction \ensuremath{z} in intervals of jet \ensuremath{p_{\mathrm{T}}} for prompt \ensuremath{\Ppsi{(2S)}} production with statistical (black error bars), correlated efficiency (empty blue boxes), and uncorrelated unfolding (shaded blue boxes) uncertainties, compared to predictions obtained from \pythia~8.309.\label{fig:Results_P2_P}}
\end{figure}

\begin{figure}
    \centering
    \includegraphics[width=0.397\textwidth]{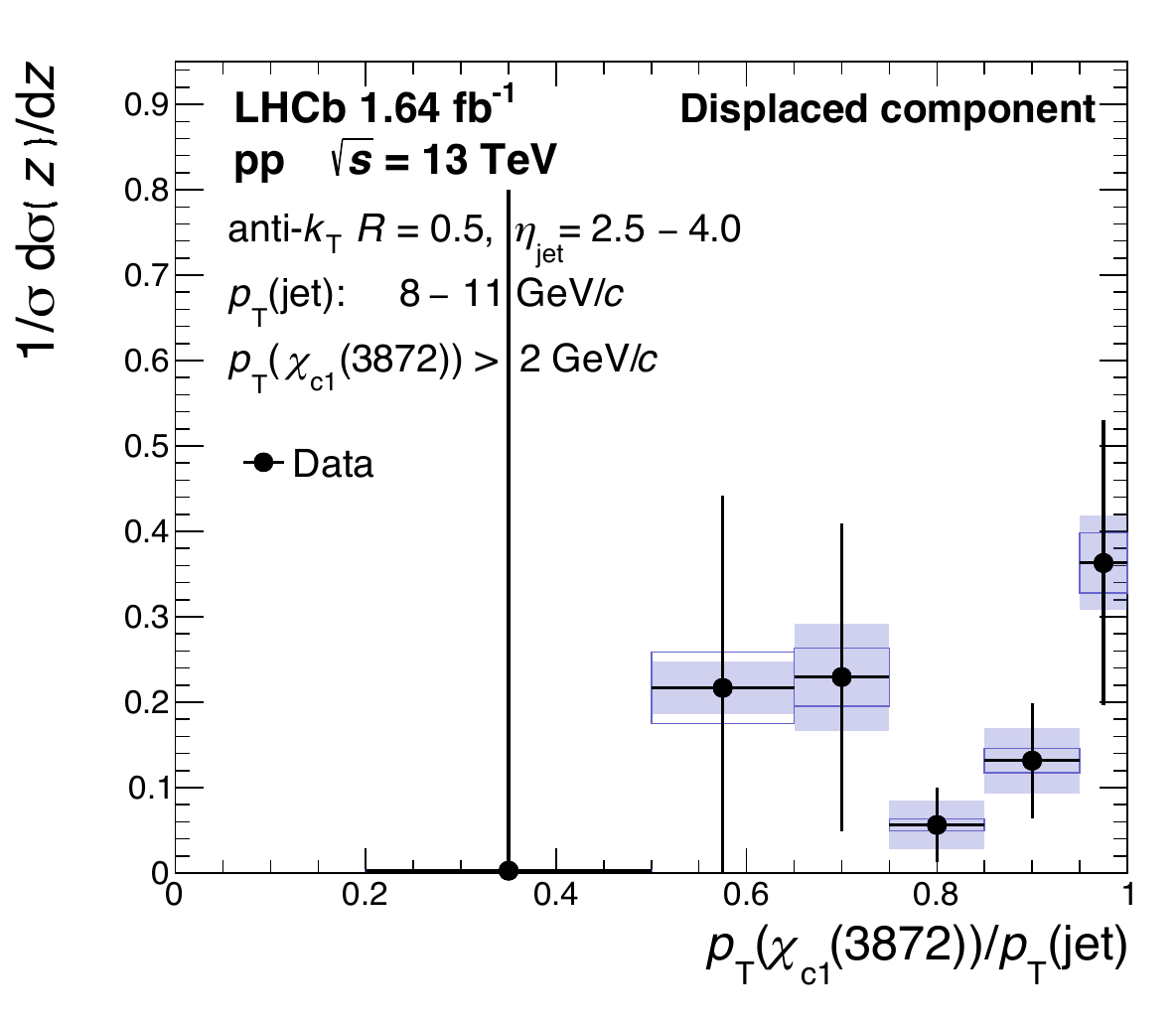}
    \includegraphics[width=0.397\textwidth]{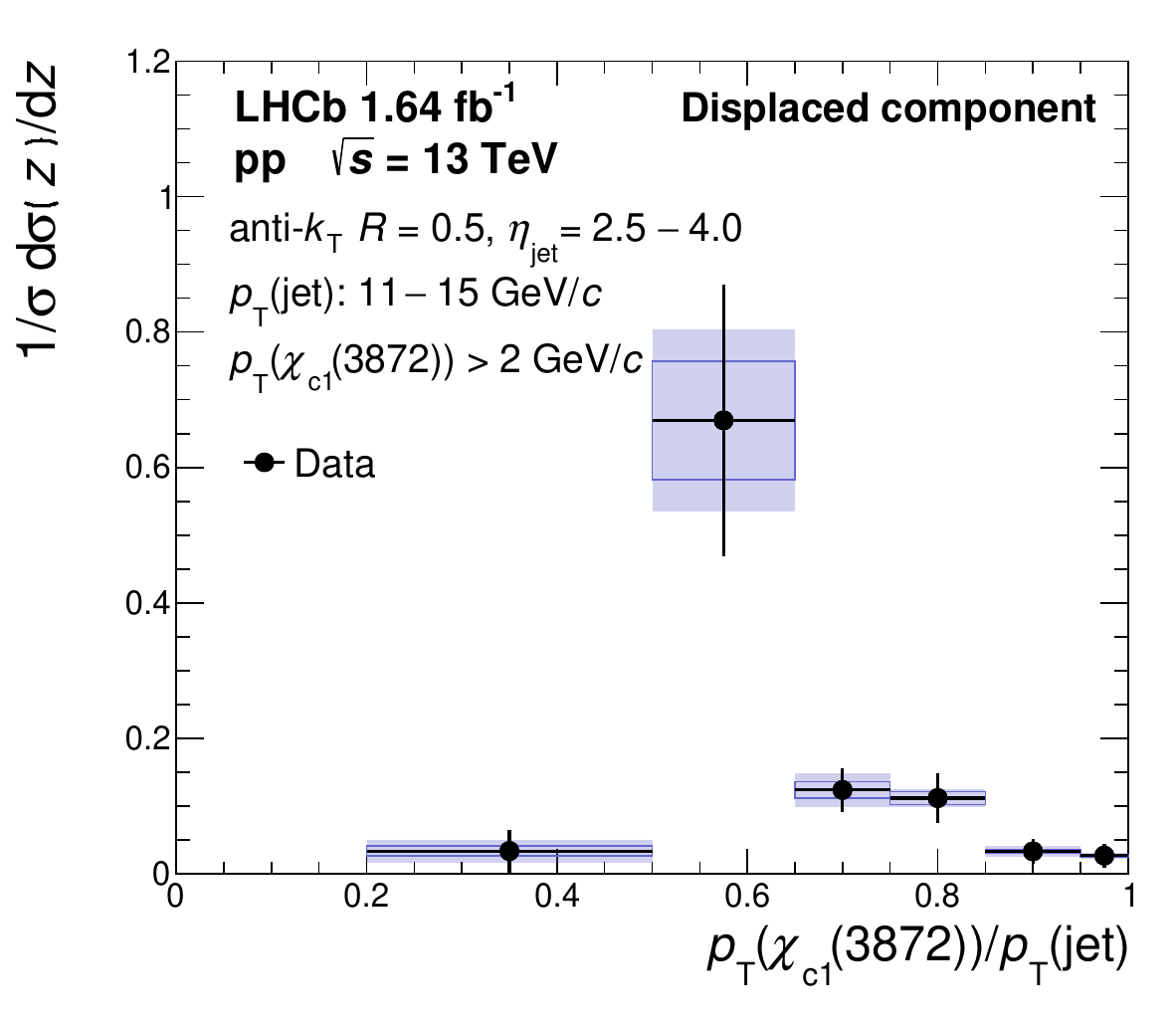}
    \\
    \includegraphics[width=0.397\textwidth]{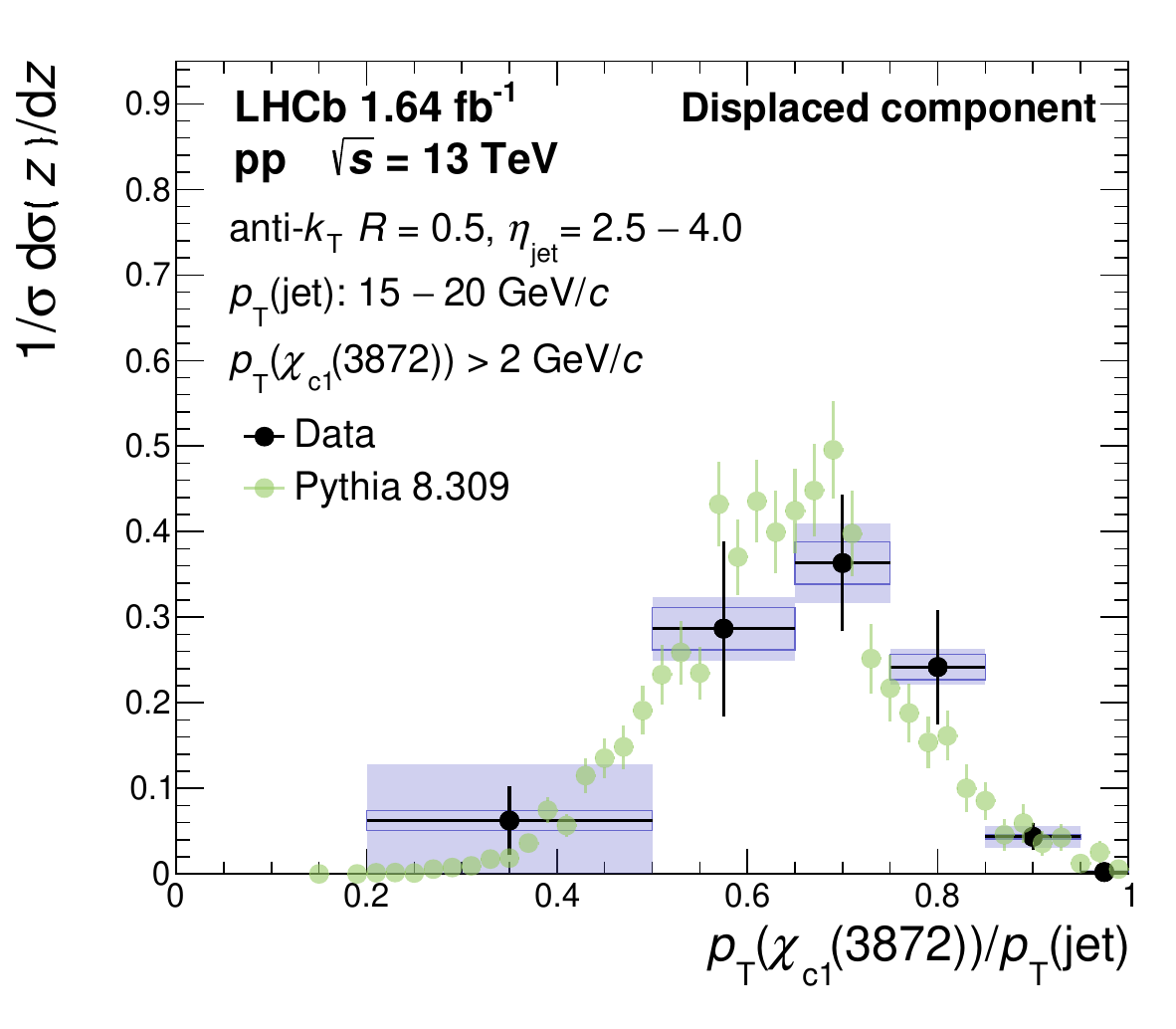}
    \includegraphics[width=0.397\textwidth]{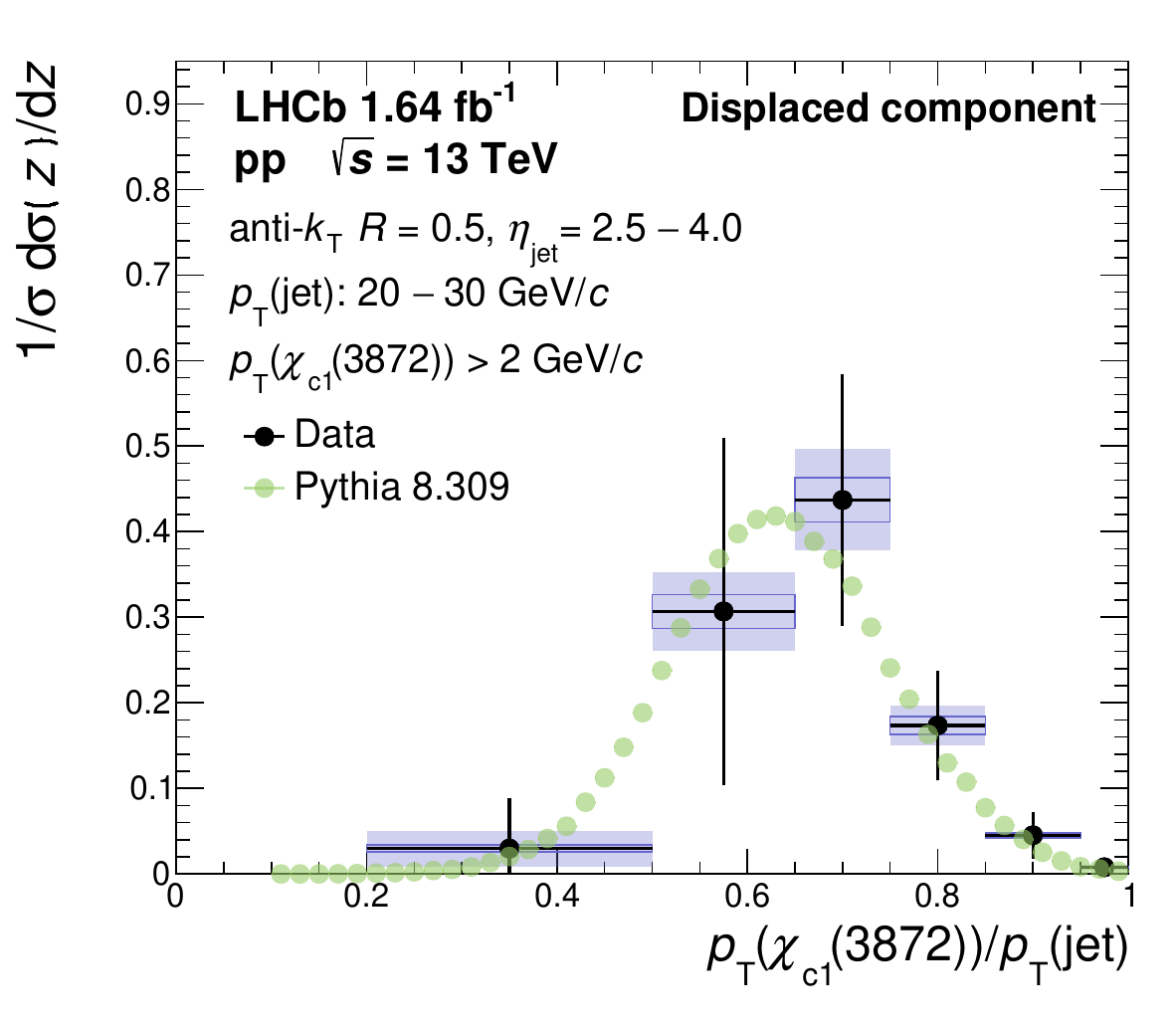}
    \caption{Distributions of the transverse momentum fraction \ensuremath{z} in intervals of jet \ensuremath{p_{\mathrm{T}}} for displaced \ensuremath{\Pchi_{c1}(3872)} production with statistical (black error bars), correlated efficiency (empty blue boxes), and uncorrelated unfolding (shaded blue boxes) uncertainties, compared to predictions obtained from \pythia~8.309.\label{fig:Results_X_NP}}
\end{figure}

\begin{figure}
    \centering
    \includegraphics[width=0.397\textwidth]{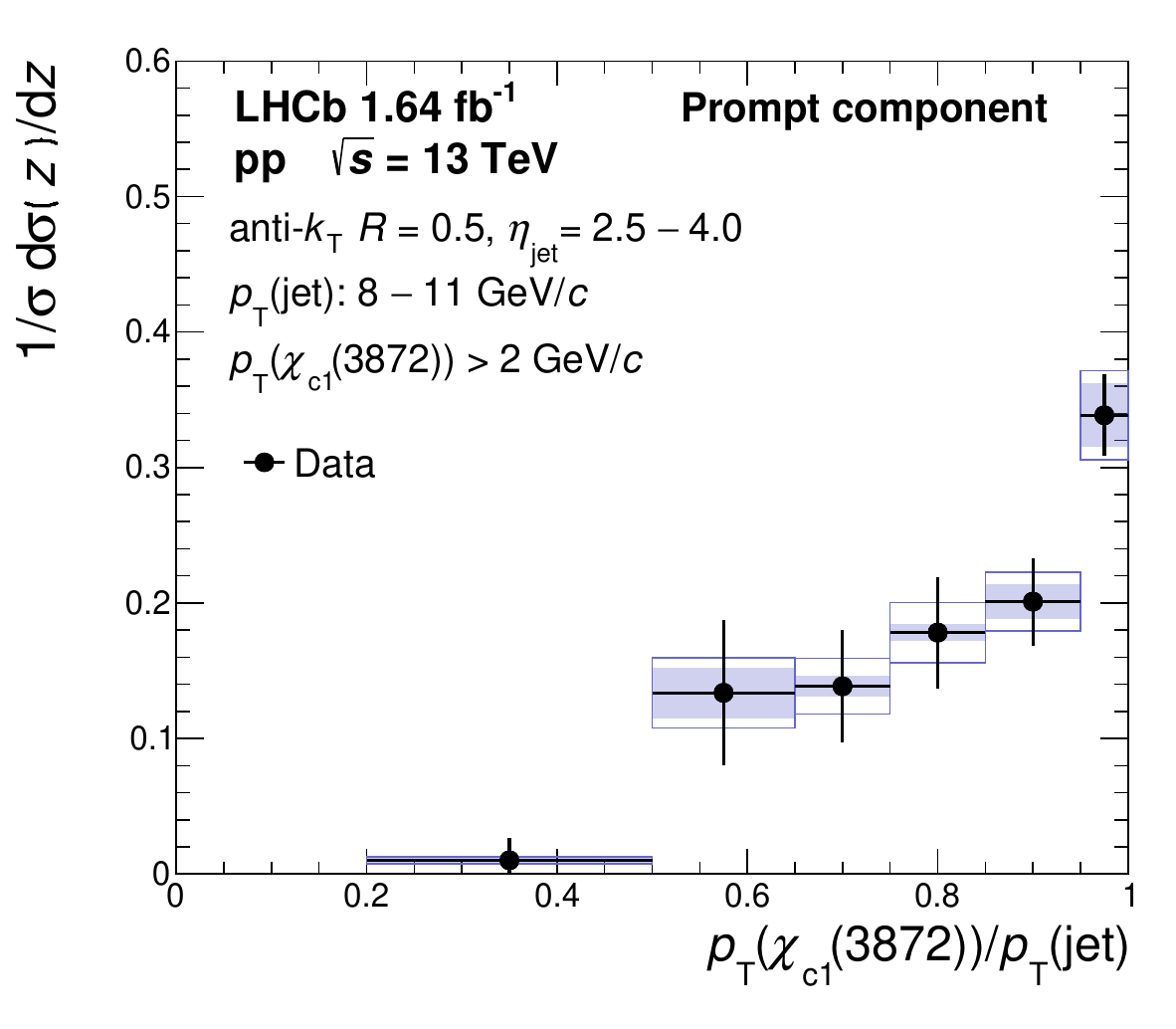}
    \includegraphics[width=0.397\textwidth]{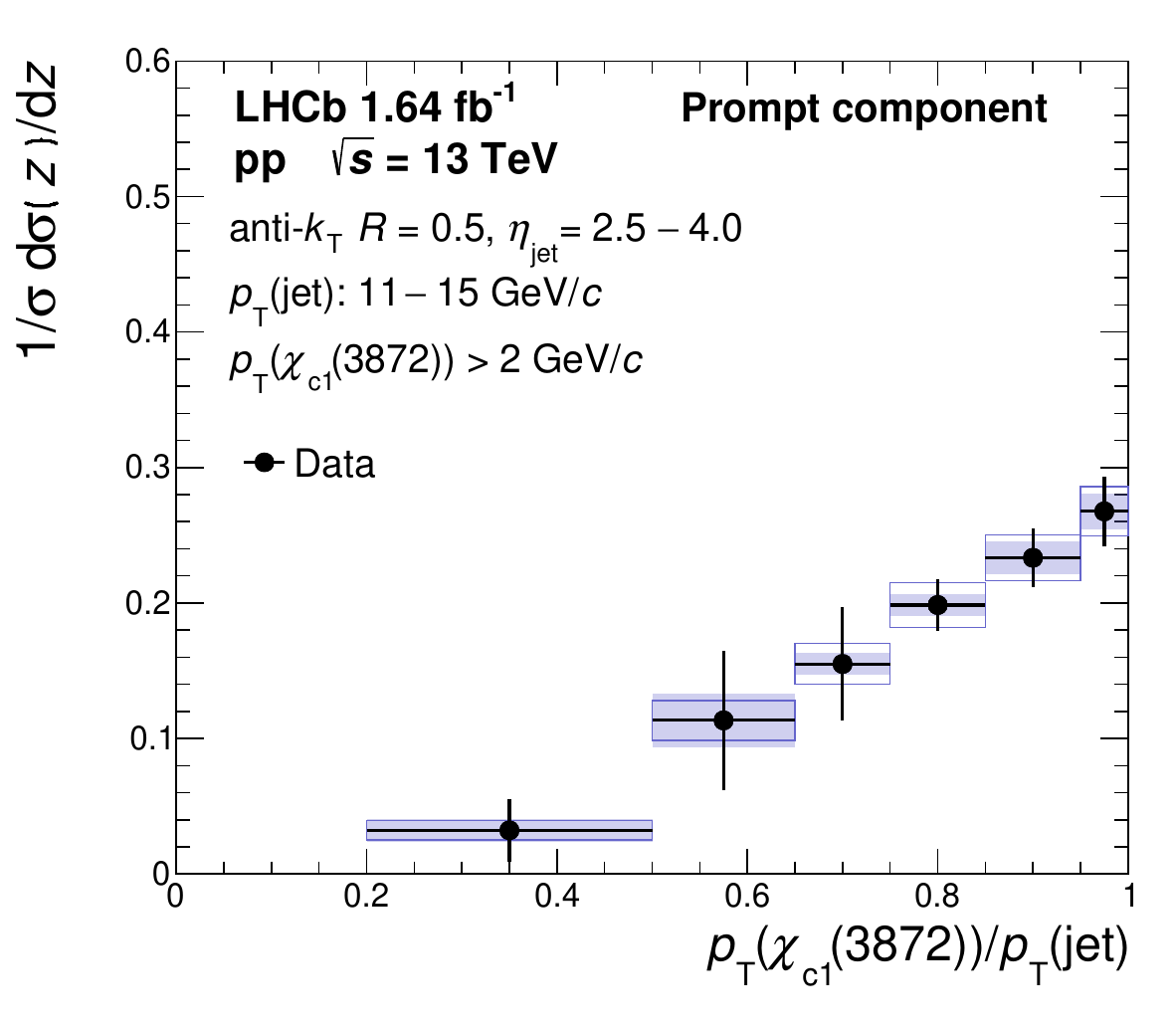}
    \\
    \includegraphics[width=0.397\textwidth]{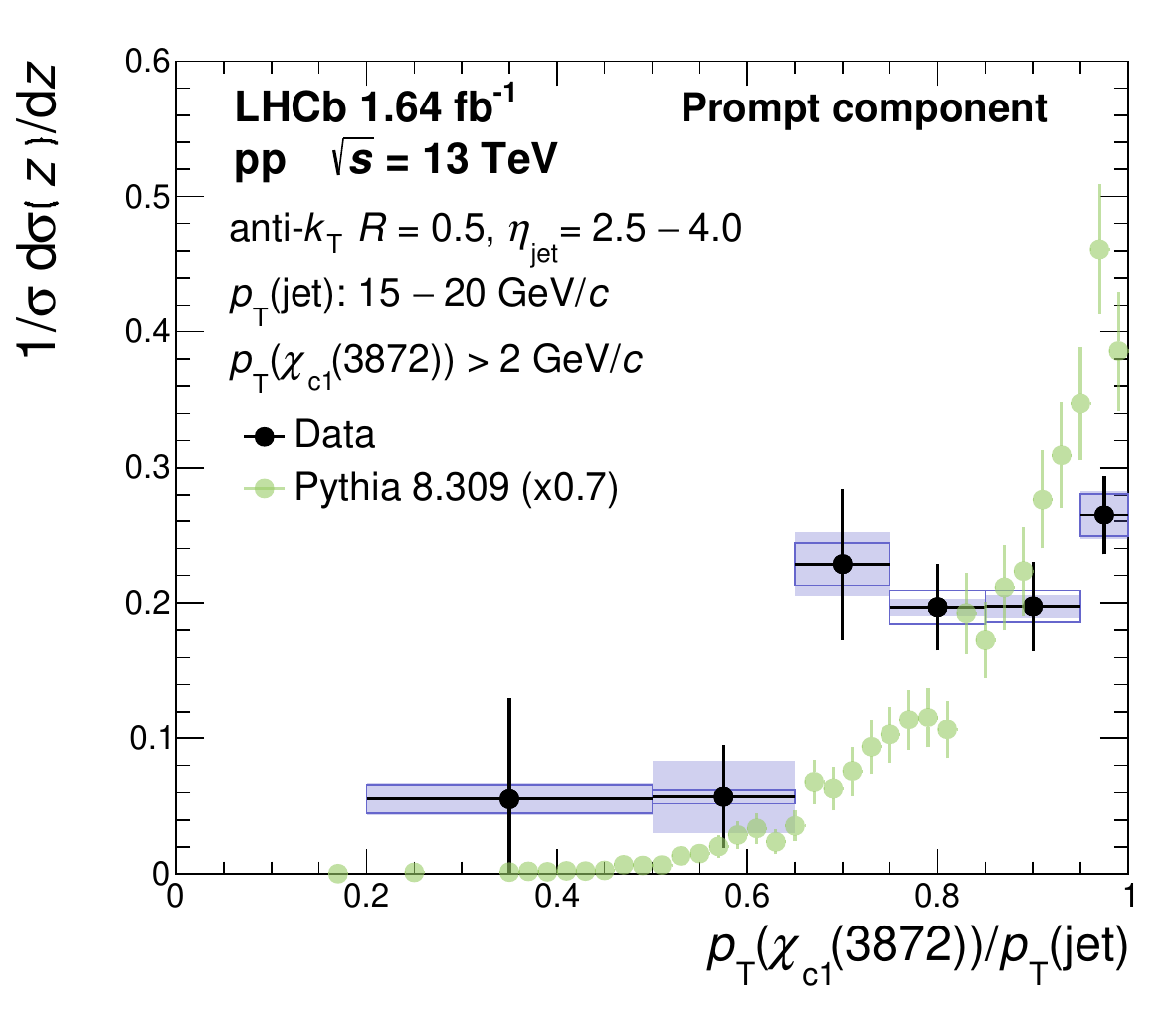}
    \includegraphics[width=0.397\textwidth]{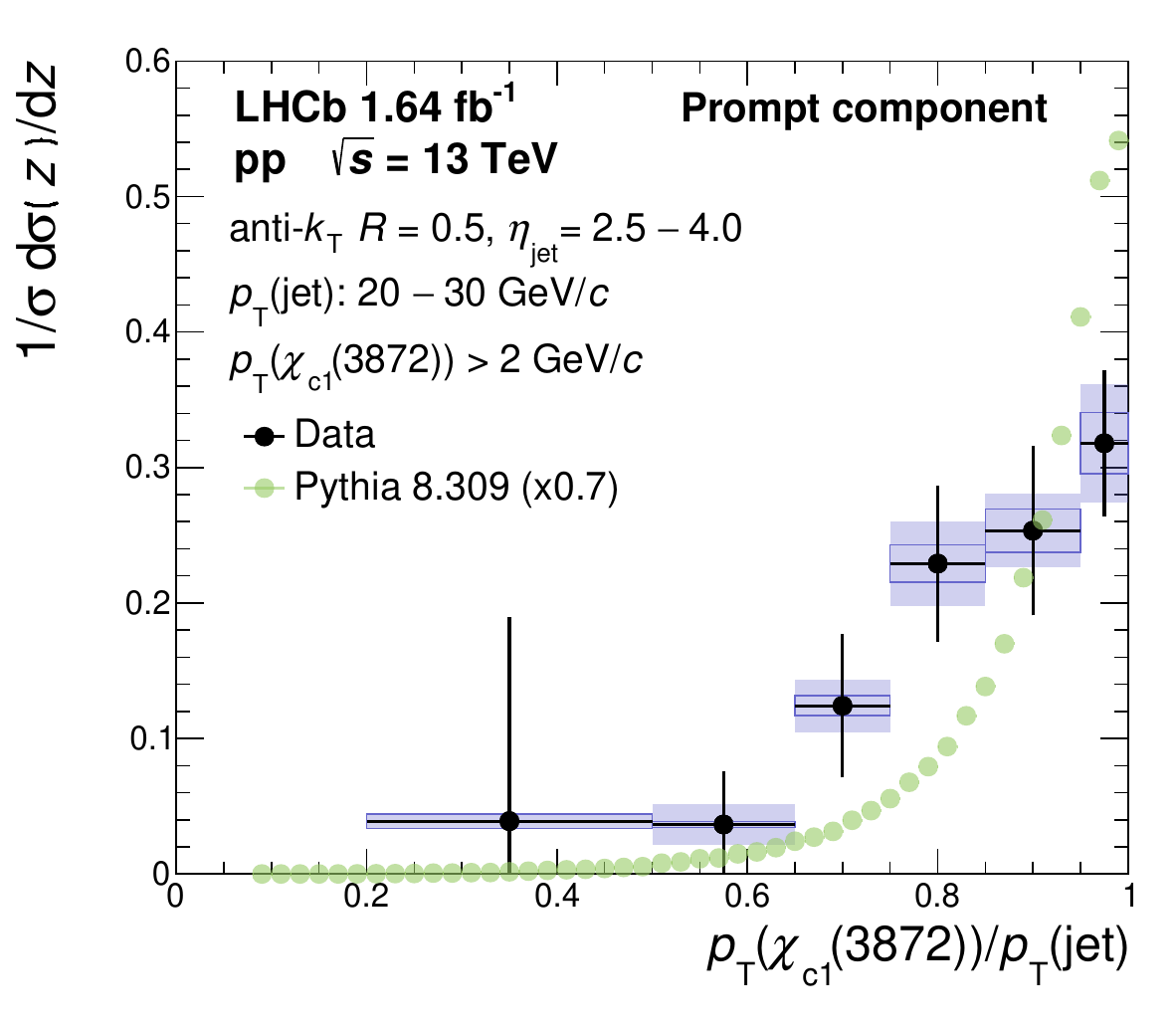}
    \caption{Distributions of the transverse momentum fraction \ensuremath{z} in intervals of jet \ensuremath{p_{\mathrm{T}}} for prompt \ensuremath{\Pchi_{c1}(3872)} production with statistical (black error bars), correlated efficiency (empty blue boxes), and uncorrelated unfolding (shaded blue boxes) uncertainties, compared to predictions obtained from \pythia~8.309.\label{fig:Results_X_P}}
\end{figure}

\Cref{fig:Results_P2_NP,fig:Results_P2_P} display the normalized \zt distributions in $\pt(\jet)$ intervals for displaced and prompt \psitwos meson production, respectively. Similarly, \cref{fig:Results_X_NP,fig:Results_X_P} provide the same distributions for the \theX meson production. The \zt distribution for each interval of $\pt(\jet)$ is normalized to unity. 
\par
\begin{figure}[htb]
    \centering
    \includegraphics[width=0.397\textwidth]{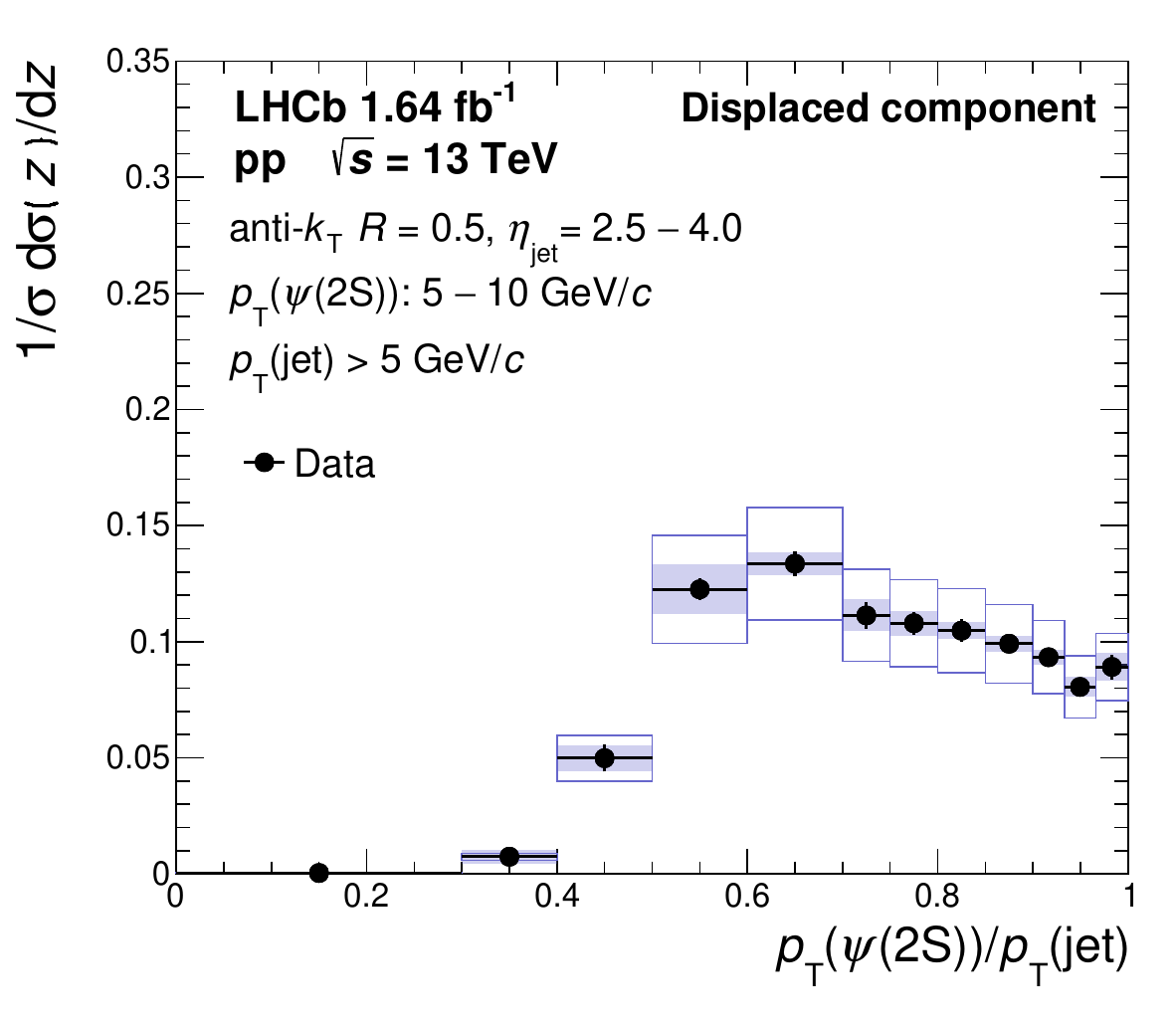}
    \includegraphics[width=0.397\textwidth]{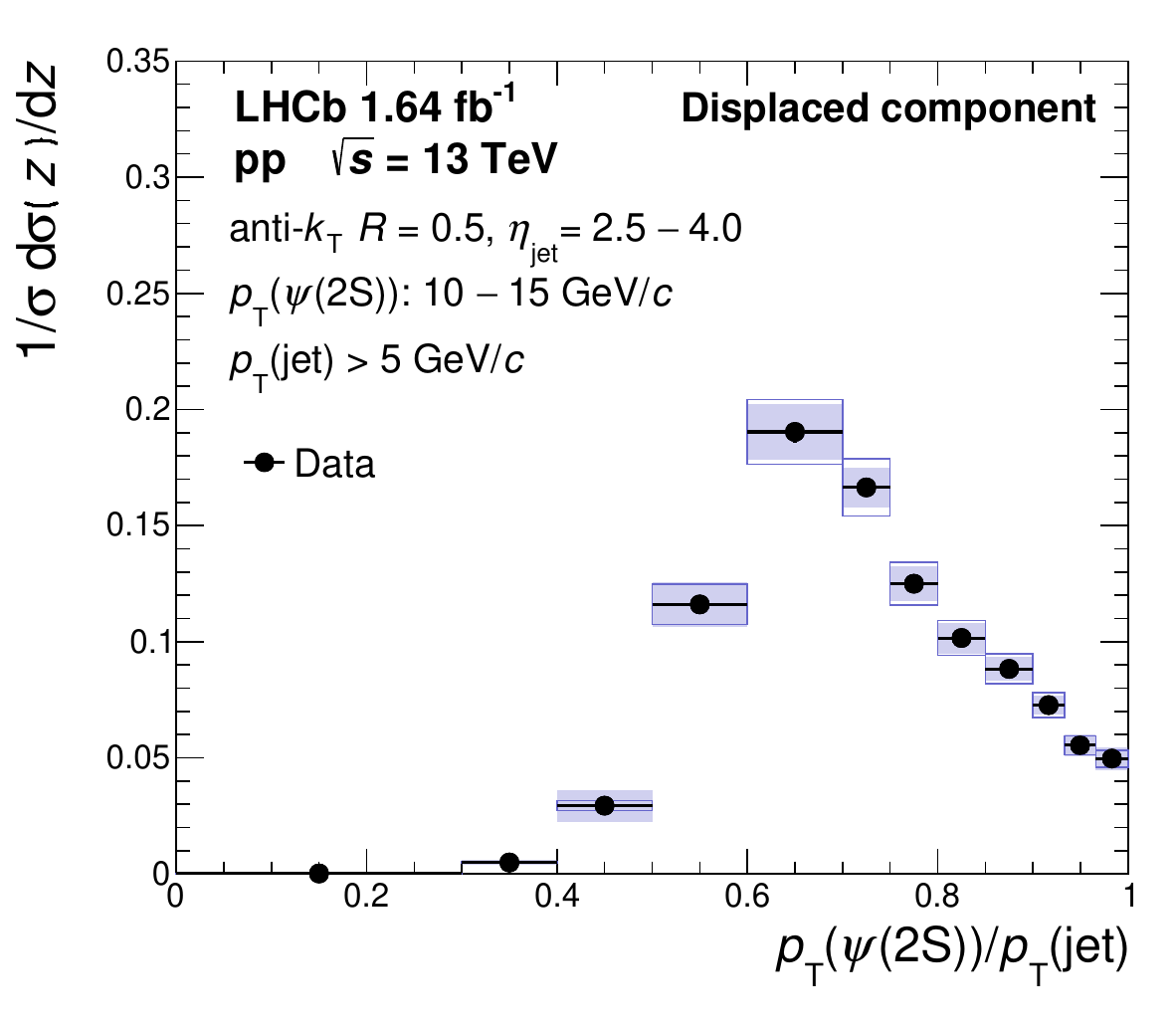}
    \\
    \includegraphics[width=0.397\textwidth]{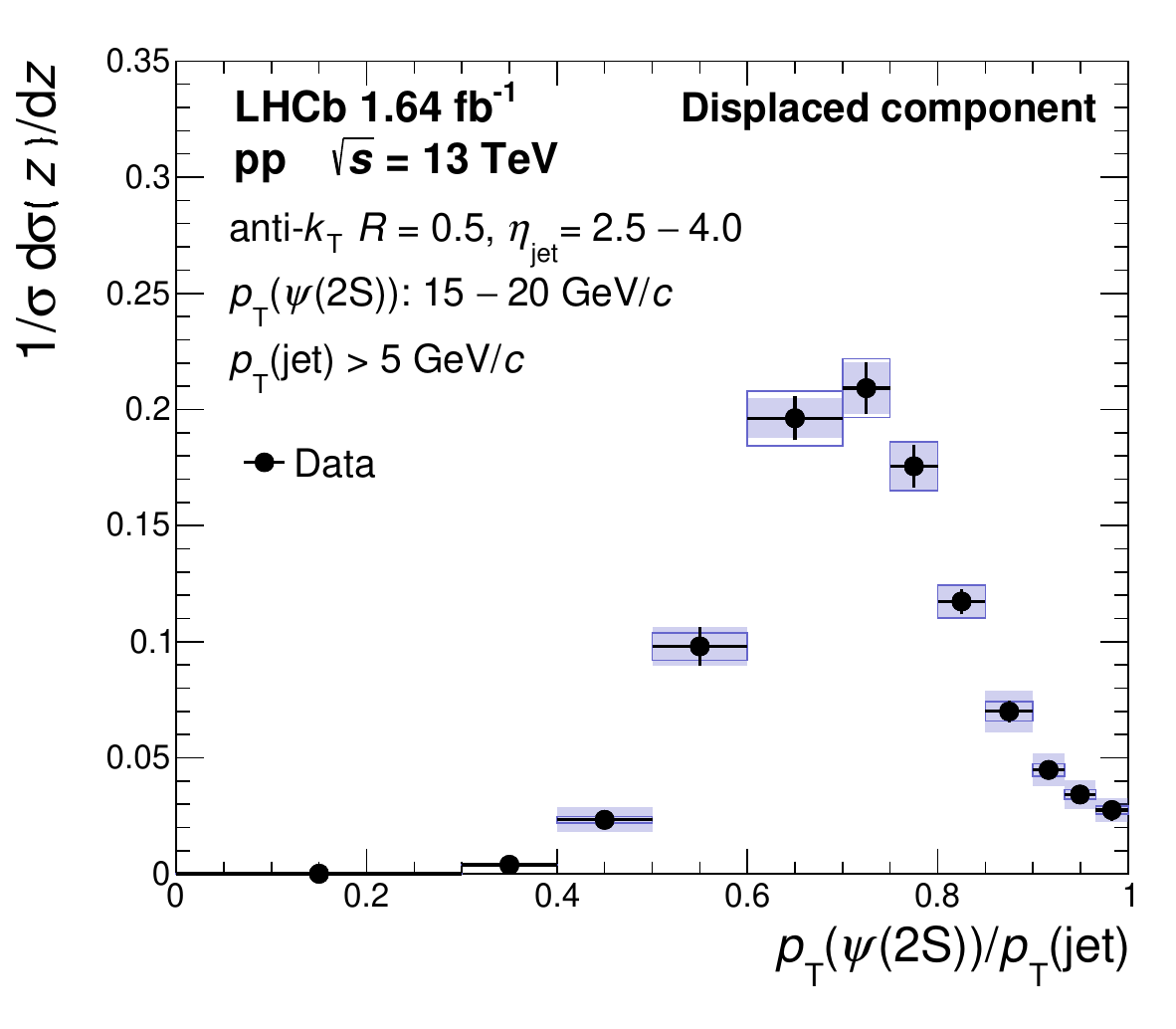}
    \includegraphics[width=0.397\textwidth]{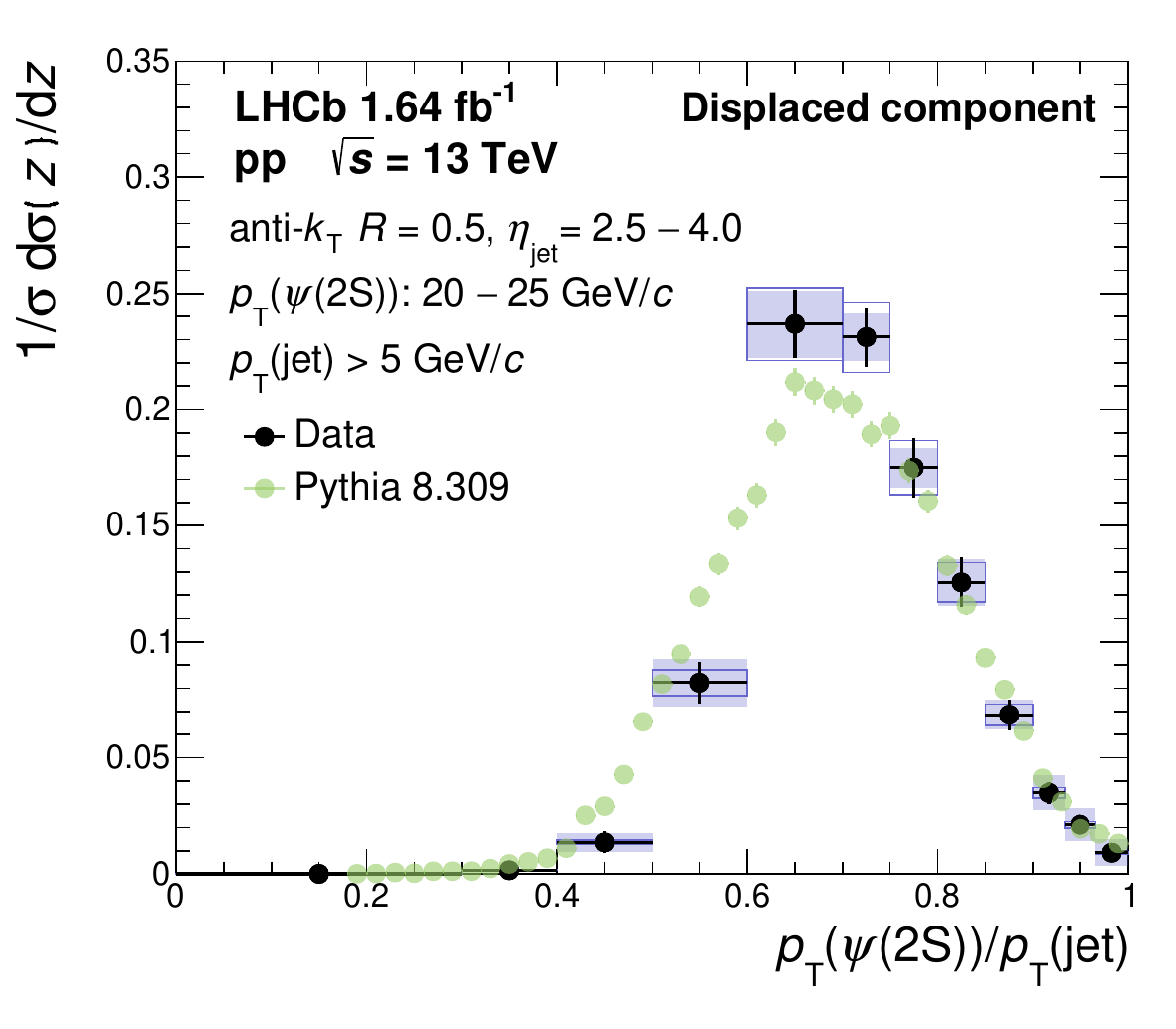}
    \\
    \includegraphics[width=0.397\textwidth]{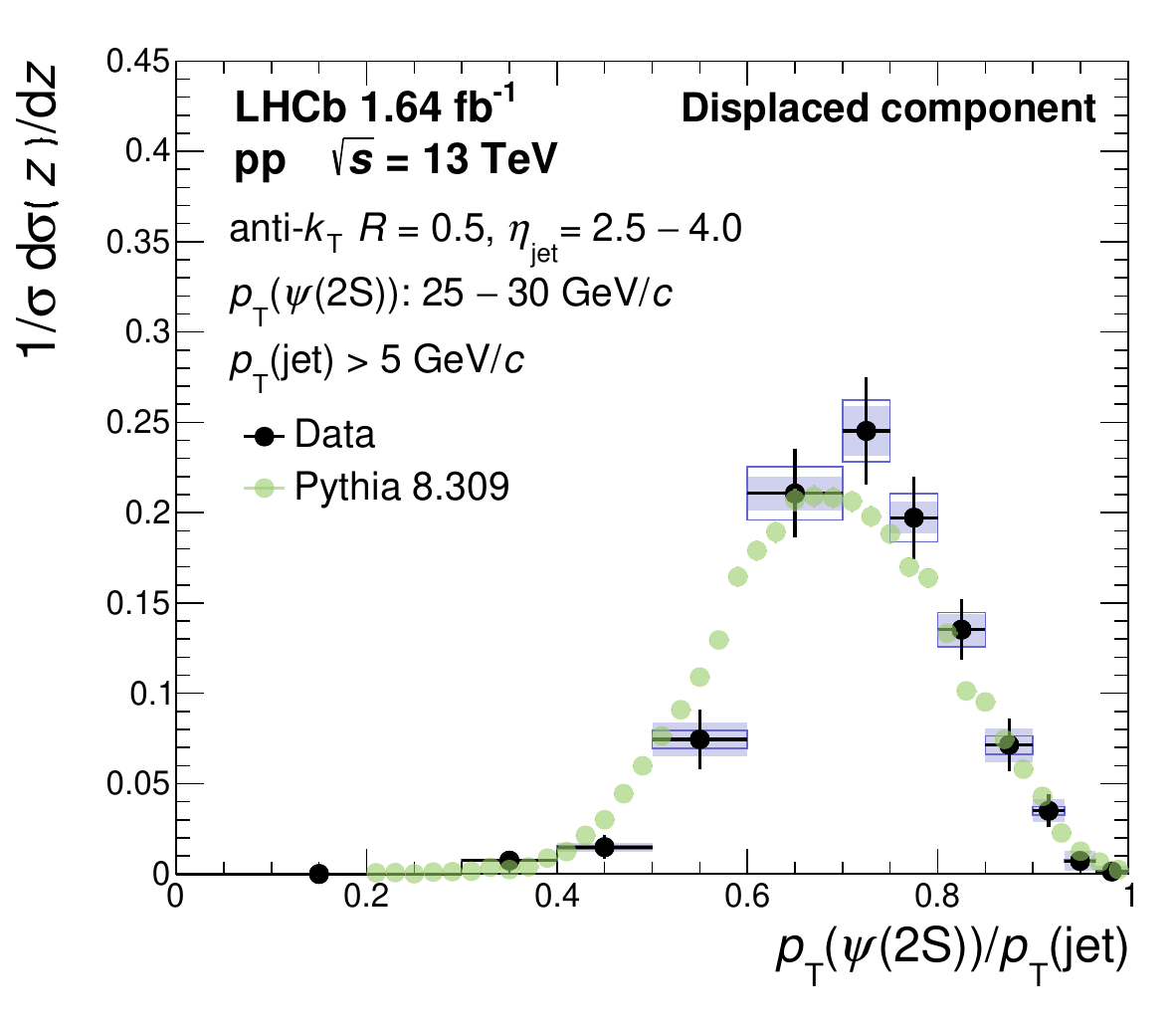}
    \includegraphics[width=0.397\textwidth]{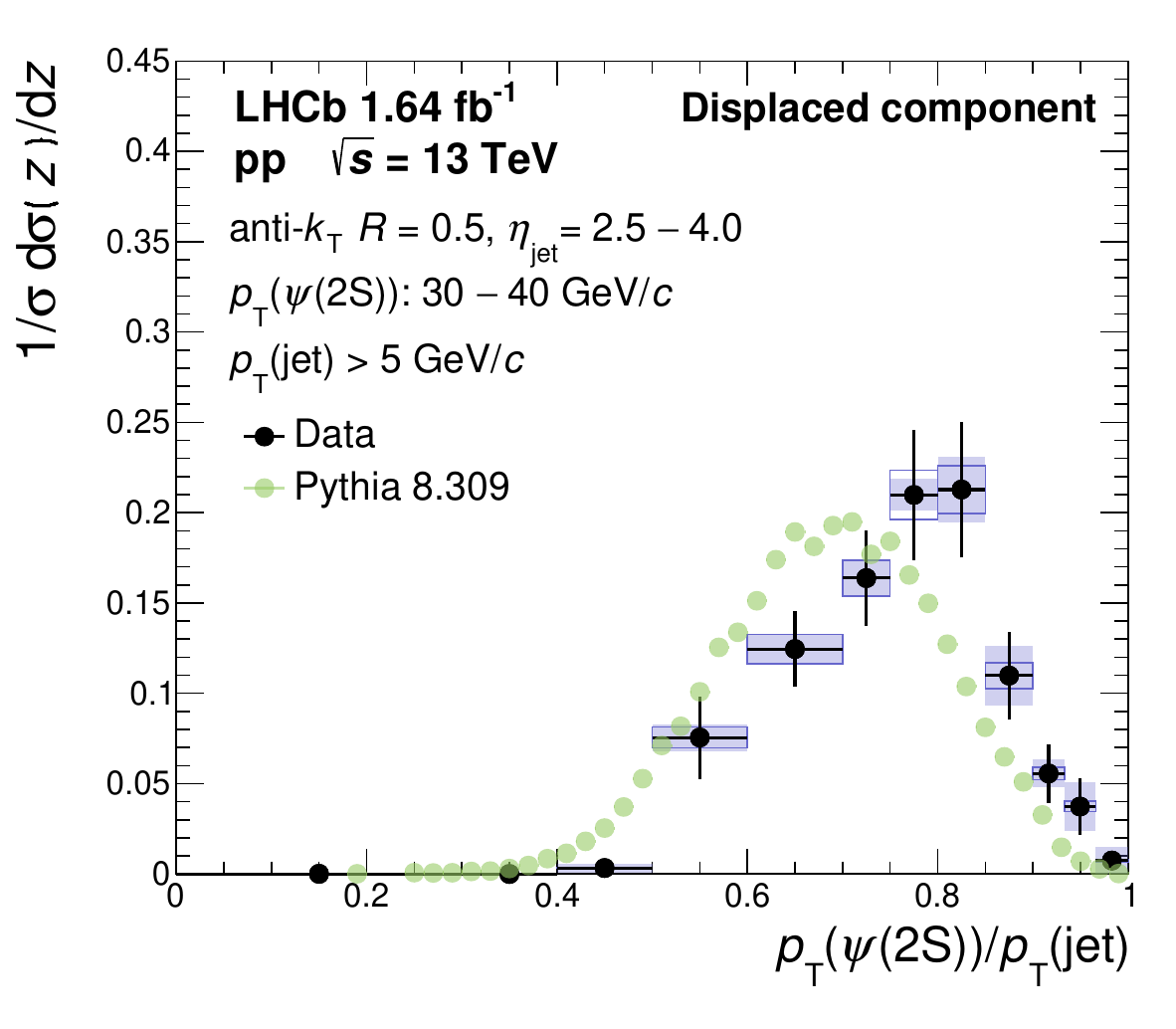}
    \caption{Distributions of the transverse momentum fraction \ensuremath{z} in intervals of \ensuremath{\Ppsi{(2S)}} \ensuremath{p_{\mathrm{T}}} for displaced \ensuremath{\Ppsi{(2S)}} production with statistical (black error bars), correlated efficiency (empty blue boxes), and uncorrelated unfolding (shaded blue boxes) uncertainties, compared to predictions obtained from \pythia~8.309.\label{fig:Results_P2_NP_TagpT}}
\end{figure}
\begin{figure}[htb]
    \centering
    \includegraphics[width=0.397\textwidth]{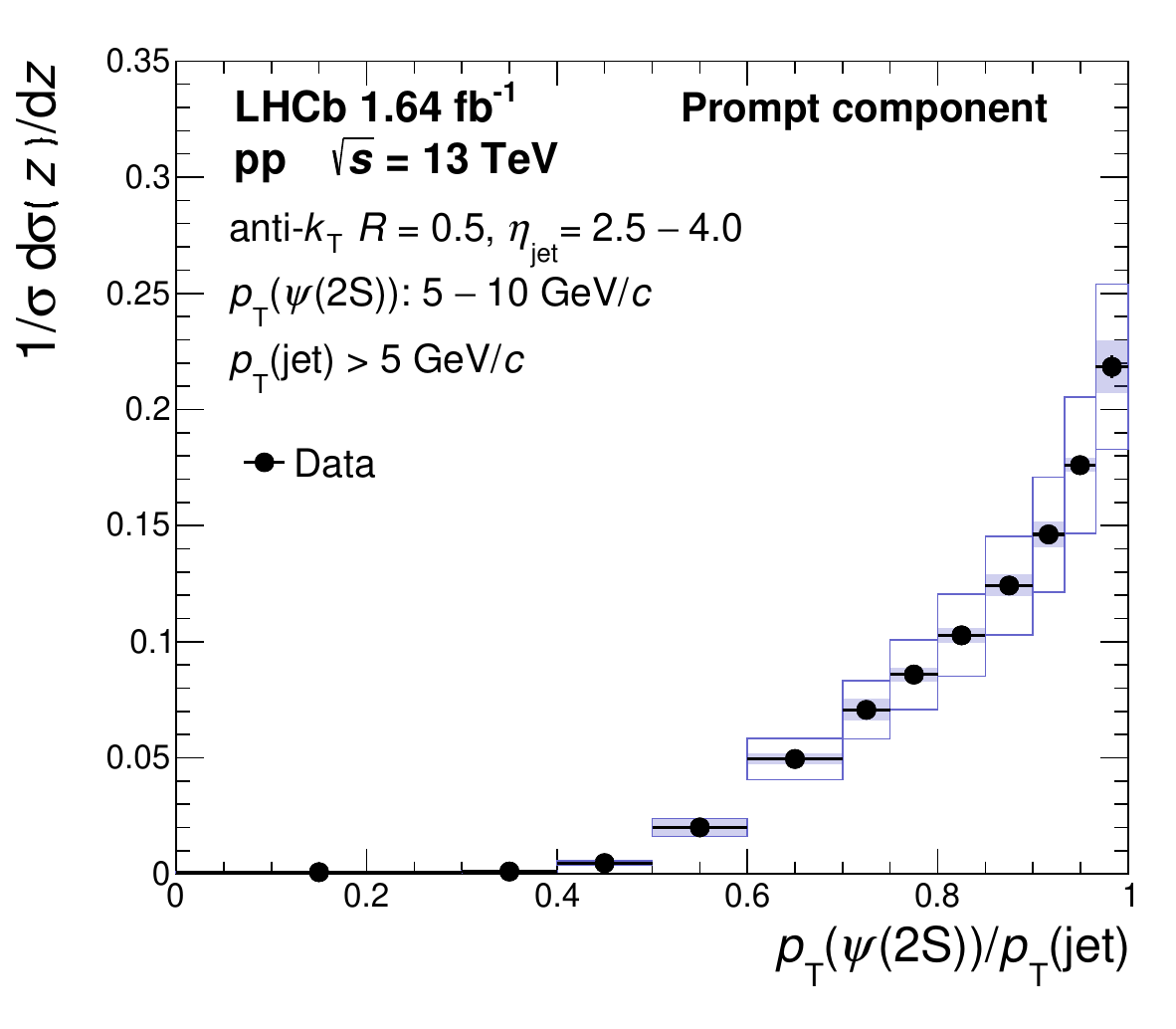}
    \includegraphics[width=0.397\textwidth]{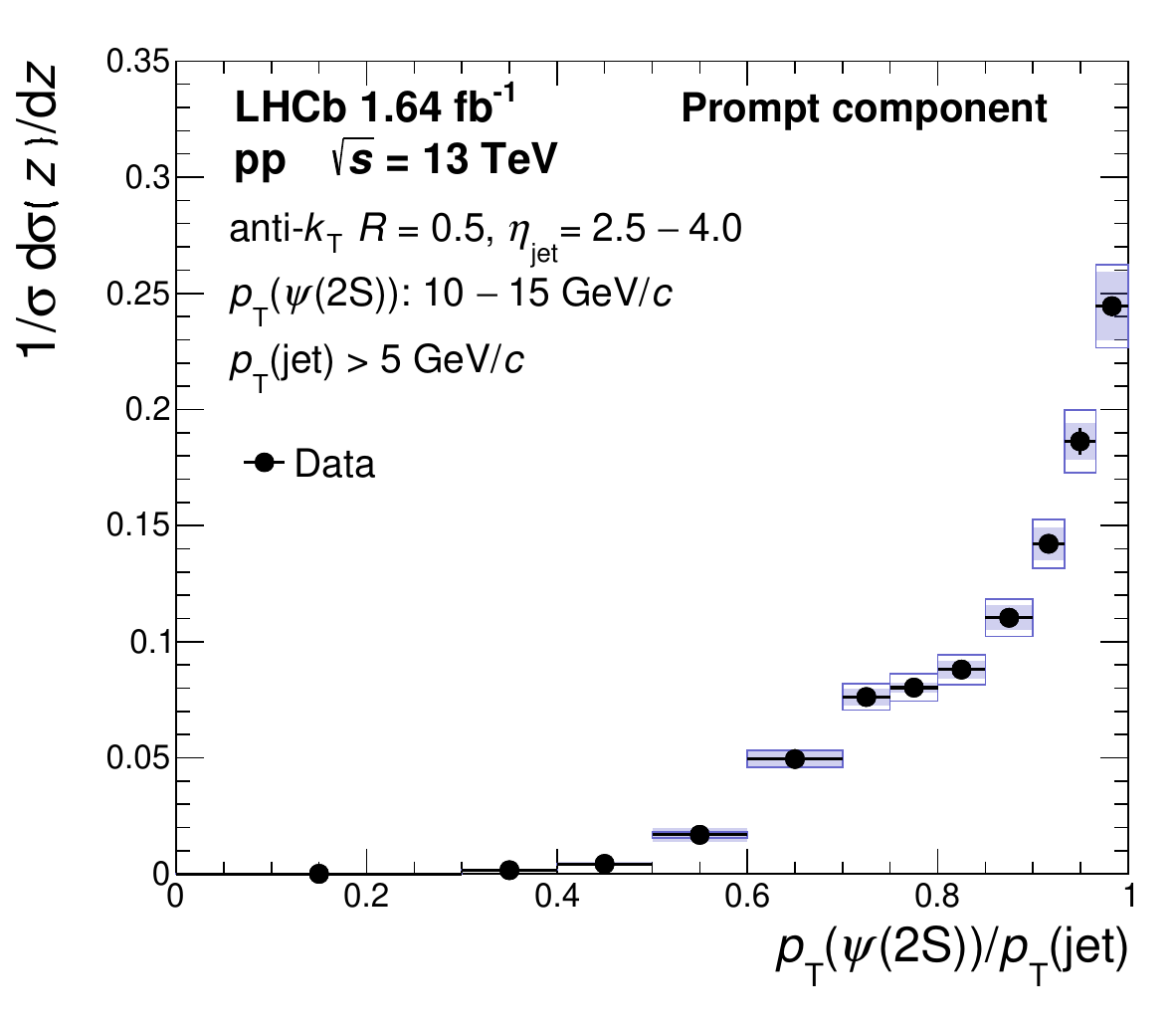}
    \\
    \includegraphics[width=0.397\textwidth]{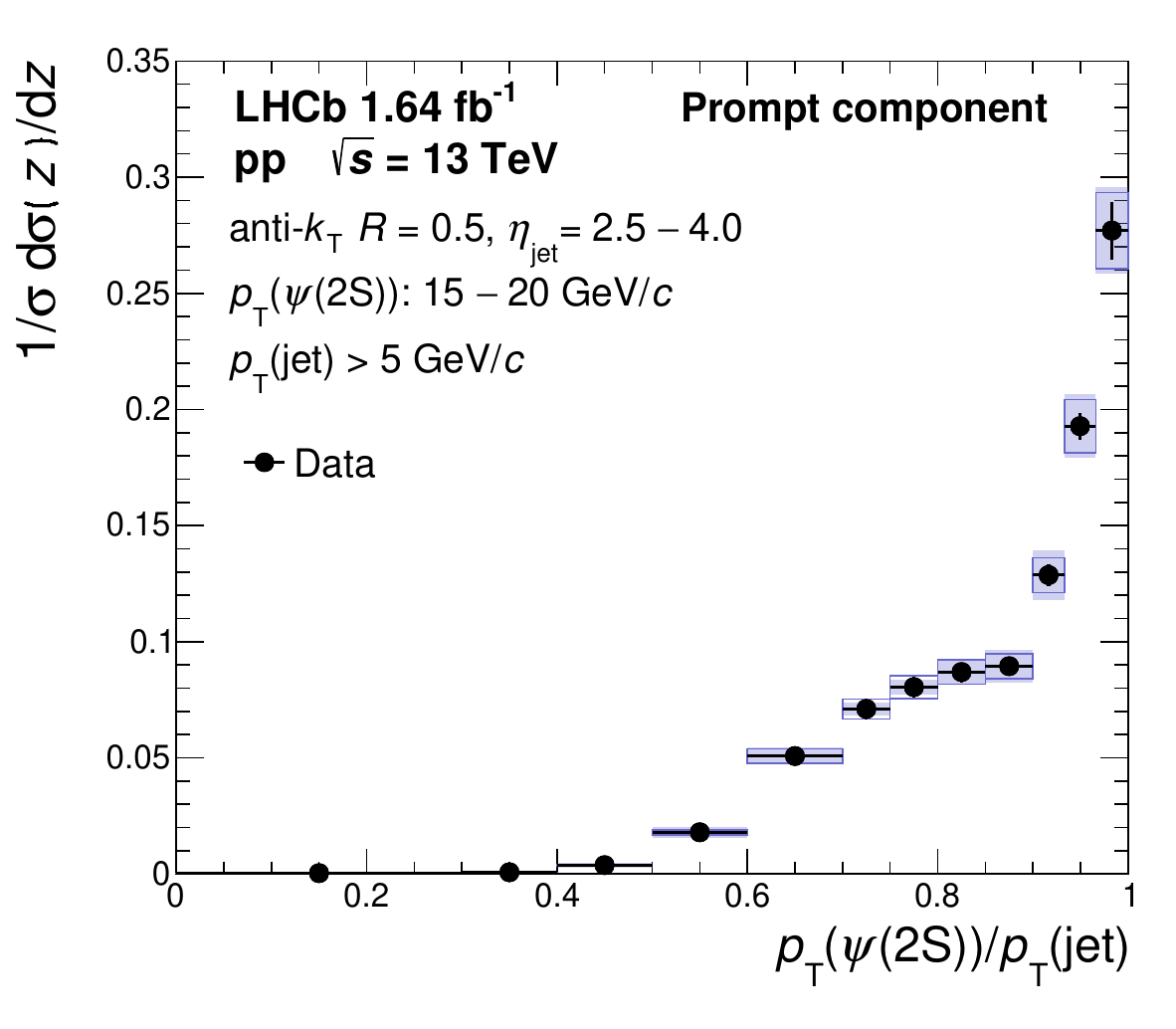}
    \includegraphics[width=0.397\textwidth]{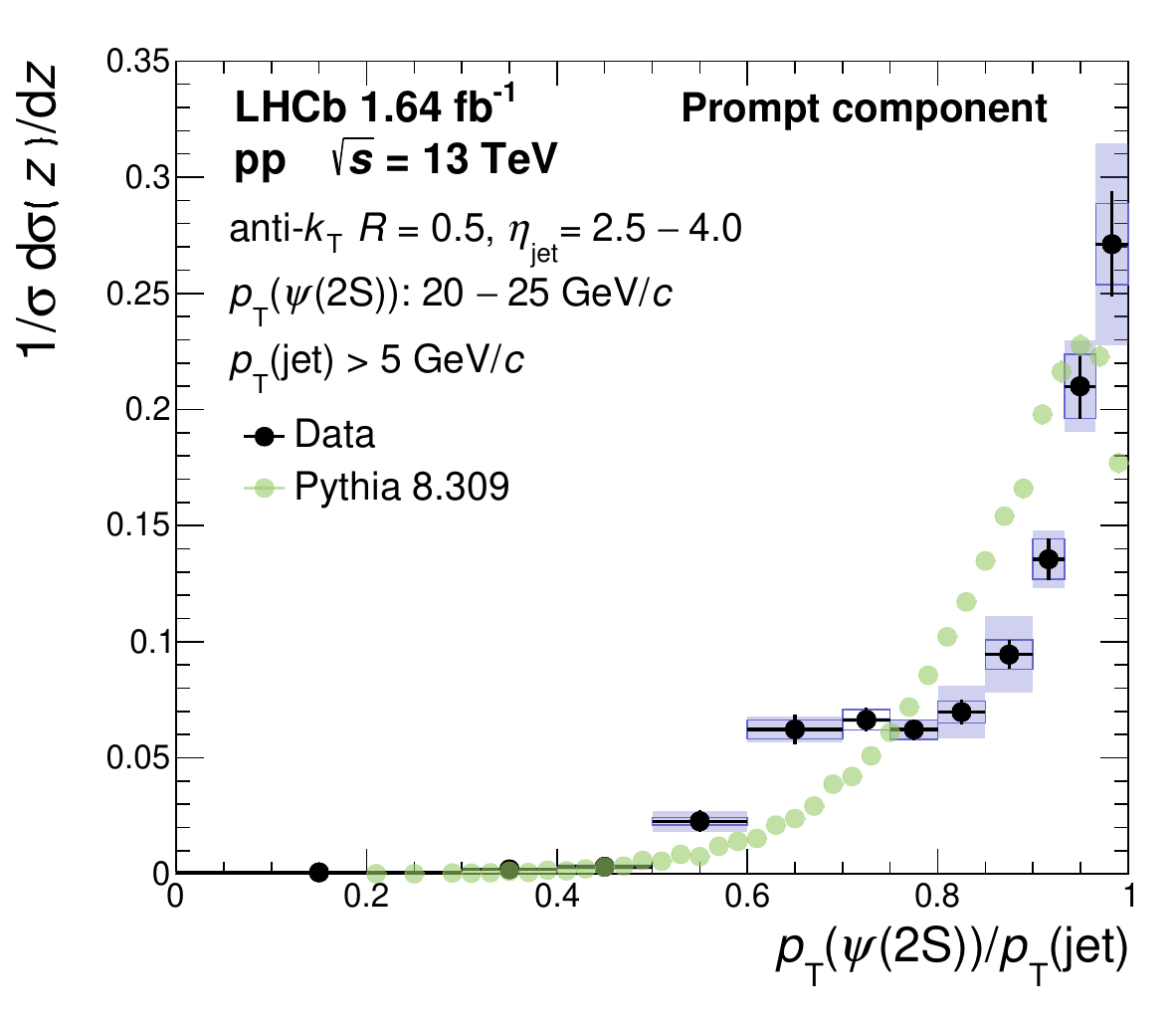}
    \\
    \includegraphics[width=0.397\textwidth]{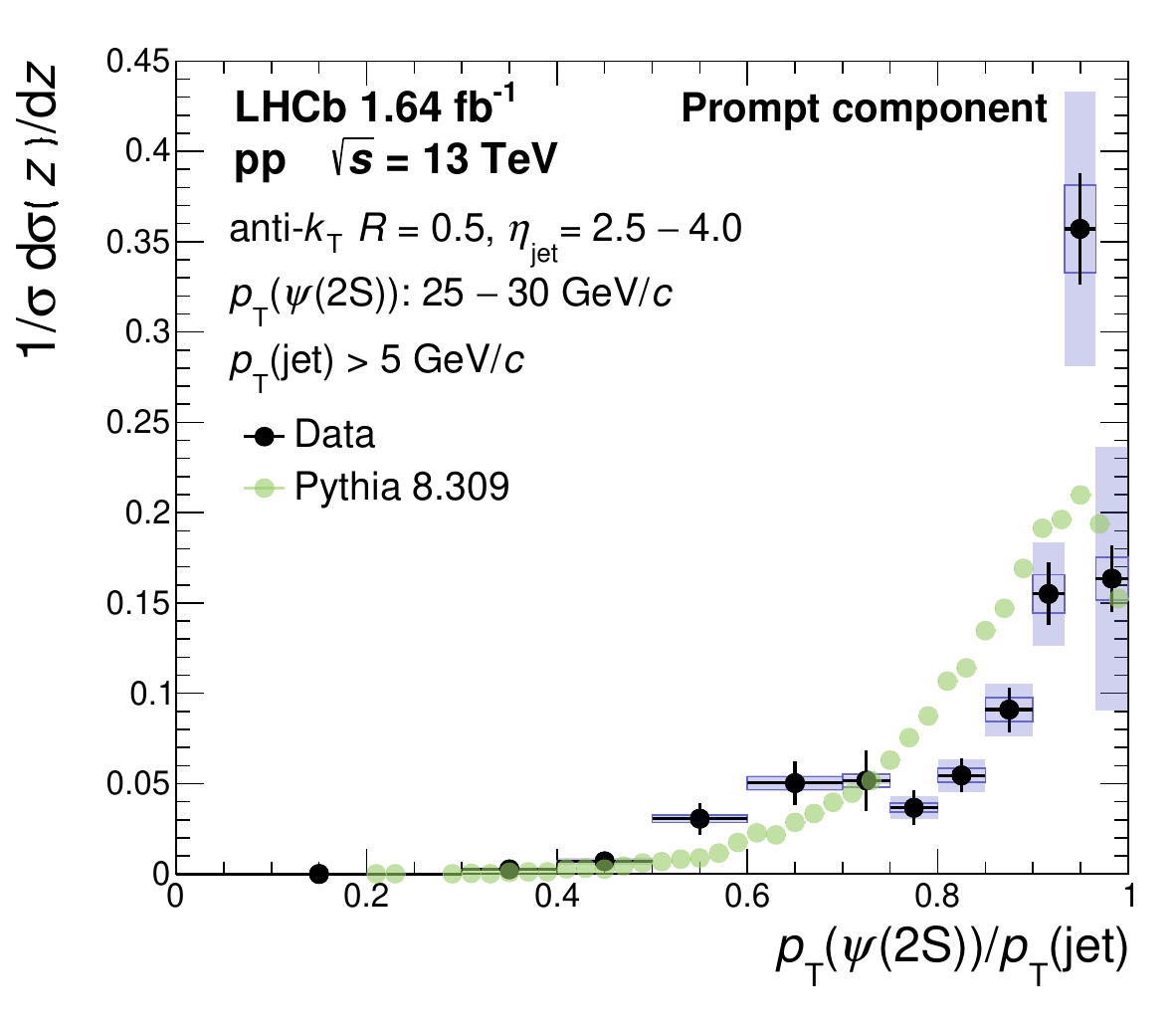}
    \includegraphics[width=0.397\textwidth]{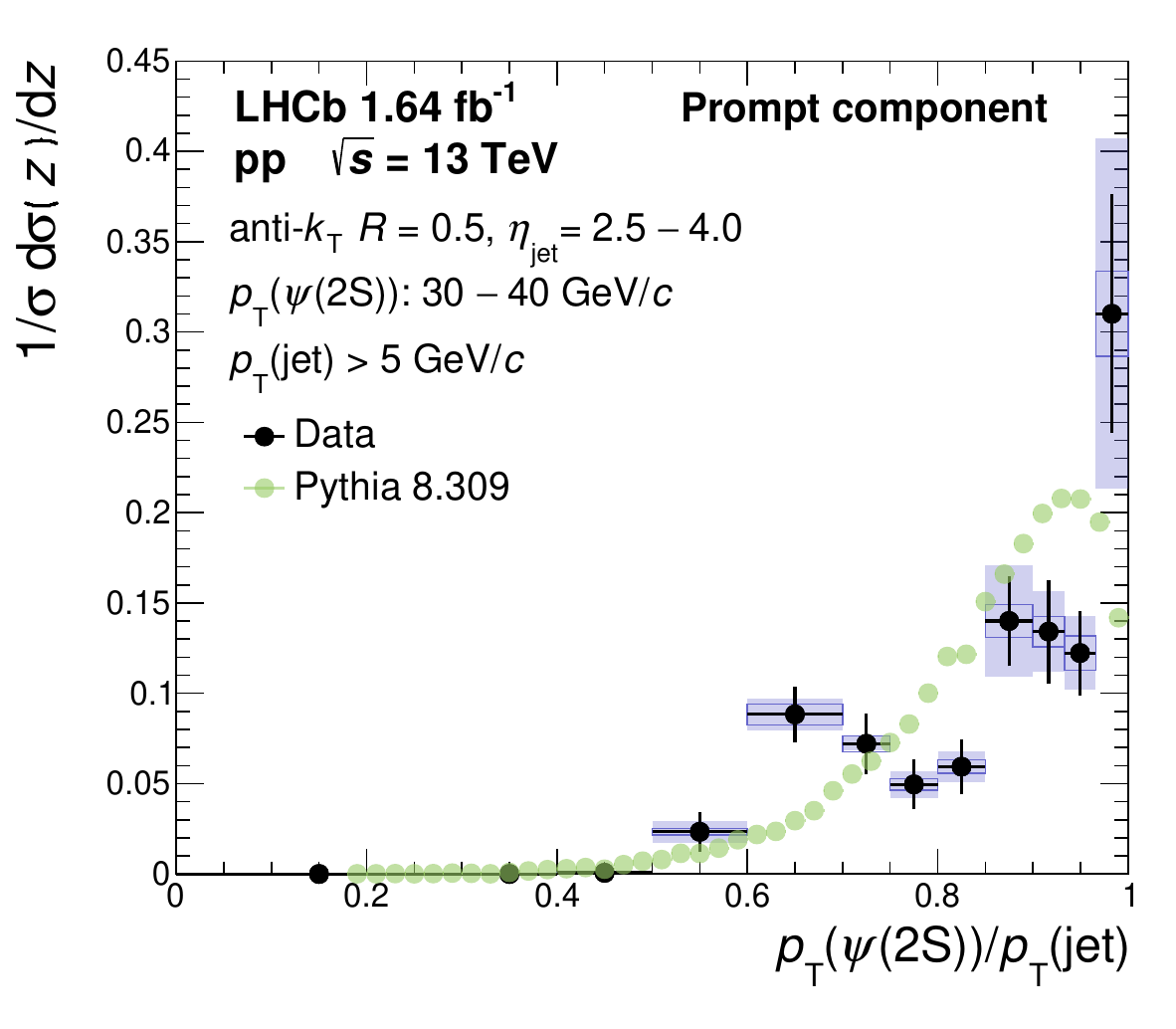}
    \caption{Distributions of the transverse momentum fraction \ensuremath{z} in intervals of \ensuremath{\Ppsi{(2S)}} \ensuremath{p_{\mathrm{T}}} for prompt \ensuremath{\Ppsi{(2S)}} production with statistical (black error bars), correlated efficiency (empty blue boxes), and uncorrelated unfolding (shaded blue boxes) uncertainties, compared to predictions obtained from \pythia~8.309.\label{fig:Results_P2_P_TagpT}}
\end{figure}
\begin{figure}[htb]
    \centering
    \includegraphics[width=0.397\textwidth]{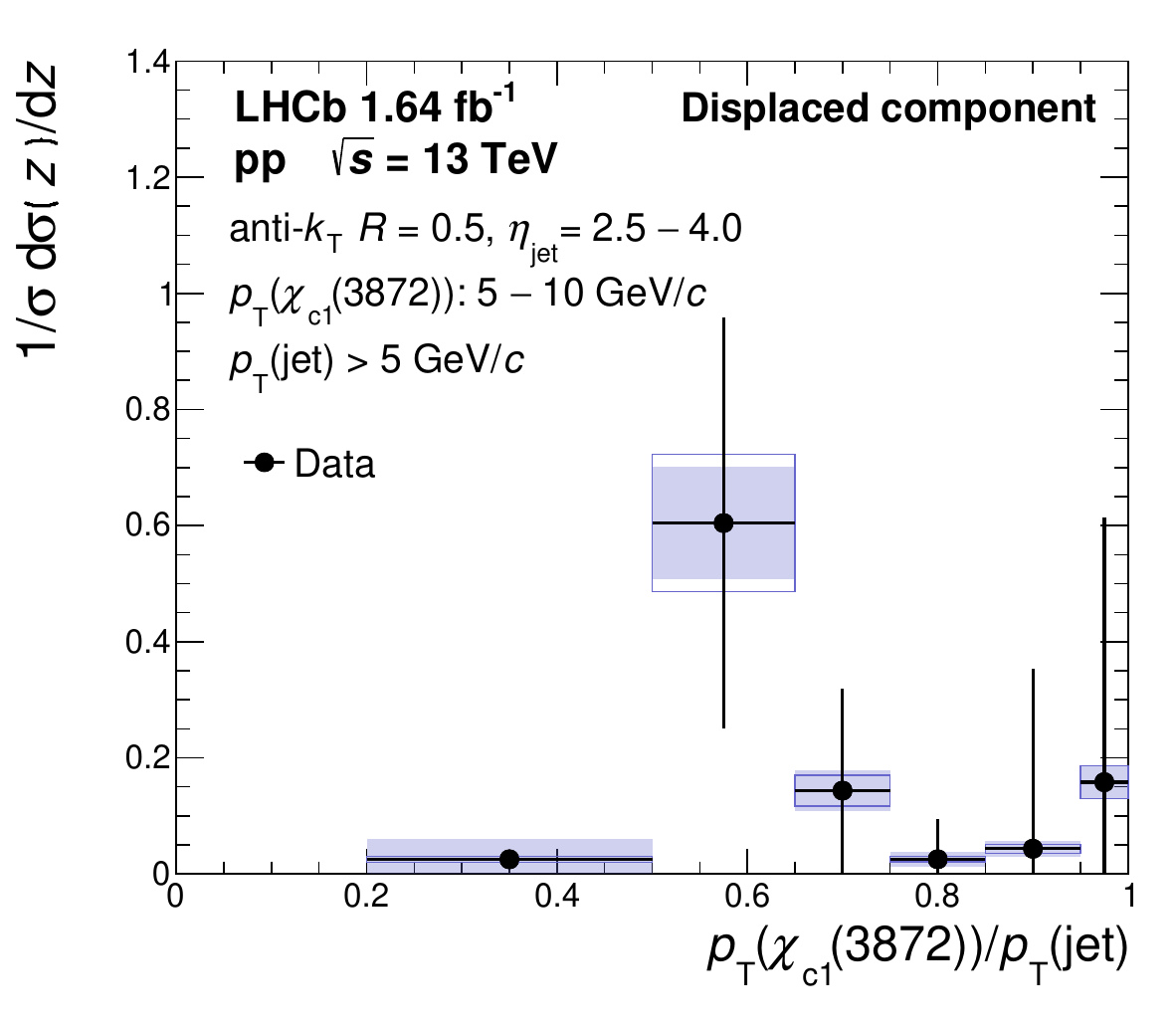}
    \includegraphics[width=0.397\textwidth]{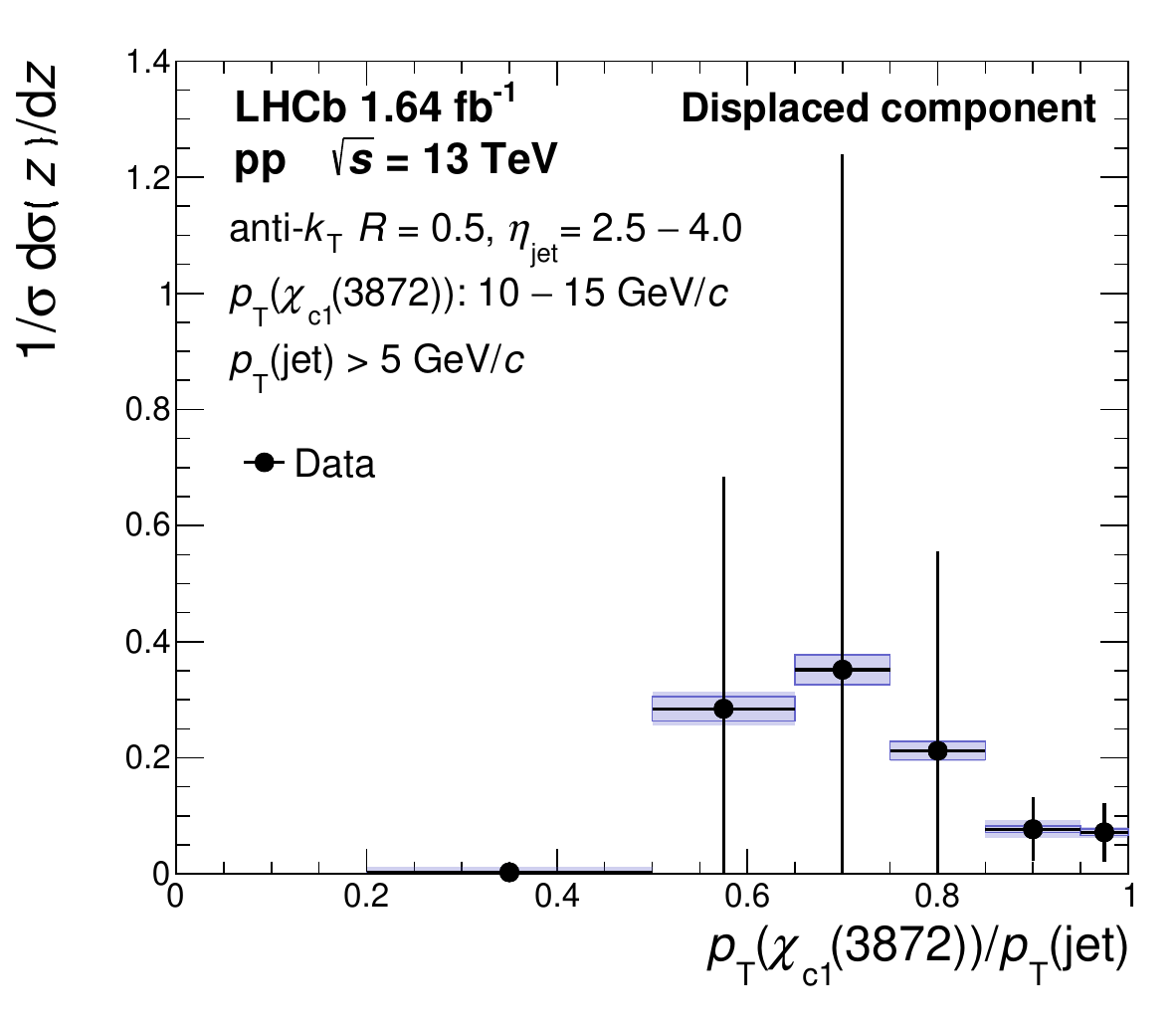}
    \\
    \includegraphics[width=0.397\textwidth]{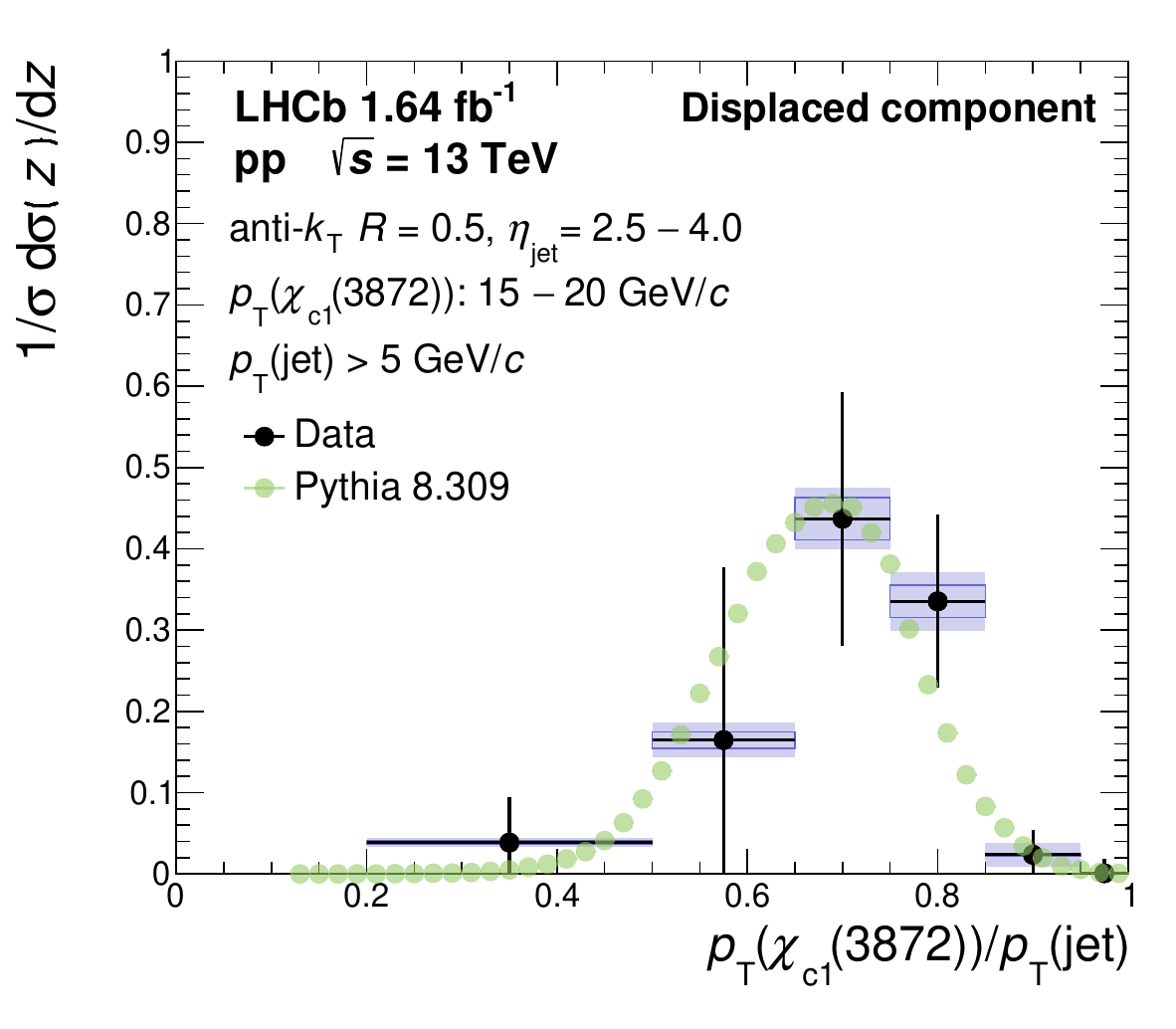}
    \includegraphics[width=0.397\textwidth]{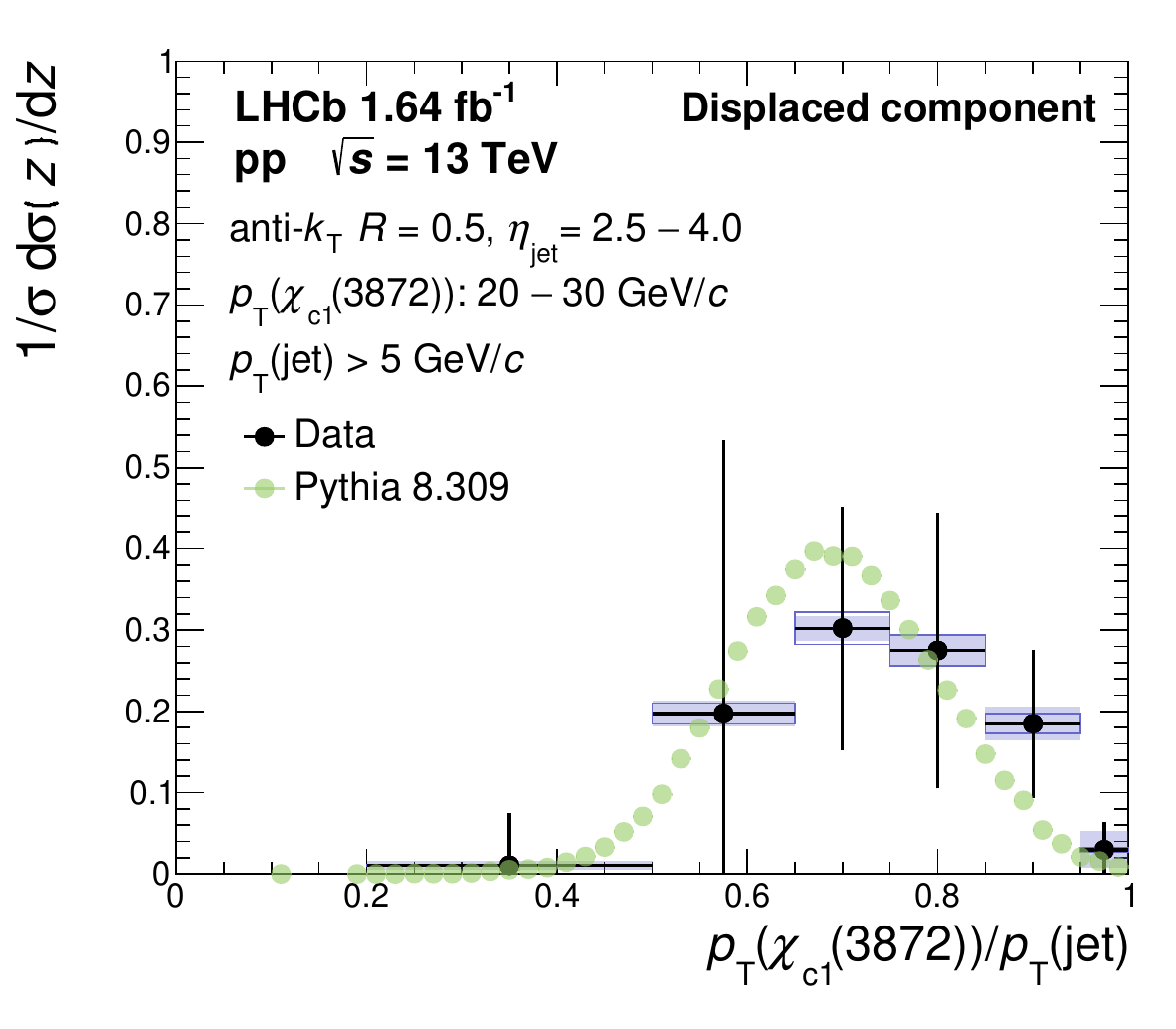}
    \caption{Distributions of the transverse momentum fraction \ensuremath{z} in intervals of \ensuremath{\Pchi_{c1}(3872)} \ensuremath{p_{\mathrm{T}}} for displaced \ensuremath{\Pchi_{c1}(3872)} production with statistical (black error bars), correlated efficiency (empty blue boxes), and uncorrelated unfolding (shaded blue boxes) uncertainties, compared to predictions obtained from \pythia~8.309.\label{fig:Results_X_NP_TagpT}}
\end{figure}
\begin{figure}[htb]
    \centering
    \includegraphics[width=0.397\textwidth]{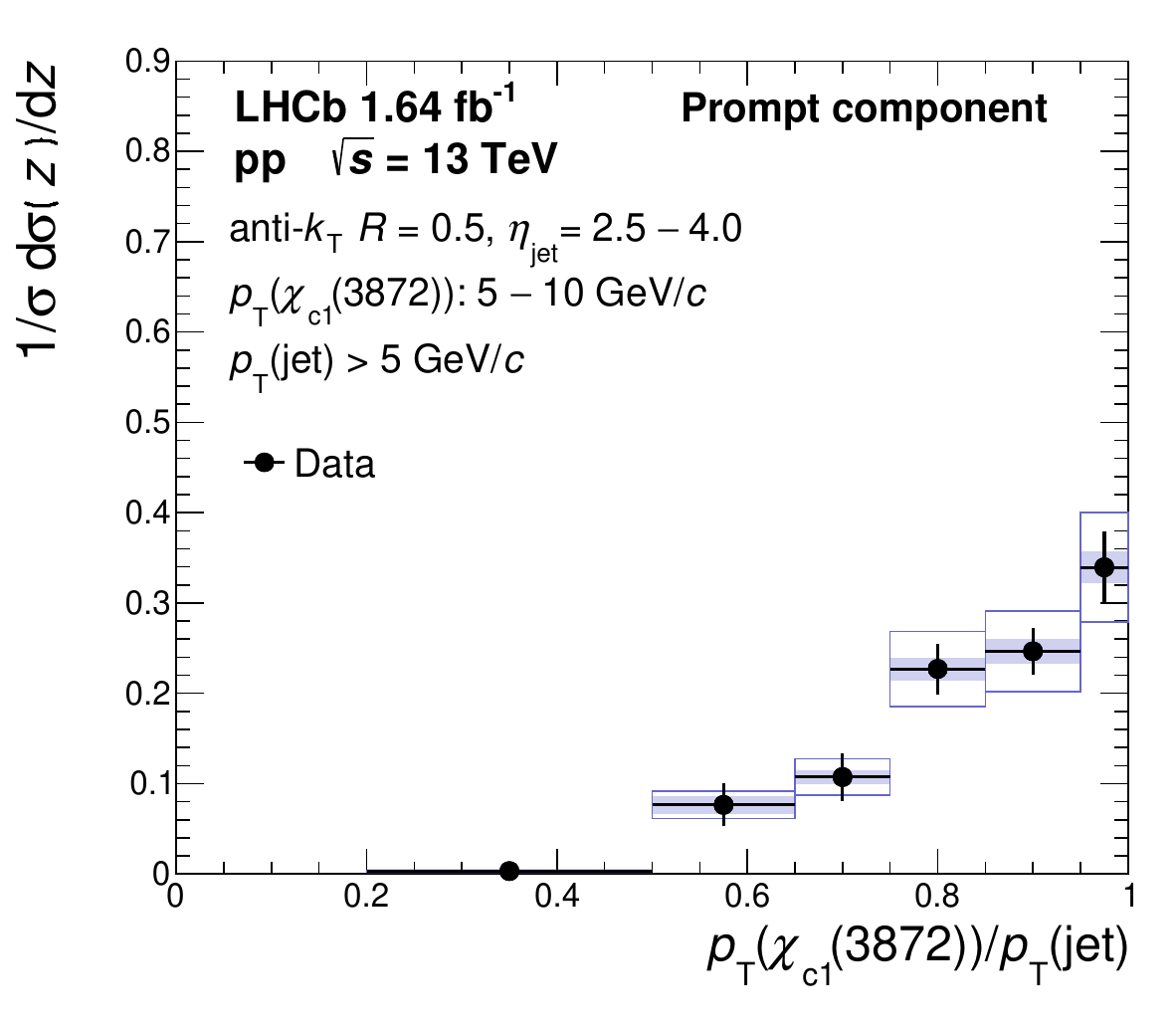}
    \includegraphics[width=0.397\textwidth]{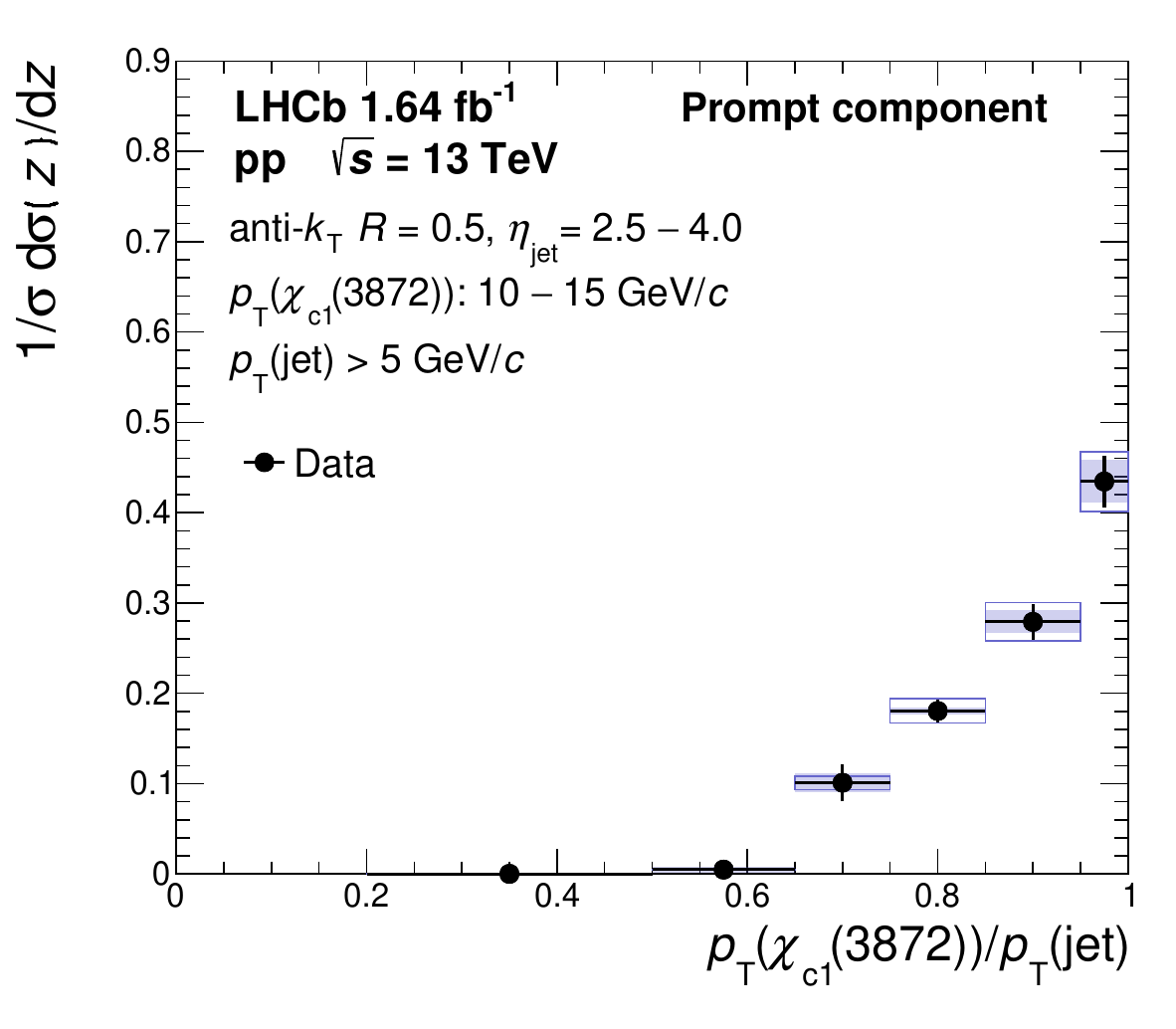}
    \\
    \includegraphics[width=0.397\textwidth]{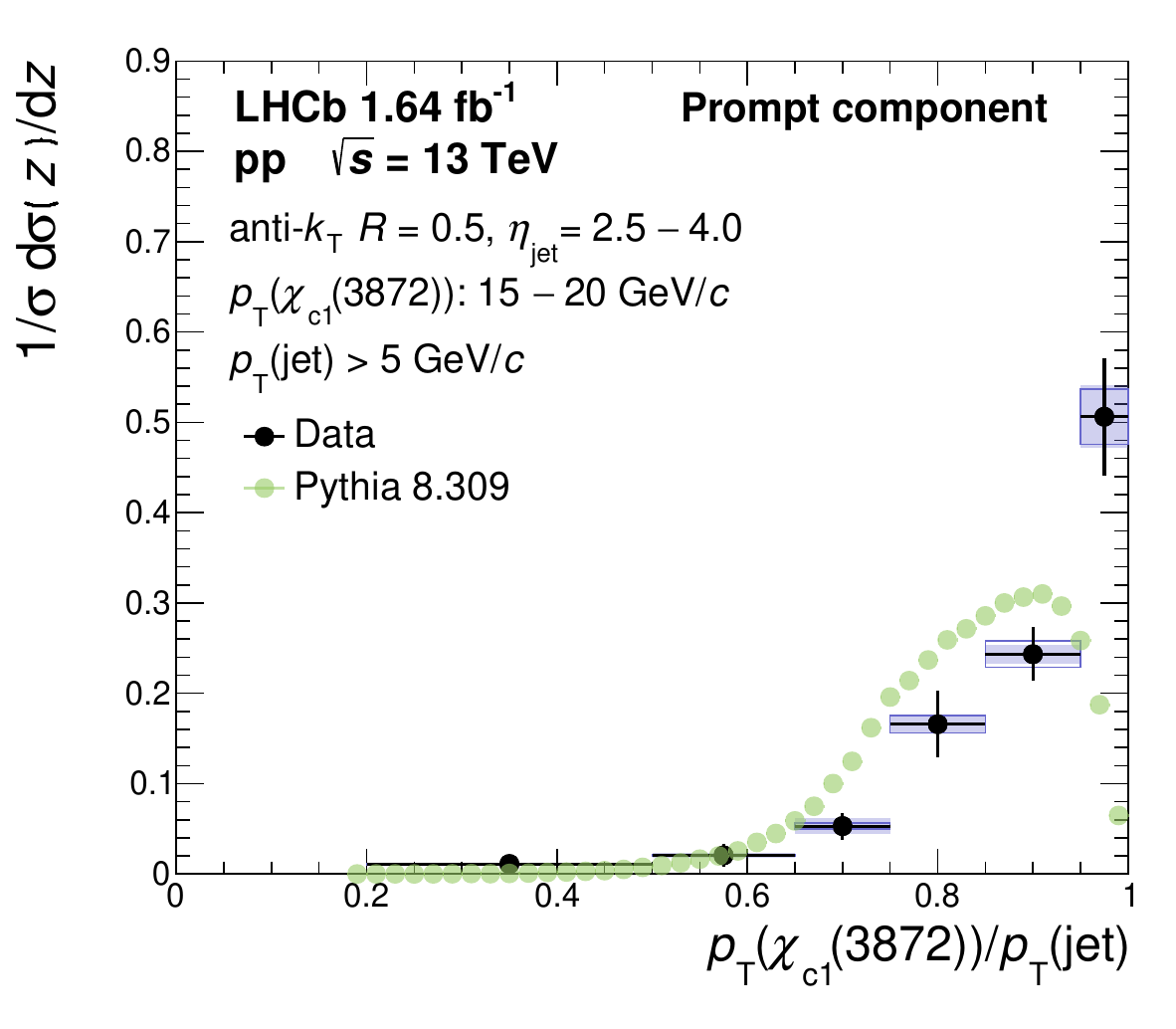}
    \includegraphics[width=0.397\textwidth]{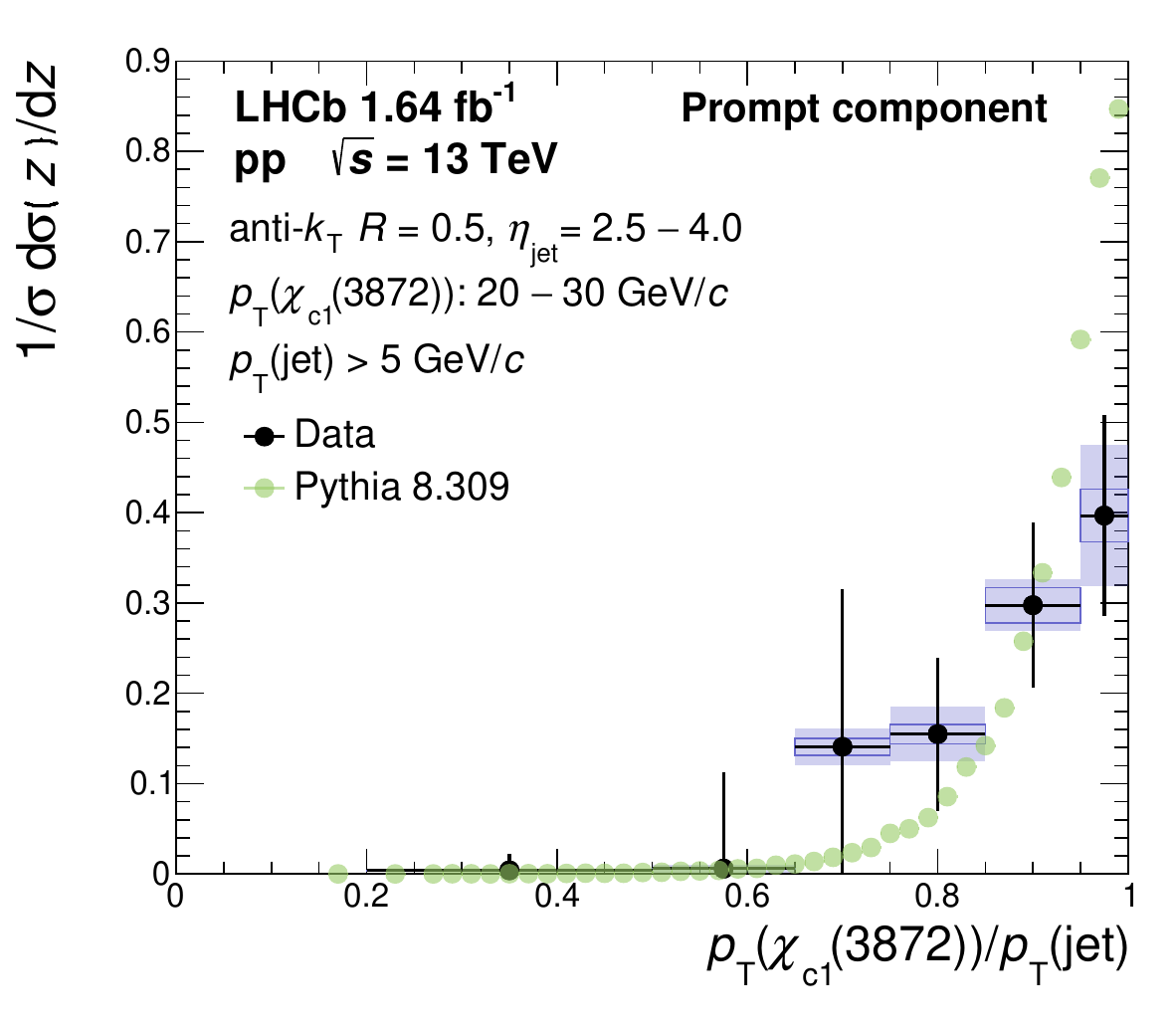}
    \caption{Distributions of the transverse momentum fraction \ensuremath{z} in intervals of \ensuremath{\Pchi_{c1}(3872)} \ensuremath{p_{\mathrm{T}}} for prompt \ensuremath{\Pchi_{c1}(3872)} production with statistical (black error bars), correlated efficiency (empty blue boxes), and uncorrelated unfolding (shaded blue boxes) uncertainties, compared to predictions obtained from \pythia~8.309.\label{fig:Results_X_P_TagpT}}
\end{figure}

A complementary set of results shows the fragmentation function differential in the $\pt(\qtag)$ of the quarkonium. \Cref{fig:Results_P2_NP_TagpT,fig:Results_P2_P_TagpT} show the results for prompt and displaced \psitwos meson production, respectively,  
and \cref{fig:Results_X_NP_TagpT,fig:Results_X_P_TagpT} show the results for prompt and 
displaced \theX meson production, respectively. In these complementary distributions, the \zt numerator, $\pt(\qtag)$, effectively remains fixed while the denominator, $\pt(\jet)$, is no longer limited. This is in contrast to \cref{fig:Results_P2_NP,fig:Results_P2_P,fig:Results_X_NP,fig:Results_X_P}, where the numerator is not limited but the denominator is effectively fixed. This creates a subset of different probed kinematic regimes, as jets softer than the tag \pt window are not present in this representation. 

\section{Conclusions}\label{sec:con}
This paper reports the first study of \psitwos and \theX production in jets. 
The analysis is based on proton-proton collision data collected by the LHCb experiment at the center-of-mass-energy of $13\tev$ in 2016, corresponding to an integrated luminosity of $1.64\invfb$.
The differential distributions of the transverse momentum ratio between the quarkonium and the jet are measured with respect to both the jet and the quarkonium \pt. These distributions are further separated into prompt and displaced quarkonium production. 
\par
The \zt distributions from the displaced production of the \psitwos and \theX mesons, as shown in \cref{fig:Results_P2_NP,fig:Results_X_NP,fig:Results_P2_NP_TagpT,fig:Results_X_NP_TagpT}, fall off rapidly as \zt approaches one. This production of quarkonia originates from the decay of \bquark hadrons and is consequently accompanied by other decay products, which are typically clustered into the jet. 
The depleted yield below $\zt = 0.4$ is in part due to the fiducial limitations of this analysis. These distributions are well described by the \pythia~8.309 event generator, similarly to the results for displaced \jpsi meson production in jets~\cite{LHCb-PAPER-2016-064,CMS:2021puf}.

New insights can be gained from the \zt distributions for prompt production of \psitwos and \theX mesons as they differ strikingly from predictions, as shown in \cref{fig:Results_P2_P,fig:Results_P2_P_TagpT,fig:Results_X_P,fig:Results_X_P_TagpT}. 
For distributions differential in both $\pt(\jet)$ and $\pt(\qtag)$ two components appear to be  visible. This is particularly evident in the \psitwos meson case, which has a larger sample size and thus finer binning than the \theX results. 
The prompt \psitwos distributions reveal a distinct component with $\zt \approx 1$ (isolated production) and another component centered around $\zt \approx 0.7$ (nonisolated production), which is more similar in shape to the displaced production in \cref{fig:Results_P2_NP}. The evolution of this pattern is best visible for the normalized \zt distributions in \ensuremath{p_{\mathrm{T}}}(\psitwos) intervals. Here, the component of nonisolated \psitwos yield is strongest at low \ensuremath{p_{\mathrm{T}}}(\psitwos), while for the highest three \ensuremath{p_{\mathrm{T}}}(\psitwos) intervals, the ratio of nonisolated yield and isolated yield appears to be roughly constant.
Since the composition of different production mechanisms of quarkonia is dependent on the quarkonium \pt, the representation of the fragmentation function in $\pt(\jet)$ intervals is driven by the quarkonium \pt composition in each $\pt(\jet)$ interval, \textit{e.g.} the highest $\pt(\jet)$ interval shows no yield at high \zt as the kinematic reach does not allow measuring \psitwos mesons above 40\gevc with sufficient abundance. 
The \theX results only hint at similar trends, but the sample sizes are too small to make any conclusive statements.
\par
The exact nature of these two components in \zt for prompt \psitwos and \theX production is not clear at this point. The data show, however, for the first time an isolated prompt production of quarkonia as expected from the NRQCD description of quarkonia production in the initial hard scattering. To explore the nature of the second component, which contains a significant fraction of prompt quarkonia production,  further development of theoretical models is required.

\section*{Acknowledgements}
%
%
\noindent We express our gratitude to our colleagues in the CERN
accelerator departments for the excellent performance of the LHC. We
thank the technical and administrative staff at the LHCb
institutes.
We acknowledge support from CERN and from the national agencies:
CAPES, CNPq, FAPERJ and FINEP (Brazil); 
MOST and NSFC (China); 
CNRS/IN2P3 (France); 
BMBF, DFG and MPG (Germany); 
INFN (Italy); 
NWO (Netherlands); 
MNiSW and NCN (Poland); 
MCID/IFA (Romania); 
MICIU and AEI (Spain);
SNSF and SER (Switzerland); 
NASU (Ukraine); 
STFC (United Kingdom); 
DOE NP and NSF (USA).
We acknowledge the computing resources that are provided by CERN, IN2P3
(France), KIT and DESY (Germany), INFN (Italy), SURF (Netherlands),
PIC (Spain), GridPP (United Kingdom), 
CSCS (Switzerland), IFIN-HH (Romania), CBPF (Brazil),
and Polish WLCG (Poland).
We are indebted to the communities behind the multiple open-source
software packages on which we depend.
Individual groups or members have received support from
ARC and ARDC (Australia);
Key Research Program of Frontier Sciences of CAS, CAS PIFI, CAS CCEPP, 
Fundamental Research Funds for the Central Universities, 
and Sci. \& Tech. Program of Guangzhou (China);
Minciencias (Colombia);
EPLANET, Marie Sk\l{}odowska-Curie Actions, ERC and NextGenerationEU (European Union);
A*MIDEX, ANR, IPhU and Labex P2IO, and R\'{e}gion Auvergne-Rh\^{o}ne-Alpes (France);
AvH Foundation (Germany);
ICSC (Italy); 
Severo Ochoa and Mar\'ia de Maeztu Units of Excellence, GVA, XuntaGal, GENCAT, InTalent-Inditex and Prog. ~Atracci\'on Talento CM (Spain);
SRC (Sweden);
the Leverhulme Trust, the Royal Society
 and UKRI (United Kingdom).

\newpage
\addcontentsline{toc}{section}{References}
\bibliographystyle{LHCb}
\bibliography{main,standard,LHCb-PAPER,LHCb-CONF,LHCb-DP,LHCb-TDR}

\newpage
\centerline
{\large\bf LHCb collaboration}
\begin
{flushleft}
\small
R.~Aaij$^{36}$\lhcborcid{0000-0003-0533-1952},
A.S.W.~Abdelmotteleb$^{55}$\lhcborcid{0000-0001-7905-0542},
C.~Abellan~Beteta$^{49}$,
F.~Abudin{\'e}n$^{55}$\lhcborcid{0000-0002-6737-3528},
T.~Ackernley$^{59}$\lhcborcid{0000-0002-5951-3498},
A. A. ~Adefisoye$^{67}$\lhcborcid{0000-0003-2448-1550},
B.~Adeva$^{45}$\lhcborcid{0000-0001-9756-3712},
M.~Adinolfi$^{53}$\lhcborcid{0000-0002-1326-1264},
P.~Adlarson$^{80}$\lhcborcid{0000-0001-6280-3851},
C.~Agapopoulou$^{13}$\lhcborcid{0000-0002-2368-0147}
% ...

{\footnotesize \it

% ...
$^{13}$Universit{\'e} Paris-Saclay, CNRS/IN2P3, IJCLab, Orsay, France\\
% ...
$^{36}$Nikhef National Institute for Subatomic Physics, Amsterdam, Netherlands\\
% ...
$^{45}$Instituto Galego de F{\'\i}sica de Altas Enerx{\'\i}as (IGFAE), Universidade de Santiago de Compostela, Santiago de Compostela, Spain\\
% ...
$^{49}$Physik-Institut, Universit{\"a}t Z{\"u}rich, Z{\"u}rich, Switzerland\\
% ...
$^{53}$H.H. Wills Physics Laboratory, University of Bristol, Bristol, United Kingdom\\
% ...
$^{55}$Department of Physics, University of Warwick, Coventry, United Kingdom\\
% ...
$^{58}$School of Physics and Astronomy, University of Glasgow, Glasgow, United Kingdom\\
$^{59}$Oliver Lodge Laboratory, University of Liverpool, Liverpool, United Kingdom\\
% ...
$^{67}$Syracuse University, Syracuse, NY, United States\\
% ...
$^{80}$Department of Physics and Astronomy, Uppsala University, Uppsala, Sweden, associated to $^{58}$\\
% ...
}
\end{flushleft}

\end{document}